\documentclass[fleqn,usenatbib]{mnras}

\usepackage{newtxtext,newtxmath}

\usepackage[T1]{fontenc}

\DeclareRobustCommand{\VAN}[3]{#2}
\let\VANthebibliography\thebibliography
\def\thebibliography{\DeclareRobustCommand{\VAN}[3]{##3}\VANthebibliography}

\usepackage{graphicx}	
\usepackage{amsmath}	
\usepackage {threeparttable}
\usepackage{siunitx}
\usepackage{xcolor}
\usepackage{url} 
\usepackage{ulem}
\usepackage{tabularx}

\title[ALMA observations of $z\sim10$ candidates]{ALMA observations of pre-\textit{JWST} $z\sim10$ galaxy candidates: A CO($J=9\text{--}8$) line from a ULIRG at $z=2.54$ and revisit of the photometric redshifts with \textit{JWST} photometry}

\author[S. Arai et al.]{
Suzuka Arai$^{1}$\thanks{E-mail: suzukaarai@ruri.waseda.jp},
Yuma Sugahara$^{1,2}$,
Akio K. Inoue$^{1,2}$,
Takuya Hashimoto$^{3,4}$,
Ken Mawatari$^{1,2}$,\newauthor{}
Yi W. Ren$^{1}$,
Steven L. Finkelstein$^{5,6}$,
John R. Weaver$^{7}$\thanks{Brinson Prize Fellow},
Rebecca L. Larson$^{8}$,
Seiji Fujimoto$^{9,10}$,\newauthor{}
Yuichi Harikane$^{11}$,
Takahiro Morishita$^{12,13}$,
Yoichi Tamura$^{14}$,
Andreas Faisst$^{12}$,
Charles Steinhardt$^{15}$,\newauthor{}
Nima Chartab$^{12}$,
Larry D. Bradley$^{8}$,
and David B. Sanders$^{16}$
\\
$^{1}$Department of Pure and Applied Physics, School of Advanced Science and Engineering, Faculty of Science and Engineering, Waseda University,\\ 3-4-1 Okubo, Shinjuku, Tokyo 169-8555, Japan\\
$^{2}$Waseda Research Institute for Science and Engineering, Faculty of Science and Engineering, Waseda University, 3-4-1 Okubo, Shinjuku,\\ Tokyo 169-8555, Japan
\\
$^{3}$ Division of Physics, Faculty of Pure and Applied Sciences, University of Tsukuba, Tsukuba, Ibaraki 305-8571, Japan\\
$^{4}$ Tomonaga Center for the History of the Universe (TCHoU), Faculty of Pure and Applied Sciences, University of Tsukuba, Tsukuba, Ibaraki 305-8571, Japan\\
$^{5}$ Department of Astronomy, The University of Texas at Austin, Austin, TX, USA\\
$^{6}$ Cosmic Frontier Center, The University of Texas at Austin, Austin, TX, USA\\
$^{7}$ MIT Kavli Institute for Astrophysics and Space Research, 70 Vassar Street, Cambridge, MA 02139, USA\\
$^{8}$ Space Telescope Science Institute, 3700 San Martin Drive, Baltimore, MD 21218, USA\\
$^{9}$ David A. Dunlap Department of Astronomy and Astrophysics, University of Toronto, 50 St. George Street, Toronto, Ontario, M5S 3H4, Canada\\
$^{10}$ Dunlap Institute for Astronomy and Astrophysics, 50 St. George Street, Toronto, Ontario, M5S 3H4, Canada\\
$^{11}$ Institute for Cosmic Ray Research, The University of Tokyo, 5-1-5 Kashiwanoha, Kashiwa, Chiba 277-8582, Japan\\
$^{12}$ IPAC, California Institute of Technology, MC 314-6, 1200 E. California Boulevard, Pasadena, CA 91125, USA\\
$^{13}$ Astronomical Institute, Tohoku University, 6-3 Aramaki, Aoba-ku, Sendai 980-8578, Japan\\
$^{14}$ Department of Physics, Graduate School of Science, Nagoya University, Furo, Chikusa, Nagoya, Aichi 464-8602, Japan\\
$^{15}$ Department of Physics and Astronomy, University of Missouri, 701 S. College Ave., Columbia, MO 65203\\
$^{16}$ Institute for Astronomy, University of Hawaii at Manoa, 2680 Woodlawn Dr., Honolulu, HI  906822, USA\\
}

\date{Accepted XXX. Received YYY; in original form ZZZ}

\pubyear{2026}

\begin{document}
\label{firstpage}
\pagerange{\pageref{firstpage}--\pageref{lastpage}}
\maketitle

\begin{abstract}
We present Atacama Large Millimetre/submillimetre Array (ALMA) observations targeting the [\ion{O}{iii}]\,$88\,\mu$m line for six $z\sim10$ galaxy candidates selected with the \textit{Hubble Space Telescope} and the \textit{Spitzer Space Telescope}.
We detect a line ($4.5\sigma$) and dust continuum emission ($30\sigma$) in UDS\_18697, while detecting neither robust line nor continuum emission in the remaining five objects.
The detected line in UDS\_18697 is identified as CO($J=9\text{--}8$), because follow-up \textit{James Webb Space Telescope} (\textit{JWST}) NIRSpec observations have confirmed the redshift as $z=2.54$.
UDS\_18697 is classified as an ultra luminous infrared galaxy (ULIRG) with far-infrared (FIR) luminosity of $L_\mathrm{FIR}\approx1.1\times10^{12}\,L_\odot$, assuming a dust temperature of $T_\mathrm{d}\approx42\,$K, estimated using a physically-motivated method.
We find that UDS\_18697 follows the $L_\mathrm{FIR}-L'_\mathrm{CO}$ relation for local and $z>2$ galaxies, albeit being slightly brighter in CO($J=9\text{--}8$).
Also, based on the follow-up NIRSpec observations and spectral energy distribution fitting using \textit{JWST}/NIRCam photometry, we found that most of our targets are suggested to be low-$z$ interlopers.
Motivated by these redshift misclassifications, we investigate colour--colour selection criteria for high-$z$ galaxies using \textit{JWST} spectroscopic survey catalogues.
We find that elevating a colour threshold tracing the Lyman break is crucial for constructing a robust high-$z$ sample, particularly for wide field surveys such as Euclid Deep Fields and Roman High-Latitude Wide-Area Survey. 
\end{abstract}




\section{Introduction}
Investigating galaxies in the Epoch of Reionisation (EoR) is key to understanding galaxy formation and cosmic reionisation.
The first step of the EoR investigation is to select high-$z$ galaxies from numerous objects appearing in imaging data, requiring photometric techniques that effectively capture galaxy spectral features.
One of the most common methods to select high-$z$ galaxy candidates is the so-called "Lyman break method" \citep{Steidel92}, which makes use of a sharp spectral drop at the Lyman limit ($\lambda_\mathrm{rest}=912\,$\AA) due to the high opacity of neutral interstellar medium (ISM) to hydrogen ionising photons within a galaxy.
This method can also capture another strong break at $\lambda_\text{rest}=1216\,$\AA, which corresponds to the edge of the Gunn-Peterson Trough.
This feature originates from the Ly~$\alpha$ absorption by neutral hydrogen in the intergalactic medium (IGM), emerging at $z\gtrsim6$.
\par

Over the past few decades, the Lyman break method has been used for extending the most distant redshift with a combination of deep imaging of space telescopes and ground-based telescopes.
The Subaru telescope has found numerous Lyman break galaxies (LBGs) with its exceptionally wide field of view.
The Great Optically Luminous Dropout Research Using Subaru HSC \citep[GOLDRUSH;][]{Ono18}, for instance, identified approximately 4 million LBGs at $z\sim4\text{--}6$ over $300\,\mathrm{deg}^2$ sky \citep{Harikane22a}.
The \textit{Hubble Space Telescope} (\textit{HST}) has dramatically improved our understanding of the early Universe by identifying LBGs up to $z\sim10$ with its unprecedented depth at near-infrared (NIR) wavelengths \citep[e.g.,][see also \citealt{Stark16} for a review]{Zheng12,Oesch14,Schmidt14,Bouwens15,Finkelstein15,Ishigaki15,Kawamata16}.
With the \textit{HST} filter combination, the accessible photometric redshift (photo-$z$) was limited up to $z\sim10$, where the Ly~$\alpha$ break is redshifted to $\sim1.3\,\mu$m.
The advent of the \textit{James Webb Space Telescope} (\textit{JWST}) broke this redshift barrier.
With its superb sensitivity and wavelength coverage spanning from optical to infrared, \textit{JWST} has revealed a large number of galaxies at $z\gtrsim9$ within the first year of its operation (e.g., \citealt{Castellano22, Naidu22-LBG,Finkelstein22b, Adams23,Atek23,donnan23,finkelstein23,Harikane23,Morishita23}).
Many subsequent efforts have been made to identify more distant galaxies and to advance the observational frontier towards the first galaxies (e.g., \citealt{Willott24,castellano25,Gandolfi25,Harikane25,Perez-Gonzalez25}).
\par

Although the Lyman break method has successfully identified numerous high-$z$ galaxies, it occasionally allows low-$z$ contaminants to be selected as high-$z$ candidates because various types of galaxies can make similar dropout features \citep{Rodighiero23}.
Major contaminants are quiescent galaxies with a prominent Balmer break that can be erroneously regarded as a Lyman break \citep[e.g.,][]{sato24,Harikane25} and dusty star-forming galaxies (DSFGs) that have strongly attenuated continua from ultraviolet (UV) to optical, resulting in non-detections in shorter-wavelength filters \citep[e.g.,][]{Zavala23}.
Another source of contamination is galaxies with extremely strong emission lines (extreme emission line galaxies; EELGs), which can mimic LBG-like spectral energy distributions (SEDs) based on photometry by boosting broadband flux densities \citep[e.g.,][]{arrabalharo23, Gandolfi25}.
A recent study making use of medium bands indicated that including medium band photometry can efficiently remove low-$z$ EELGs \citep{Adams25,Asada25}.
Certainly, the contamination fraction strongly depends on the adopted selection criteria and field-to-field variations, but a high contamination rate resulting from broad band-only selection can lead to an overestimation of $\sim0.6$ dex in the UV luminosity function at $z\sim15$ \citep{Asada25}.
Given this situation, spectroscopic follow-up observations are necessary to discuss properties of high-$z$ galaxies.
\par

Before the completion of cosmic reionisation, the Ly~$\alpha$ emission line, which has been used to determine spectroscopic redshift of $z>3$ galaxies, is absorbed by neutral hydrogen in the IGM, making the line detection difficult.
Therefore, the spectroscopic confirmation of the Ly~$\alpha$ line with ground-based 8--10\,m telescopes was expensive in the pre-\textit{JWST} era \citep[e.g.,][]{Vanzella11,Zitrin15}.
Alternatively, far infrared (FIR) fine structure lines such as [\ion{C}{ii}]$158\,\mu$m and [\ion{O}{iii}]$88\,\mu$m have been used to spectroscopically confirm redshifts of $z\gtrsim6$ galaxies.
The Atacama Large Millimeter/submillimeter Array (ALMA) offers us high sensitivity and a unique spectral window to detect FIR fine structure lines in galaxies in the EoR.
Consistent with a theoretical prediction that the [\ion{O}{iii}]$\,88\,\mu$m line is presumably the brightest emission line tracing \ion{H}{ii} regions at $z\gtrsim7$ \citep{Inoue14}, the line was successfully detected from a galaxy at $z=7.21$ \citep{Inoue16}.
Since then, [\ion{O}{iii}]$\,88\,\mu$m observations have spectroscopically confirmed galaxies in the EoR \citep[e.g.,][]{Carniani17,Hashimoto18, Marrone18,Tamura19, Hashimoto19,Hashimoto19b,Harikane20,Wong22, Witstok22,Ren23,Algera24, Algera25b,Knudsen25}.
Furthermore, the [\ion{O}{iii}]\,$88\,\mu$m line has also been detected even in \textit{JWST}-discovered $z>10$ galaxies such as GHZ2 \citep{zavala24}, GS-z11-0 \citep{witstok25}, and GS-z14-0 \citep{schouws25,carniani25}.
\par

In this paper, we present ALMA follow-up observations of the [\ion{O}{iii}]\,$88\,\mu$m line for $z\sim10$ candidate galaxies selected with the \textit{\textit{Spitzer} Space Telescope} (\textit{\textit{Spitzer}}), \textit{\textit{HST}}, and ground-based telescopes.
This paper is organised as follows:
Section~\ref{sec:data} describes our sample, ALMA observations, and photometry.
ALMA data analysis, its results, and investigation of physical properties based on the detected lines are presented in Section~\ref{sec:analysis&results}.
In Section~\ref{sec:SEDfitting}, we perform SED fitting to investigate properties of our targets.
We discuss the effect of photometry on an LBG selection and the accuracy of colour--colour selection in Section~\ref{sec:discussion}, and we summarise the paper in Section~\ref{sec:conclusion}.
Throughout this paper, we adopt the flat $\Lambda$CDM cosmology with parameters $\Omega_{\mathrm{M}} =0.310$, $\Omega_{\Lambda}=1-\Omega_\mathrm{M}$ and $H_0=67.7\,{\mathrm{km\,s^{-1}\,Mpc^{-1}}}$ \citep{planck20}.
All magnitudes are provided in the AB system \citep{oke83}.

\section{Data}
\label{sec:data}
\subsection{Target Selection}
\label{sec:samples}
We constructed a sample of six $z\sim10$ galaxy candidates for ALMA follow-up observations of the [\ion{O}{iii}]\,$88\,\mu$m line, aiming at spectroscopic confirmation of their redshifts.
We selected the six targets from $z\sim10$ candidates presented in the literature:  the Cosmic Assembly Near-Infrared Deep Extragalactic Legacy Survey \citep[CANDELS;][]{finkelstein22}, the Brightest of Reionizing Galaxies Survey (BoRG[z9]; \citealp{morishita18, Bridge19}), \textit{HST} Frontier Fields \citep[HFFs:][]{Laporte15,Oesch18}, COSMOS Ultra VISTA Field \citep{Stefanon19}, Reionization Lensing Cluster Survey \citep[RELICS;][]{Salmon18}, and Cosmic Evolution Survey photometric catalogue \citep[COSMOS2020;][]{weaver22}.
\par

For our sample selection, we used the following criteria for the literature other than the COSMOS2020 catalogue: (1) a photometric redshift of $z_{\mathrm{phot}}>9.0$ with a $68\%$ uncertainty smaller than $\pm0.5$, (2) the existences of observations in at least one of \textit{Spitzer}/IRAC ch1 ([3.6]) and ch2 ([4.5]) filters, (3) no strong IRAC detection, to reduce the possibility of low-redshift contamination, and (4) \textit{HST} F160W magnitude of $H_{\mathrm{160}}<25.5\,\mathrm{mag}$.
These criteria were employed to exclude low-$z$ interlopers and to detect the [\ion{O}{iii}]\,$88\,\mu$m line in a reasonable integration time.
With these criteria, four galaxies, UDS\_18697\footnote{\citet{Boueens19} flagged UDS\_18697 (which they referred to as UDS910-13) as a $z<9$ source based on a $4\sigma$ detection in the ground-based FourStar $J2$ band.}, COSMOS\_20646, UDS\_7815\footnote{UDS\_7815 was not included in our initial sample. After the proposal was accepted, however, one of the original targets was found to be likely at low-$z$ and was therefore replaced with this galaxy.}, and J2140+0241, were found from \citet{finkelstein22} and \citet{morishita18}.
Note that \cite{finkelstein22} excluded UDS\_7815 from their final $z\gtrsim9$ candidates due to a $2\sigma$ detection in the \textit{HST}/WFC3 F098M filter, implying that this object is at $z\lesssim7$.
However, the galaxy remained in our ALMA observation targets because the follow-up F098M observation of \citet{finkelstein22} was actually conducted during \textit{HST} Cycle 27, which overlapped with our ALMA observation of the object.
From the COSMOS2020 catalogue\footnote{Note that the COSMOS2020 catalogue does not include \textit{HST} filters other than \textit{HST}/ACS F814W filter.} \citep{weaver22}, targets were selected by the following criteria: (1) $9<z_{\mathrm{phot}}<10$, (2) no significant detection blueward of $1.2\,\mu$m, and (3) no significant detection at $24\,\mu$m.
These criteria ensure that galaxies drop blueward of the Lyman break at $z\sim9\text{--}10$ and are not likely to be a low-$z$ dusty galaxy.
These criteria were satisfied by two galaxies, COSMOS-z10-1 and COSMOS-z10-2.
In summary, the six galaxies at $z_{\mathrm{phot}} \sim 10$ have been selected for our sample. The target information is listed in Table~\ref{tab:observation & target}.
With this procedure, we constructed the most promising sample of $z\sim10$ galaxy candidates from the highest quality imaging data obtained before the emergence of \textit{JWST}.

\subsection{ALMA Observation}
\label{sec:obervation}
The ALMA Band 7 observations targeting the [\ion{O}{iii}]\,$88\,\mu$m line at $z\sim10$ were carried out during Cycles 7 and 8 (\#2019.1.00397.S, \#2021.1.00389.S, PI: T. Hashimoto), investing in total of $\sim14$ hours of on source time.
Observation dates, central frequencies, on-source times, and array configurations are shown in Table~\ref{tab:observation & target}.
\par

In Cycle 7, UDS\_18697, COSMOS\_20646, UDS\_7815, and J2140+0241 were observed using four tuning sets to search for the line across a wide frequency range.
Each tuning set had four Spectral Windows (SPWs) with a $1.875\,{\mathrm {GHz}}$ bandwidth of the frequency division mode (FDM), giving a $\sim27\,{\mathrm {GHz}}$ bandwidth in total.
This corresponds to a redshift range of $\Delta z \sim0.8\text{--} 0.9$ at $z\sim10$, covering $1\sigma$ uncertainties of their photometric redshift. 
An exception is UDS\_7815, where a large photo-$z$ uncertainty ($\Delta z=1.53$) reduces a covering fraction to 60.9\%.
Compact array configurations (C43-2, C43-3, and C43-4), yielding an average beam size of $0\farcs48\times0\farcs42$, were used to cover the entire emission from the targets rather than to resolve them, because the main objective of the observations was to detect the [\ion{O}{iii}]\,$88\,\mu$m line.
\par
In Cycle 8, observations of COSMOS-z10-1 and COSMOS-z10-2 were conducted using six tuning sets for each object.
All SPWs were set to have $1.875\,{\mathrm {GHz}}$ of the FDM, resulting in a total bandwidth of $\sim34\,{\mathrm {GHz}}$.
This frequency range corresponds to a redshift range of $\Delta z\approx1.10\text{--}1.20$, covering $\sim70\%$ and $\sim80\%$ of the entire photo-$z$ probability distribution of COSMOS-z10-1 and COSMOS-z10-2, respectively. 
The C43-2 and C43-3 array configurations were used, yielding a beam size of $0\farcs74\times0\farcs66$.
Note that the large atmospheric absorption at $323-327\,\mathrm{GHz}$ is not taken into account to calculate the photo-$z$ covering fractions for both Cycles 7 and 8 observations.
The photo-$z$ range and frequency coverage for all targets are presented in Appendix~\ref{appendix:full data}.

\begin{table*}
\caption{Summary of ALMA Observations}
\label{tab:observation & target}
\renewcommand{\arraystretch}{1.2}
    \begin{tabular}{lccccccccc}
    \hline
    $^\blacklozenge$Object & R.A. & Dec.  & $^{\star}M_{\mathrm{UV}}$ & $z_{\mathrm{phot}}$ & $^\ddagger$Ref. & Date & Cent. Freq. & Time & Array Config.\\
     & & & [mag] & &  & [YYYY/MM/DD] & [GHz] & [min.]\\ 
    \hline
    \multicolumn{10}{c}{Cycle 7 (2019.1.00397.S)} \\
    UDS\_18697 & $34.255616$ & $-5.166554$ & $-22.25^{+0.12}_{-0.10}$ & $^\dagger9.89^{+0.16}_{-0.15}$ & M24, F22 & 2019/10/04-10/11 & $304.60$ &  $108.8$ & C43-2, C43-3, C43-4\\ 
    COSMOS\_20646 & $150.081822$ & $2.262757$ & $-22.26^{+0.29}_{-0.04}$ & $^\dagger9.80^{+0.10}_{-0.46}$ &  M24, F22 & 2019/10/06-10/17 & $319.91$ &  $155.3$ & C43-2, C43-3, C43-4\\ 
    UDS\_7815 & $34.392791$ & $-5.259930$ & $-21.71^{+0.15}_{-0.15}$ & $^\dagger10.03^{+0.98}_{-0.55}$ &  M24, F22 & 2019/10/30-11/07 & $309.46$ &  $124.1$ & C43-2\\ 
    J2140+0241 & $324.885438$ & $2.685170$ & $-22.6^{+0.4}_{-0.2}$ & $^\Diamond9.99^{+0.41}_{-0.25}$ & M18 & 2019/10/07-10/10 & $308.20$ & $103.2$ & C43-2, C43-3, C43-4\\ \hline
    \multicolumn{10}{c}{Cycle 8 (2021.1.00389.S)} \\
    COSMOS-z10-1 & $149.356072$ & $2.711451$ & $-22.82^{+0.06}_{-0.06}$ & $^\dagger9.82^{+0.13}_{-0.11}$ & W22 & 2022/04/02-05/14 & $312.33$ &  $170.7$ & C43-2, C43-3\\
    COSMOS-z10-2 & $149.412575$ & $1.844460$ &$-22.60^{+0.12}_{-0.12}$ & $^\dagger9.56^{+0.12}_{-0.14}$ & W22 & 2022/04/02-09/22 & $325.46$ &  $174.8$ & C43-2, C43-3\\ \hline
    \end{tabular}
    \par\raggedright
    $\blacklozenge$: UDS\_18697 and J2140+0241 are magnified by a factor of $\mu=1.35$ and $1.5^{+0.7}_{-0.3}$, respectively.
    $\star$: Calculated based on $z_{\mathrm{phot}}$.
    $\dagger$: Follow-up spectroscopic observations and/or SED fitting with a new NIRCam photometry suggest redshifts of $z\sim2\text{--}3$.
    $\Diamond$: As reported by \citet{morishita18}, the SED fitting estimated a $\sim35\%$ probability of being at $z<7$. The uncertainties and the value listed here correspond to the 16th, 50th, and 84th percentiles of the probability distribution of $z>7$ solution.
    $\ddagger$: References, M18: \citet{morishita18}, F22: \citet{finkelstein22}, W22: \citet{weaver22}, M24: \citet{merlin24}.
\end{table*}

\subsection{Optical and Near Infrared Photometric Data}
\label{sec: photometry}
For the SED fitting analysis, we used available \textit{JWST}/NIRCam and \textit{HST} photometric data from the ASTRODEEP-JWST catalogue, which contains objects detected in stack images of NIRCam F356W and F444W in the well-known extragalactic fields \citep{merlin24}.
The catalogue provides photometric data of four \textit{HST}/ACS filters (F435W, F606W, F775W, F814W), four \textit{HST}/WFC3 filters (F105W, F125W, F140W, F160W), and eight \textit{JWST}/NIRCam filters (F090W, F115W, F150W, F200W, F277W, F356W, F410M, F444W).
We searched for catalogue counterparts within 1 arcsec from the target galaxies\footnote{ASTRODEEP-JWST catalogue coordinates are aligned to Gaia stars.}. 
Out of our six target galaxies, UDS\_18697, COSMOS\_20646, and UDS\_7815 were found in the catalogue
with offsets of 0.20, 0.13, and 0.09 arcsec, respectively, from positions reported in \citet{finkelstein22}.
The others were not found just because they have not been observed with NIRCam.
Regarding COSMOS-z10-1 and COSMOS-z10-2, we also checked the COSMOS2025 catalogue \citep{cosmos-web}.
However, neither is included as both objects lie outside the NIRCam footprint.
Table~\ref{tab:photometry} shows the flux densities of the three objects.

\begin{table}
    \caption{Photometry from the ASTRODEEP-JWST catalogue.}
    \label{tab:photometry}
    \begin{flushleft}
    \renewcommand{\arraystretch}{1.2}
    \begin{tabular}{lccc}
    \hline
         Filter & UDS\_18697 & COSMOS\_20646 & UDS\_7815 \\
         \hline
         F435W & $-4.49\pm18.75$ & --- & $64.94\pm37.51$ \\
         F606W & $-3.72\pm9.10$ & $3.73\pm5.59$ & $52.52\pm17.74$ \\
         F814W & $2.64\pm9.92$ & $-2.58\pm5.48$ & $60.52\pm17.63$ \\
         F125W & $45.53\pm14.41$ & $69.94\pm15.61$ & $117.28\pm30.62$ \\
         F140W & $98.20\pm32.90$ & $133.43\pm13.07$ & $201.74\pm26.49$ \\
         F160W & $177.11\pm21.06$ & $221.08\pm19.96$ & $281.81\pm26.62$ \\\hline
         F090W & $21.71\pm12.54$ & $-4.77\pm7.19$ & $76.47\pm19.81$ \\
         F115W & $32.50\pm12.01$ & $44.22\pm6.79$ & $76.61\pm19.82$ \\
         F150W & $141.34\pm11.44$ & $173.93\pm6.24$ & $194.70\pm16.86$ \\
         F200W & $386.98\pm9.24$ & $473.27\pm5.77$ & $219.33\pm14.47$ \\
         F277W & $806.77\pm10.16$ & $851.93\pm4.72$ & $233.64\pm10.38$ \\
         F356W & $1502.35\pm11.22$ & $1117.49\pm4.57$ & $234.40\pm9.61$ \\
         F410M & $1952.90\pm18.81$ & $1385.44\pm3.50$ & $238.11\pm16.87$ \\
         F444W & $2313.09\pm12.70$ & $1544.05\pm2.70$ & $263.46\pm12.64$ \\
         \hline
    \end{tabular}
    \par\raggedright
    \textit{Note:} Flux densities are given in $\mathrm{nJy}$.
\end{flushleft}
\end{table}

\section{Analyses \& Results}
\label{sec:analysis&results}
\subsection{ALMA Data Analysis}
The ALMA data were reduced and calibrated using the Common Astronomy Software Applications (CASA) package \citep{mullin07} pipeline versions 5.6.1-8 (UDS\_18697, COSMOS\_20646, UDS\_7815, and J2140+0241), 6.2.1.7 (COSMOS-z10-1 and COSMOS-z10-2), and 6.4.1.12 (one observation of COSMOS-z10-2).
\par

To search for emission lines, we first created dirty cubes using the CASA task \texttt{tclean}. 
Spectral binning was performed to resample six native channels into one, resulting in a velocity width of $\sim 44$--$46\,{\mathrm {km\,s^{-1}}}$ per channel. 
We used natural weighting to maximise sensitivity for a point source.
For a continuum-detected source, we subtracted it to make a pure line cube using the CASA task \texttt{uvcontsub}, assuming that the continuum is constant over the observed frequency range.
We then created clean cubes for objects with line emissions down to triple the mean of the root mean square (rms) level of each cube.
During the clean procedure, targets were masked in cases with nearby objects to avoid misidentification of sidelobes as real objects.
For objects exhibiting an emission line, moment-0 maps of the line were created with the CASA task \texttt{immoments}, and line fluxes were then measured by applying the CASA task \texttt{imfit} to the moment-0 maps.
\par

For dust continuum emission, we created continuum images using the CASA task \texttt{tclean} with natural weighting.
For objects with line detections, we excluded only channels within $\sim2$ FWHM of the line rather than excluding whole SPWs that contain the line.
The continuum flux density was measured on the continuum image with the CASA task \texttt{imfit}.
Table~\ref{tab: cube properties} shows the beam sizes and sensitivities of the created cubes and continuum images.

\subsection{ALMA Data Results}
In our ALMA data, we found two possible emission lines in UDS\_18697 ($4.46\sigma$) and COSMOS-z10-2 ($4.56\sigma$).
UDS\_18697 also exhibits strong dust continuum emission ($29.9\sigma$).
Neither emission line nor dust continuum emission was found in the other four objects.
Measurements of the line and the continuum, including $3\sigma$ upper limits, are summarised in Table~\ref{tab:ALMA measurments}.
In the following, we describe the ALMA observational results in detail.
\par

\subsubsection{UDS\_18697}\label{sec:UDS}
UDS\_18697 exhibits an emission line around $293.3\,\mathrm{GHz}$.
The moment-0 map is shown in the top right panel of Figure~\ref{fig:continuum&moment-0}, and a 1D spectrum taken at the peak of the moment-0 map is shown in the top panel of Figure~\ref{fig:spectrum}.
The integration frequency range for the moment-0 map is 292.90–293.70 GHz, corresponding to twice the full-width-at-half-maximum (FWHM) of the line.
We obtained the line flux of $0.397\pm0.139\,{\mathrm {Jy\,km\,s^{-1}}}$ and a peak $S/N$ of $4.46$ on the moment-0 map, which means a marginal detection of the line.
The reliability of the line based on clump search analysis is discussed in Appendix~\ref{appendix:line search}.
The line centre frequency and the line FWHM measured from the 1D spectrum are $293.27\pm0.03\,\mathrm{GHz}$ and $399.48\pm 41.05\,\mathrm{km\,s^{-1}}$, respectively.
\par

The obtained line centre frequency would correspond to a redshift of $z=10.57$ if the line were [\ion{O}{iii}]\,$88\,\mu$m.
After our ALMA observations, however, follow-up \textit{JWST}/NIRSpec PRISM spectroscopy (GO1758, PI: S. Finkelstein) have detected multiple optical emission lines to reveal that this object is at $z=2.54$ (R. Larson et al. in prep.).
Hence, we identify the marginally detected line as CO($J=9\text{--}8$), rather than [\ion{O}{iii}]\,$88\,\mu$m.
In this case, the redshift indicated by the emission line is $z_{\mathrm{CO}}=2.5356\pm0.0004$\, which is consistent with the \textit{JWST}/NIRSpec observations (R. Larson et al. in prep.).
Although the detection remains marginal, given that the line is co-spatial with the position of UDS\_18697 and consistent with the redshift from PRISM spectroscopy, we consider that the line is not likely to be a spurious.
\par

UDS\_18697 is also significantly detected in the dust continuum emission.
The top left panel of Figure~\ref{fig:continuum&moment-0} shows the dust continuum image.
The flux density is $368\pm 19\,{\mathrm{\mu Jy}}$ with the peak significance level of $29.9\sigma$.
Since the size of the continuum emitting region is larger than the image beam size ($0\farcs42\times0\farcs37$), we deconvolved the beam from the continuum emission with \texttt{imfit}, and derived the beam-deconvolved size of $\approx0\farcs214\times0\farcs144$, corresponding to $1.76\,\mathrm{kpc}$ $\times$ $1.19\,\mathrm{kpc}$ at $z=2.54$. 
Note that the cosmic microwave background (CMB) has a negligible effect on both the line and continuum emission at this redshift and frequency \citep{daCunha13}.
\par

We found slight spatial offsets between the dust and CO line emission peaks and source position in the ASTRODEEP-JWST catalogue, which is measured in a NIRCam F356W+F444W stacked image.
The offsets are 0.16\,arcsec for the CO line and 0.19\,arcsec for the dust continuum, both of which are larger than an expected ALMA positioning accuracy of 0.10\,arcsec.
Such spatial offsets between centroid of UV/optical emission and that of dust and CO emission have been reported for submillimetre galaxies (SMGs) at $z\sim3$ \citep[e.g.,][]{Chen2015,Chen2017,CalistroRivera2018}.
These offsets are likely to be a consequence of strong dust attenuation that affects UV/optical morphologies rather than physical separation \citep{Hodge2025}.
Given that the significant dust continuum emission is detected in UDS\_18697, the observed offsets may reflect the dust obscuration.

\subsubsection{COSMOS-z10-2}
\begin{figure}
\begin{tabular}{c}
    \includegraphics[width=1.0\linewidth]{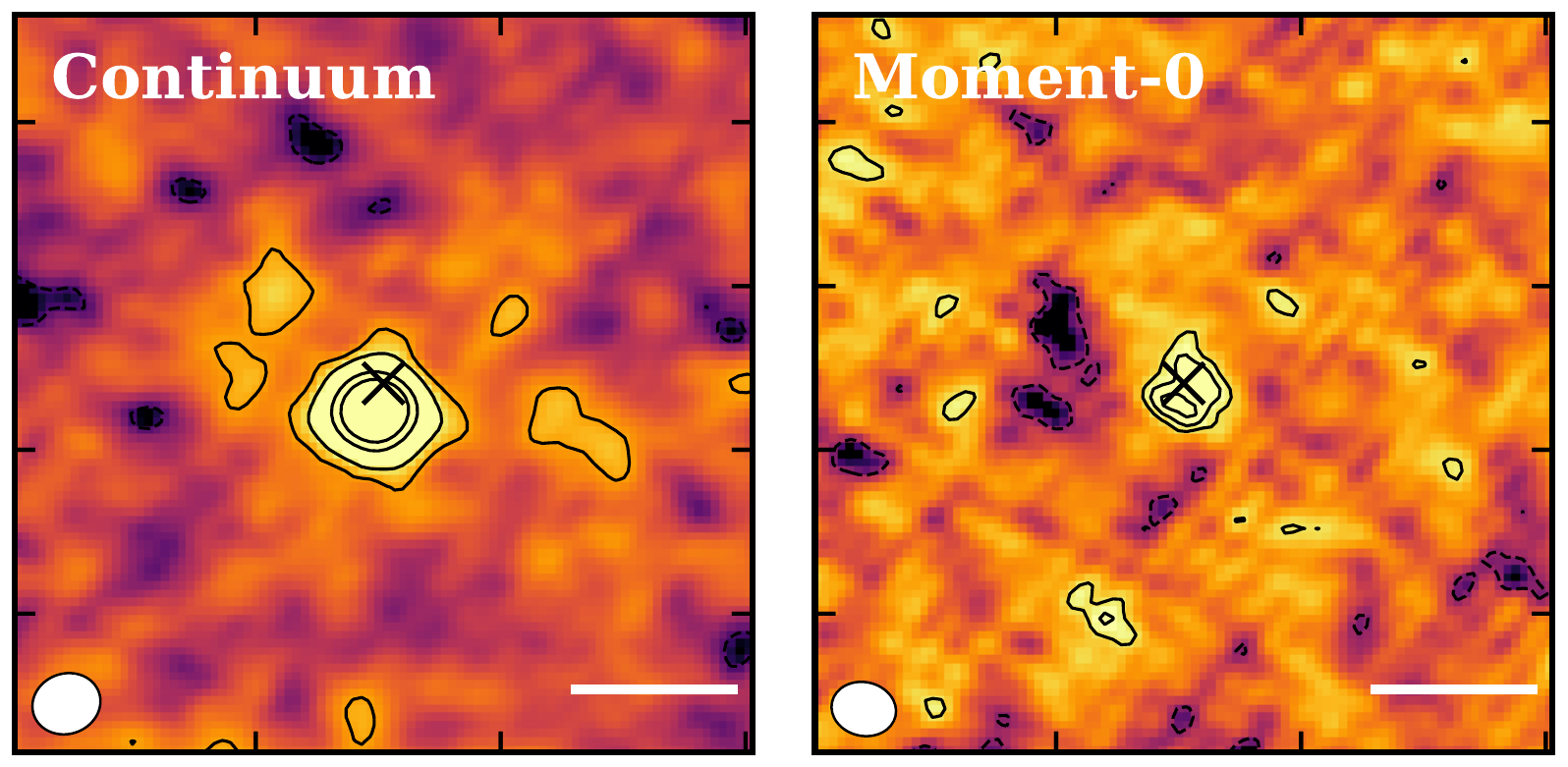}
    \\
    \includegraphics[width=1.0\linewidth]{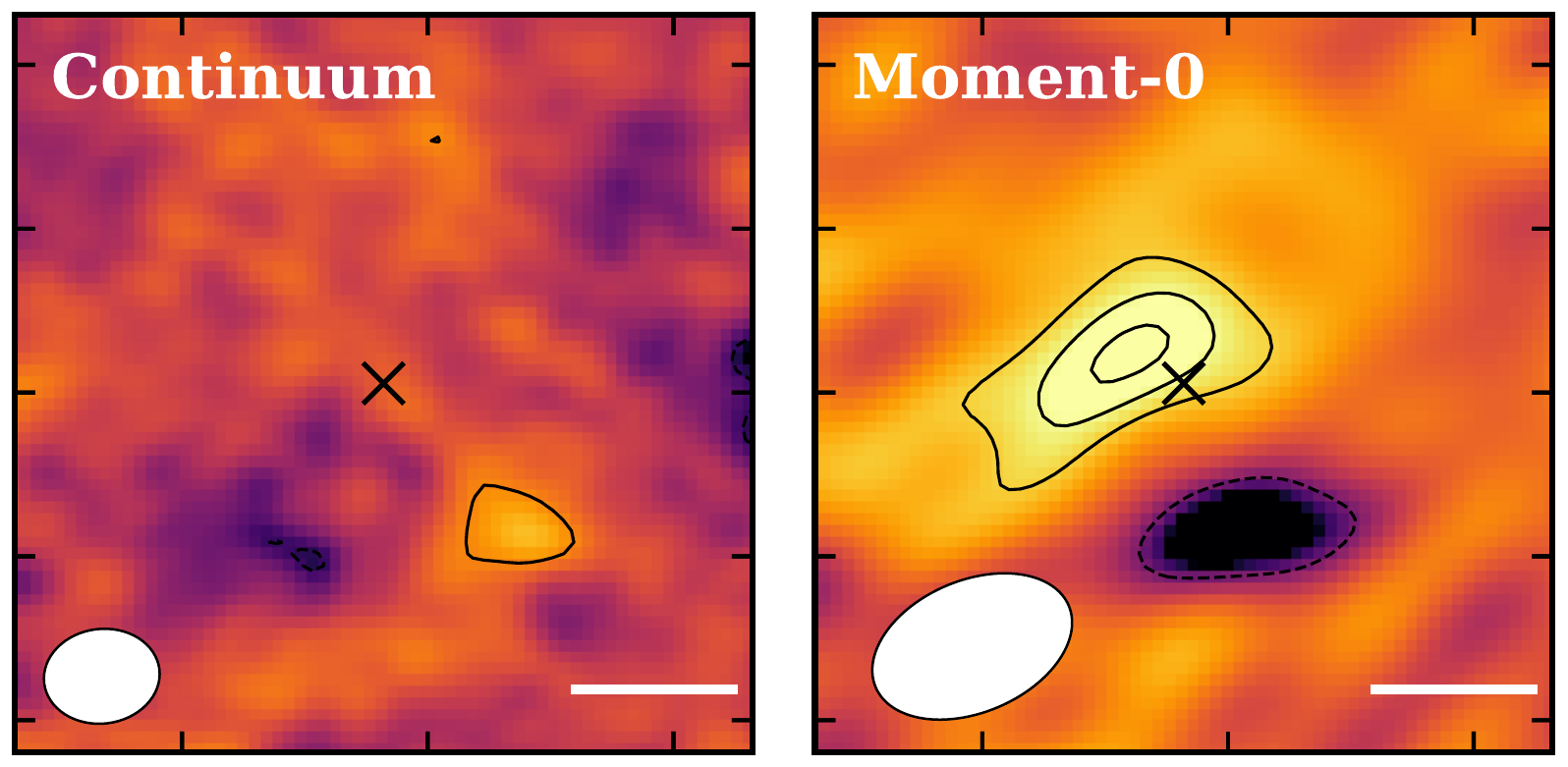}
\end{tabular}
    \caption{The continuum images and moment-0 maps of UDS\_18697 (top row) and COSMOS-z10-2 (bottom row), along with the ALMA beam size (white ellipse) and the object position (black cross).
        \textit{Left}:
        The 4\farcs5$\times$4\farcs5 band 7 continuum emission.
        Contours show significance level of $(2,6,10,14)\times\sigma$.
        Negative contour level is $-2\sigma$ and shown in dashed lines.
        \textit{Right}:
        The 4\farcs5$\times$4\farcs5 moment-0 maps.
        Beam sizes are 0\farcs40$\times$0\farcs33 and 1\farcs23$\times$0\farcs81 for UDS\_18697 and COSMOS-z10-2, respectively.
        Contours are drawn as $\pm N\times\sigma$, where $N =2,3,4$.
        Negative contours are shown in dashed lines.
        The scale bars in the lower-right corner indicate $1.0\,\mathrm{arcsec}$ for both continuum images and moment-0 maps.
        }
    \label{fig:continuum&moment-0}
\end{figure}

\begin{figure}
\begin{tabular}{c}
    \includegraphics[width=1.00\linewidth]{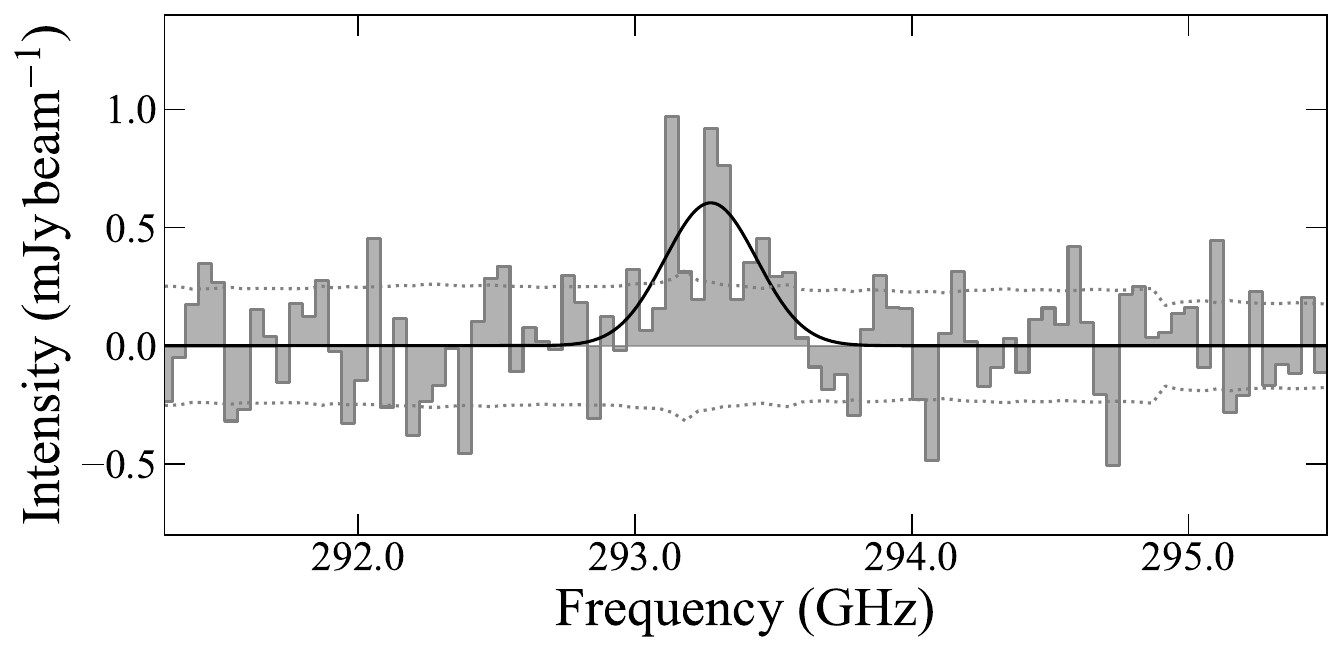}
    \\
    \includegraphics[width=1.00\linewidth]{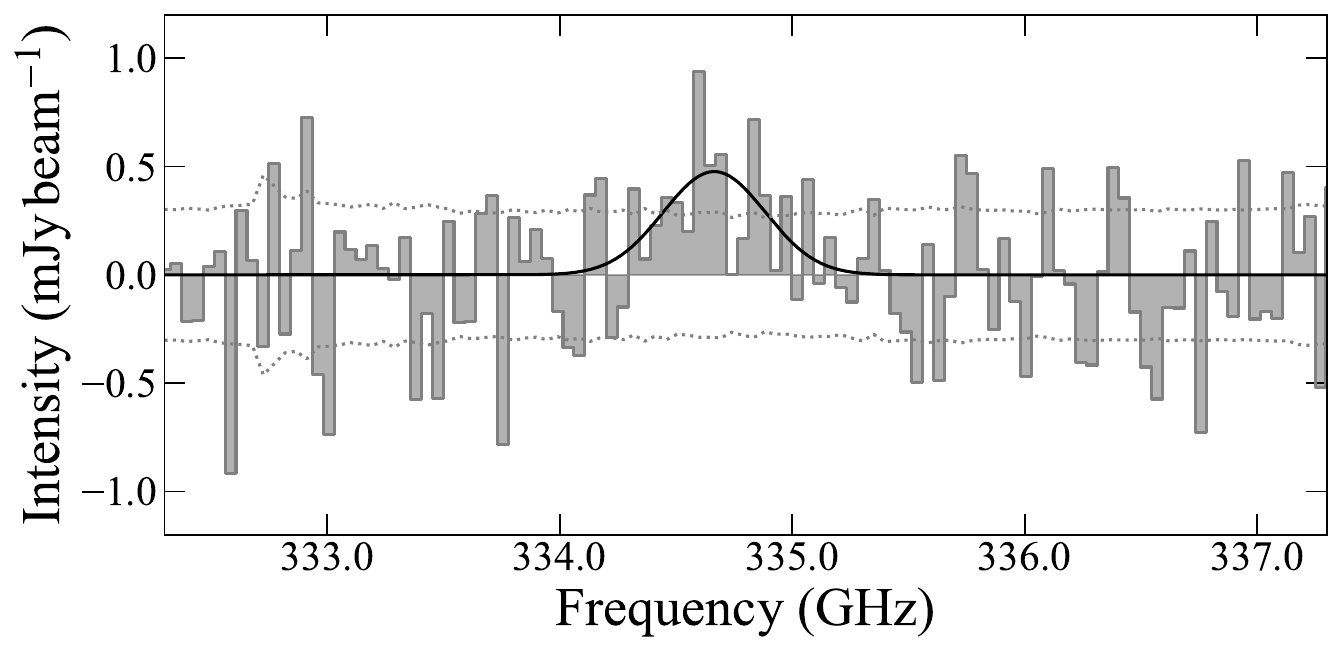}
    
\end{tabular}
    \caption{Extracted line spectra of UDS\_18697 (top panel) and COSMOS-z10-2 (bottom panel) along with the best-fit Gaussian functions (solid line). The rms levels are shown in the dotted lines.
        }
    \label{fig:spectrum}
\end{figure}

COSMOS-z10-2 shows a marginal line signal around $334.6\,\mathrm{GHz}$.
The moment-0 map and a 1D spectrum at the peak position are shown in the bottom right panel of Figure~\ref{fig:continuum&moment-0} and the bottom panel of Figure~\ref{fig:spectrum}, respectively.
A frequency range used for the moment-0 map is $334.13\text{--}335.07\,\mathrm{GHz}$, corresponding to twice the FWHM of the line.
COSMOS-z10-2 exhibits an spatial offset of $0.37\,$arcsec between the peak position of the line emission and the object position in the COSMOS2020 catalogue\footnote{The COSMOS2020 catalogue coordinates are aligned to Gaia DR2 stars.}, which is somewhat larger than an expected positional accuracy of $0.30$ arcsec.
A peak significance level on the moment-0 map is $4.56\sigma$ and a spatially integrated line flux is $0.475 \pm 0.169\,\mathrm{Jy\,km\,s^{-1}}$.
At this significance level, the line detection remains marginal.
The central frequency measured from the 1D spectrum is $334.66\pm0.07\,\mathrm{GHz}$ with the line FWHM of $463.06\pm73.77\mathrm{km\,s^{-1}}$.
If the line were [\ion{O}{iii}]\,$88\,\mu$m, the redshift of COSMOS-z10-2 would be $z=9.138\pm0.002$, which is considerably lower than the expected redshift of $z=9.56$.
\par

Dust continuum emission was not found in COSMOS-z10-2 (the bottom left panel of Figure~\ref{fig:continuum&moment-0}), and consequently a $3\sigma$ upper limit of $<37\,\mathrm{\mu Jy}$ was adopted for the flux density, assuming that the size of the object is sufficiently smaller than the beam size.
\par

\par

After the ALMA observations, a \textit{JWST}/NIRSpec follow-up spectroscopy was obtained (GO2659, PI: J. R. Weaver), indicating that this object is at $z\sim1.7$ based on the Balmer break (J. Weaver et al. in prep.; N. Nezhad et al. in prep.).
For this redshift, the 334.6 GHz emission line can be a CO($J=8\text{--}7$) line at $z=1.755$. 
The clump search analysis gives a 10\% probability of obtaining a spurious $4.6\sigma$ signal (Appendix~\ref{appendix:line search}).
Given this non-negligible probability, together with the spatial offset and uncertain spectroscopic redshift, we do not claim a detection of the CO($J=8\text{--}7$) line.
The $4.6\sigma$ signal may therefore be a false positive, and confirming its nature requires a secure spectroscopic redshift based on emission lines.
\par

\subsubsection{Other targets}
\label{sec:non-detections}
For the remaining four objects, neither line nor continuum emission was detected.
We put $3\sigma$ upper limits on the line fluxes and continuum flux densities (Table~\ref{tab:ALMA measurments}).
These upper limits on the line fluxes were calculated from an averaged rms\footnote{The rms was computed for each channel and averaged over the full bandwidth, excluding the frequency range 322.5--327.5\,GHz, which is affected by strong atmospheric absorption.} and integrated over $400\,{\mathrm {km\,s^{-1}}}$.
We assumed that the emission lines are within the observed frequency range and that their size are sufficiently smaller than beam sizes.
The velocity width of $400\,{\mathrm {km\,s^{-1}}}$ was assumed as twice a typical FWHM of $\lesssim200\,\mathrm{km\,s^{-1}}$ in previous [\ion{O}{iii}]$\,88\,\mu$m detections of $z\gtrsim10$ galaxies \citep{zavala24,carniani25, schouws25,witstok25}.
The dust continuum upper limits were provided by assuming an unresolved source size.
We note that \textit{JWST}/NIRSpec follow-up observation (GO2659, PI: J. R. Weaver) revealed that COSMOS-z10-1 is at $z\sim2.8$ (J. Weaver et al. in prep.; N. Nezhad et al. in prep.).
\par

\begin{table*}
\caption{Beam sizes and sensitivities of created Cubes and Continuum Image}
\label{tab: cube properties}
\begin{flushleft}
\renewcommand{\arraystretch}{1.2}
    \begin{tabular}{lccccc}
    \hline
         Object & Cube Beam Size \& PA & Velocity Width & Cube Mean rms & Continuum Beam Size \& PA & Continuum rms \\
         & (FWHM) [deg] & {[${\mathrm{km\,s^{-1}}}$]} & {[${\mathrm{\mu Jy\,beam^{-1}}}$]} & (FWHM) [deg] & {[${\mathrm {\mu Jy\,beam^{-1}}}$]}\\ \hline
         UDS\_18697 & $0\farcs43\times0\farcs36,-69.01$ & $47.85$ & $20.4$ & $0\farcs42\times0\farcs37, -70.95$ & $9.811$ \\
         COSMOS\_20646 & $0\farcs41\times0\farcs39, -67.57$ & $43.84$ & $20.8$ & $0\farcs41\times0\farcs33,\ \ 69.18$ & $8.609$ \\
         UDS\_7815 & $0\farcs65\times0\farcs59, -77.78$ & $43.84$ & $22.6$ & $0\farcs65\times0\farcs59, -85.43$ & $9.187$ \\
         J2140+0241 & $0\farcs46\times0\farcs38,\ \  58.78$ & $43.84$ & $29.6$ & $0\farcs41\times0\farcs38,\ \ 71.30$ & $11.72$ \\
         COSMOS-z10-1 & $0\farcs69\times0\farcs64,\ \ 73.08$& $43.86$ & $26.5$ & $0\farcs78\times0\farcs73,-79.09$ & $10.81$ \\
         COSMOS-z10-2 & $0\farcs65\times0\farcs54,\ \ 83.42$ & $41.95$ & $35.5$ & $0\farcs71\times0\farcs58, -83.17$ & $16.31$ \\ \hline
    \end{tabular}
\end{flushleft}
\end{table*}

\begin{table*}
\caption{Properties of the line and dust continuum}
\label{tab:ALMA measurments}
\begin{flushleft}
\renewcommand{\arraystretch}{1.2}
    \begin{tabular}{lcccc}
        \hline
         Object& Line Flux\tnote{\dag} & Line Central Frequency & Line Width (FWHM) & Continuum Flux Density\\
          & [$\rm{Jy\,km\,s^{-1}}$] & [{\rm{GHz}}] & [$\mathrm{km\,s^{-1}}$] & [${\mathrm{\mu Jy}}$] \\ \hline
        UDS\_18697 & $0.397\pm0.139$ & $293.27\pm0.03$ & $399.48\pm41.05$ & $376\pm22$\\
        COSMOS\_20646 & $<\,0.106$ & -- & -- & $<\,25.83$\\
        UDS\_7815 & $<\,0.091$ & -- & -- & $<\,27.56$ \\
        J2140+0241 & $<\,0.117$ & -- & -- &$ <\,35.15$\\
        COSMOS-z10-1 & $<\,0.189$ & -- & -- & $<\,32.42$\\
        COSMOS-z10-2 & $0.439\pm0.157$ & $334.67\pm0.06$ & $415.42\pm63.16$ & $<\,38.39$\\ \hline
    \end{tabular}
    \par\raggedright
    \textit{Note.}
    Upper limits are $3\sigma$.
\end{flushleft}
\end{table*}

    \subsection{CO($J=9\text{--}8$) line in UDS\_18697}
The detected line in UDS\_18697 is likely to be the CO($J=9\text{--}8$) line at $z_{\mathrm{CO}}=2.536$ (Section~\ref{sec:UDS}).
Although it was not [\ion{O}{iii}]\,$88\,\mu$m that we aimed at, this serendipitous detection provides a distinctive data point within previous CO($J=9\text{--}8$) observations.
\par
In the local Universe, there is a tight correlation between far-infrared luminosity $L_\mathrm{FIR}$ and CO line luminosity $L'_{\mathrm{CO}}$ for multiple transitions ($J_\mathrm{up}=4\text{--}12$, \citealt{Liu15}).
The same correlation has been reported for CO($J=9\text{--}8$) in $z\sim2\text{--}4$ QSOs \citep{Butler23} and $z\sim2\text{--}6$ DSFGs \citep{Riechers21a,Riechers25}, although DSFGs tend to show slightly stronger CO line luminosity.
Motivated by these findings, we investigated the property of UDS\_18697 based on the detected continuum and CO($J=9\text{--}8$) line.
\par

Line luminosity $L'_{\mathrm{CO}}$ in units of $\mathrm{K\,km\,s^{-1}\,pc^2}$ are calculated as,
\begin{equation}
    L'_{\mathrm{CO}}=3.25\times10^7\left(\frac{S_{\mathrm{CO}}\Delta v}{\mathrm{Jy\,km\,s^{-1}}}\right)\left(\frac{D_L}{\mathrm{Mpc}}\right)^2\left(\frac{\nu_{\mathrm{obs}}}{\mathrm{GHz}}\right)^{-2}\left(\frac{1}{1+z}\right)^3,
\end{equation}
where $S_{\mathrm{CO}}\Delta v$ is the line flux, $D_L$ is the luminosity distance, and $\nu_\mathrm{obs}$ is the observed frequency \citep{Solomon92}.
With this equation, we obtained the line luminosity of $L'_{\mathrm{CO}(J=9\text{--}8)}=1.53\pm0.54\,\mathrm{K\,km\,s^{-1}\,pc^2}$.
\par

The galaxy SED of the FIR range is well approximated by a single temperature modified blackbody (MBB, e.g., \citealt{Inoue20}).
First, we estimated the dust temperature $T_{\rm d}$ based on a radiative-equilibrium method, which utilises a single-band FIR continuum and UV luminosity $L_\mathrm{UV}$ \citep{Inoue20,Fudamoto23}\footnote{\url{https://github.com/yfudamoto/FIS22sed.git}}.
By assuming a uniform distribution of dust and radiation sources, the dust temperature was estimated as $T_{\rm d}<70.4\,$K.
Since UDS\_18697 is not detected in the rest-UV wavelength, we can only estimate an upper limit for the dust temperature.
\par
Next, we estimated $T_{\rm d}$ based on a method by \citet{Sommovigo22}, which reproduced consistent dust temperature with those derived from multi-band FIR SED fitting across a redshift range of $z\sim0\text{--}8$.
They formulated the dust temperature as a function of redshift and metallicity $Z\,[{\rm Z_\odot}]$:
\begin{equation}
    T_\mathrm{d}=29.6\left[\frac{f'(z)E(z)}{Z}\right]^{1/6.03}\,\mathrm{K},
\end{equation}
where $f'(z)=-0.24+0.75(1+z)$ and $E(z)=\lbrack\Omega_\mathrm{M}(1+z)^3+\Omega_\Lambda\rbrack^{1/2}$.
Although the metallicity $Z$ was not constrained for UDS\_18697, $T_\mathrm{d}$ is insensitive to the metallicity as $T_\mathrm{d}\propto Z^{-1/6}$.
Therefore, metallicity was randomly picked from $Z=0\text{--}2.00\,Z_\odot$ under the uniform prior in the linear space\footnote{Although the NIRSpec data have been taken for the object, the metallicity is not yet available. A detailed analysis will be presented in R.~Larson~et~al.~in~prep.}.
The IR ($8\text{--}1000\,\mathrm{\mu}$m) and FIR ($42.5\text{--}122.5\,\mathrm{\mu}$m) luminosity were then calculated from the observed continuum flux density and an MBB with the given $T_\mathrm{d}$ and an emissivity index of $\beta=2.0$.
We repeated these steps 10,000 times and obtained the dust temperature, IR luminosity, and FIR luminosity of $T_\mathrm{d}=42.8^{+8.7}_{-3.5}\,\mathrm{K}$, $L_\mathrm{IR}=1.5^{+1.9}_{-0.5}\times10^{12}\,L_{\odot}$, and $L_\mathrm{FIR}=1.1^{+1.0}_{-0.3}\times10^{12}\,L_{\odot}$, respectively.
The derived $T_\mathrm{d}$ is consistent with a redshift--temperature relation calibrated on DSFGs at $z=1.9\text{--}6.9$ \citep{Reuter20} and a general trend of increasing dust temperature with redshift \citep[e.g.,][]{Schreiber18}.
With this FIR luminosity, UDS\_18697 is classified as an ultra luminous infrared galaxy (ULIRG).
Although the metallicity of UDS\_18697 is highly uncertain, dust temperature only ranges from $39\,\mathrm{K}$ to $51\,\mathrm{K}$, resulting in relatively small uncertainty in $L_{\mathrm{FIR}}$.

\begin{figure}
    \centering
    \includegraphics[width=1.0\linewidth]{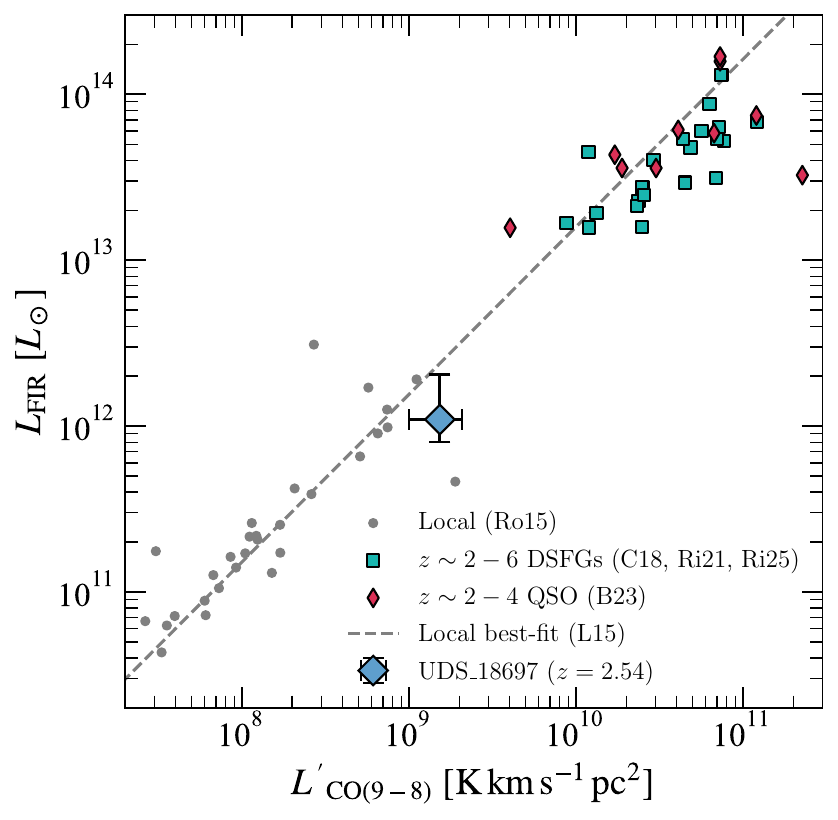}
    \caption{The FIR luminosity $L_{\mathrm{FIR}}$ and CO line luminosity $L'_\mathrm{{CO}}$ correlation.
    The value of UDS\_18697 is shown in the blue diamond.
    Green squares are DSFGs at $z\sim2\text{--}6$ taken from \citet{Canameras18}, \citet{Riechers21a} and \citet{Riechers25}.
    Red diamonds indicate QSOs at $z\sim2\text{--}4$ shown in \citet{Butler23}.
    Grey circles are local galaxies presented in \citet{Rosenberg15}.
    The dashed line shows the best-fit line for 165 local galaxies presented by \citet{Liu15}.
     $L_{\mathrm{FIR}}$ is integrated over $42.5\text{--}122.5\,\mathrm{\mu}$m.
     Sources whose $L_{\mathrm{FIR}}$ are presented in different integral wavelengths were scaled based on luminosity ratio (see text for details).
     }
    \label{fig:line_vs_FIR}
\end{figure}

Figure~\ref{fig:line_vs_FIR} shows the relation between $L_{\mathrm{FIR}}$ and $L'_{\mathrm{CO}(J=9\text{--}8)}$.
Since \citet{Canameras15} and \citet{Butler23} calculated $L_{\mathrm{FIR}}$ by integrating $8\text{--}1000\,\mathrm{\,\mu}$m and $40\text{--}120\,\mathrm{\,\mu}$m, respectively, the values are scaled to that of $42.5\text{--}122.5\,\mathrm{\,\mu}$m using the MBB luminosity ratio between different integrated wavelength.
We calculated the luminosity ratio using $T_{\mathrm{d}}=50\,\mathrm{K}$ ($45\,\mathrm{K}$) for \citet{Canameras15} (\citealt{Butler23}) based on the reported dust temperature.
Although UDS\_18697 is possibly gravitationally magnified ($\mu=1.35$) by a nearby bright galaxy, the magnification factor has a substantial uncertainty \citep{finkelstein22}.
Thus, we did not correct the luminosities.
\par
The luminosity of UDS\_18697 is one order of magnitude fainter than that of $z>2$ DSFGs previously detected in CO($J=9\text{--}8$) \citep[see][for a recent compilation]{Tadaki26}.
Also, UDS\_18697 has a CO($J=9\text{--}8$) luminosity comparable to the bright end of local galaxies \citep{Rosenberg15}.
The luminosities $L_{\mathrm{FIR}}$ and $L'_{\mathrm{CO}(J=9\text{--}8)}$ of UDS\_18697 are almost consistent with the local relation \citep{Liu15}, additionally supporting that the CO($J=9\text{--}8$) line is surely detected.
However, UDS\_18697 seems to be located slightly below the local relation; it shows brighter $L'_{\mathrm{CO}(J=9\text{--}8)}$ for the given $L_{\mathrm{FIR}}$, which is consistent with previous studies for DSFGs \citep{Riechers21a,Riechers25}.
A larger sample is necessary to statistically confirm the relation for $z\gtrsim2$ galaxies.

\section{SED fitting with NIRCam photometry}
\label{sec:SEDfitting}
Given that the detected line in UDS\_18697 was not the [\ion{O}{iii}]\,$88\mu$m line and the others were marginal or non-detections, we revisited the photometric redshift and physical properties of the selected $z\sim10$ candidates.
While our ALMA target galaxies were selected with \textit{HST} and \textit{Spitzer} photometry, deeper \textit{JWST}/NIRCam photometry is available today.
We compared two photometric datasets: \textit{JWST}/NIRCam, \textit{HST}/ACS, and \textit{HST}/WFC3 photometry available in the ASTRODEEP-JWST catalogue (hereafter, the \textit{NIRCam+HST} photometry)\footnote{ALMA continuum data were not used in the fitting as we intended to see how SED fitting results change with an increase in the rest-frame optical to near-infrared data points and high-resolution data.} and \textit{HST}/ACS, \textit{HST}/WFC3, and \textit{Spitzer} photometry presented in \citet{finkelstein22} (the \textit{Spitzer+HST} photometry).
Note that although the \textit{NIRCam+HST} and the \textit{Spitzer+HST} photometry include photometry of the same \textit{HST} filters, the flux density measurements provided in the two samples are sometimes different, probably due to different image reduction and flux density measurement methods.
An advantage of adding NIRCam photometry is accurate photometry measurement with its superb spatial resolution, as well as filling a large wavelength gap between $\lambda\sim1.6\,\mu$m (F160W) and $3.6\,\mu$m ([3.6]).
These are essentially important for constructing a robust high-$z$ sample by capturing the rest-frame UV spectral slope of $z\sim10$ galaxies.
As described in Section~\ref{sec: photometry}, the NIRCam photometry measurements are only available for UDS\_18697, COSMOS\_20646, and UDS\_7815.
Therefore, we performed SED fitting to only these three objects.
For all photometry, the minimum errors were set to $5\%$ of their flux densities to take into account the absolute flux density uncertainties \citep{Rigby23}
\par 

We used the Bayesian Analysis of Galaxies for Physical Inference and Parameter EStimation \citep[\textsc{bagpipes},][]{carnall18} for the SED fitting.
By using a code different from the photometric redshift code \textsc{eazy} \citep{brammer08} used in \citet{finkelstein22}, we will check whether fitting results highly depend on codes with the \textit{Spitzer+HST} photometry.
We adopted the dust attenuation law proposed by \citet{calzetti00} and assumed a delayed-$\tau$ star formation history.
Fitting parameters were redshift $z$, time since star formation began $t_{\mathrm{SF}}$, time scale of star formation decrease $\tau$, stellar mass $M_*$, dust attenuation at $V$-band $A_V$, metallicity $Z$, and ionisation parameter $U$.
The best-fit solutions were sought over the following parameter ranges with uniform priors: $z\in [0.0, 12.0]$, $t_\mathrm{SF}/\mathrm{Gyr}$ $\in [0.001, 3.0]$, $\tau/\mathrm{Gyr}\in [0.001, 1.0]$, log$\,(M_*/M_{\odot})\in [7.5, 11.5]$, $A_V/\mathrm{mag}\in [0.0, 6.0]$, $Z/Z_{\odot}\in [0.001, 2.0]$, and log$\,U\in [-4.0, -1.0]$.
\par

\begin{figure}
    \begin{tabular}{c}
    \includegraphics[width=\columnwidth]{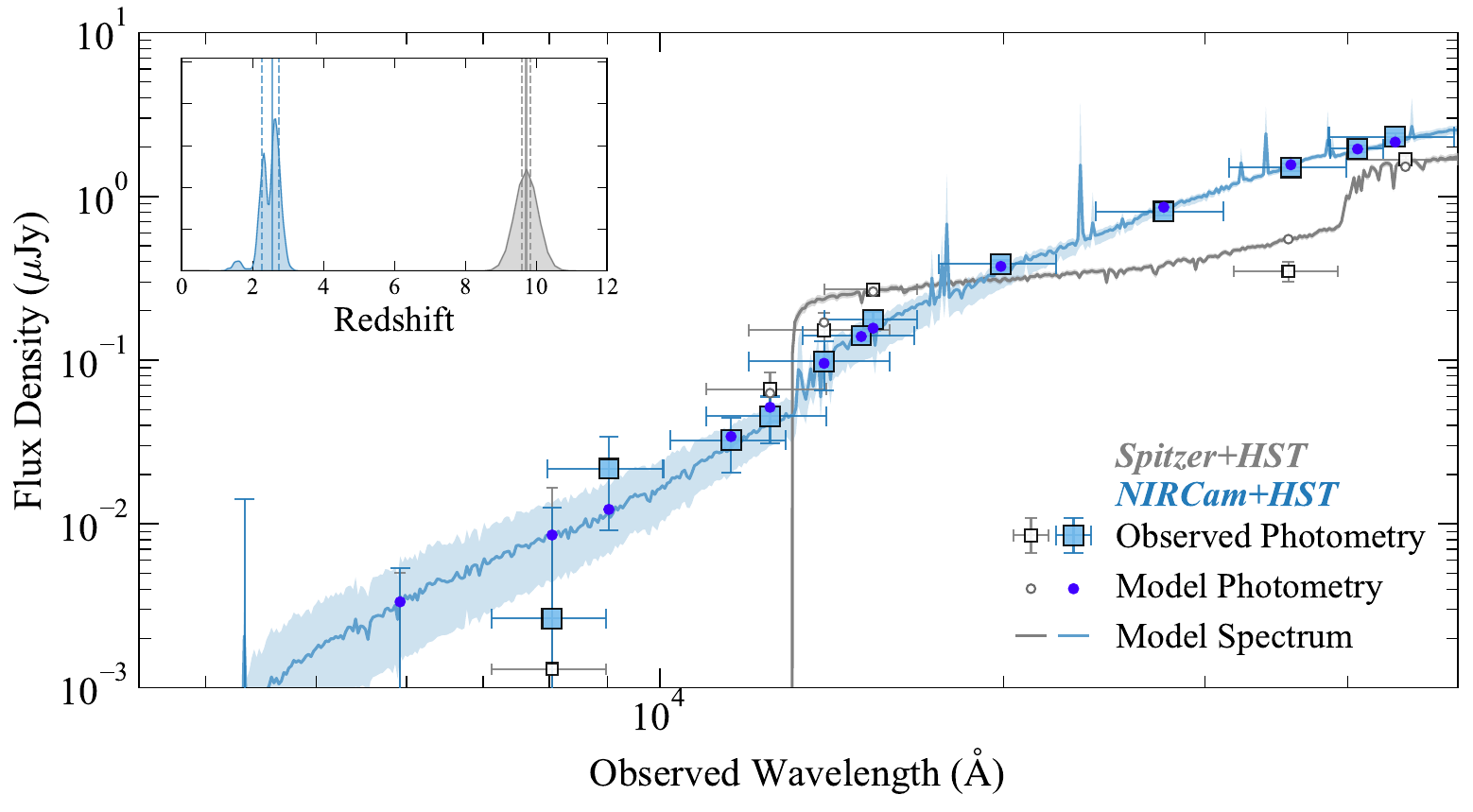}
        \\
    \includegraphics[width=\columnwidth]{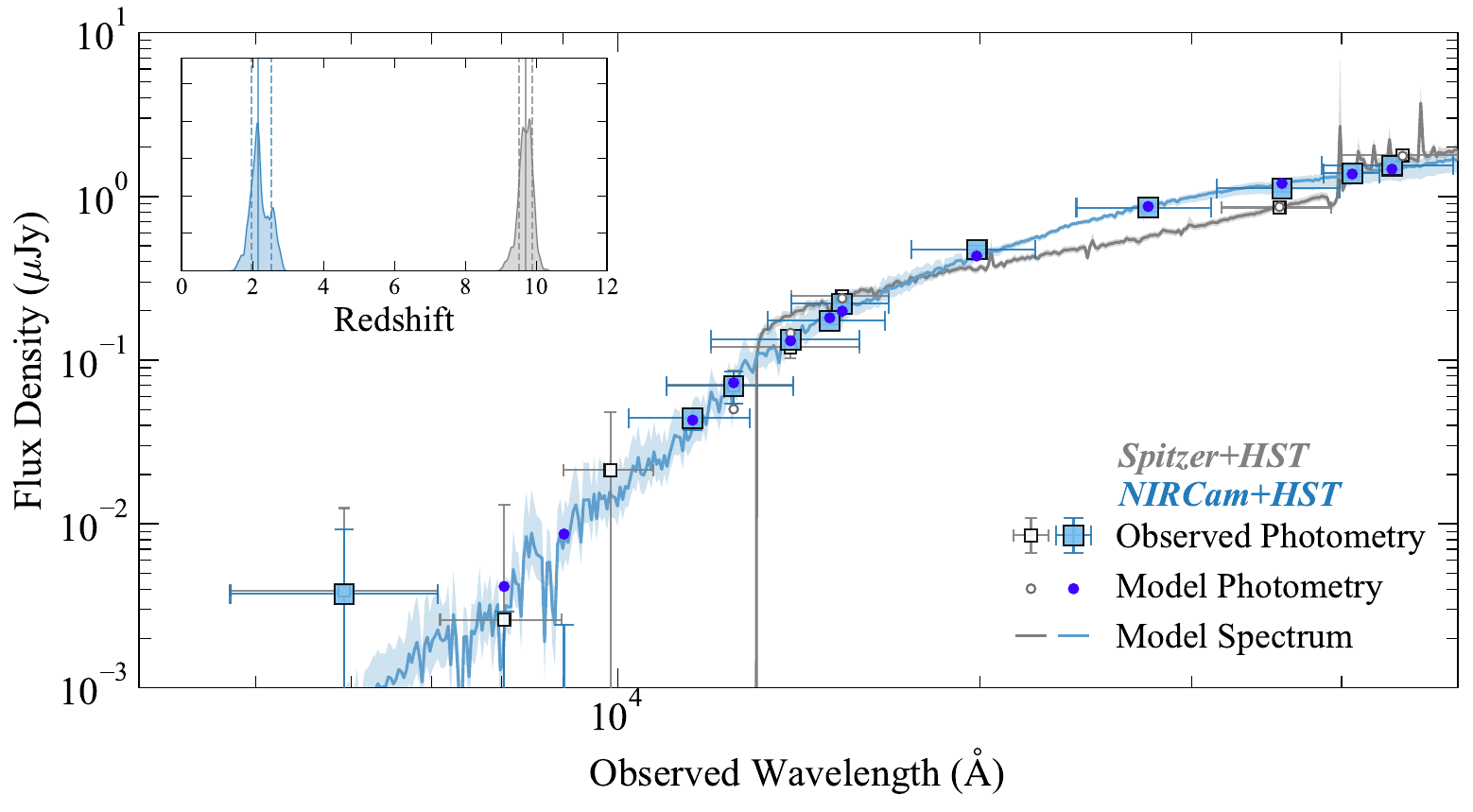}
        \\
    \includegraphics[width=\columnwidth]{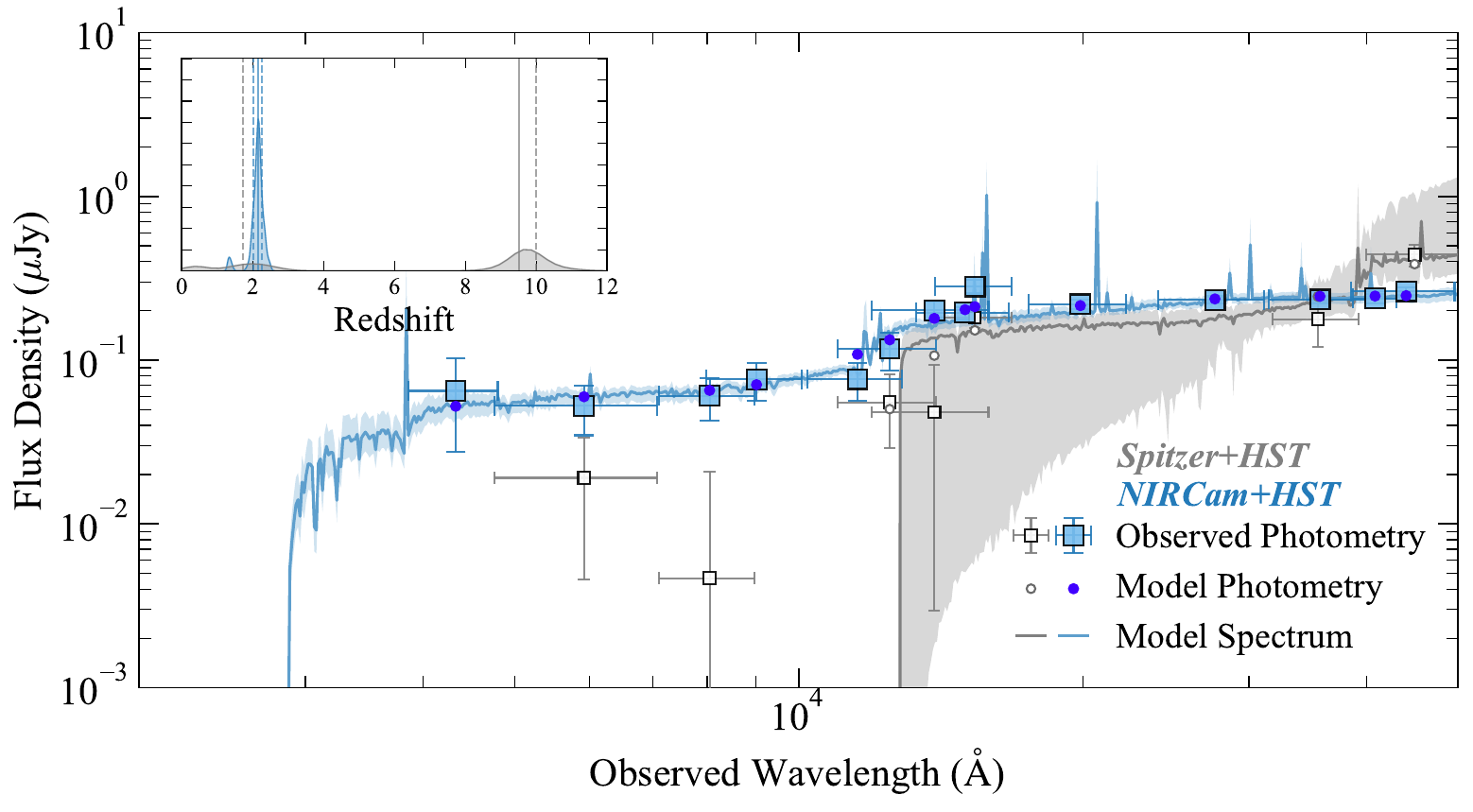}
    \end{tabular}
    \caption{SEDs of UDS\_18697 (top), COSMOS\_20646 (middle), and UDS\_7815 (bottom), together with the redshift probability distribution (top left in each panel).
    Blue lines and circles show the model spectrum and photometry, respectively, while blue squares indicate the measured photometry of the \textit{NIRCam+HST} photometry.
    The grey lines and open markers present the result of the fittings using the \textit{Spitzer+HST} photometry (symbols are the same as the \textit{NIRCam+HST} fitting).
    The solid lines and shaded regions are the best-fit spectra and their $16\text{--}84$ percentiles.
    }
    \label{fig:sed}
\end{figure}

Figure~\ref{fig:sed} shows the observed photometry and best-fit SEDs of the three galaxies.
The \textit{NIRCam+HST} and the \textit{Spitzer+HST} photometry are depicted by the blue and grey squares, respectively.
The best-fit SEDs for the two types of photometry are drawn by the blue and grey lines with $1\sigma$ uncertainties.
The best-fit parameters of the SED fitting for the \textit{NIRCam+HST} photometry are summarised in Table~\ref{tab:sed_fitting}.
This section describes fitting results, while the detailed discussions will be given in the next section.

The SEDs of UDS\_18697 are shown in the top panel of Figure~\ref{fig:sed}.
The largest difference between the \textit{NIRCam+HST} and the \textit{Spitzer+HST} photometry is found between the NIRCam and \textit{Spitzer}/IRAC photometry at $3.6\,\mu$m, resulting in different best-fit models.
The best-fit redshift for the \textit{Spitzer+HST} photometry was $z_{\mathrm{Spitzer}}=9.72^{+0.12}_{-0.11}$, which is consistent with the best-fit redshift estimated by \citet{finkelstein22}, $z_{\mathrm{EAZY}}=9.89^{+0.16}_{-0.15}$.
This consistency ensures that the best-fit redshifts are not strongly dependent on fitting codes.
However, the best-fit results for the \textit{NIRCam+HST} photometry represent that UDS\_18697 is a dusty star-forming galaxy at $z_{\mathrm{NIRCam}}=2.56^{+0.19}_{-0.30}$ with the dust attenuation of $A_V=2.81^{+0.76}_{-0.45}\,\mathrm{mag}$.
This best-fit redshift is consistent with the spectroscopic redshift of $z_{\mathrm{spec}}=2.54$ (R. Larson et al. in prep.).
The expected dusty star-forming nature is also consistent with the detection of the dust continuum emission.
Thus, the \textit{NIRCam+HST} photometry properly confirms that UDS\_18697 is a low-$z$ interloper.
\par

The SEDs of COSMOS\_20646 are shown in the middle panel of Figure~\ref{fig:sed}.
The best-fit redshift for the \textit{Spitzer+HST} photometry is $z_{\mathrm{Spitzer}}=9.70^{+0.19}_{-0.19}$, which is again consistent with the best-fit redshift $z_{\mathrm{EAZY}}=9.80^{+0.10}_{-0.46}$ estimated by \citet{finkelstein22}.
On the other hand, the SED fitting for the \textit{NIRCam+HST} photometry shows that this object is a passive and dust reddened galaxy with the age of $t_{\mathrm{SF}}=2.19^{+0.48}_{-0.58}\,\mathrm{Gyr}$ at $z_{\mathrm{NIRCam}}=2.17^{+0.34}_{-0.22}$. 
This photometric redshift is roughly consistent with that estimated based on the newest COSMOS2025 catalogue \citep[$z=2.69^{+0.11}_{-0.19}$;][]{cosmos-web}, indicating that this object is again a low-$z$ interloper.
\par

Finally, the best-fit SEDs for UDS\_7815 are shown in the bottom panel of Figure~\ref{fig:sed}.
The SED fitting using \textit{Spitzer+HST} indicates a best-fit redshift of $z_\mathrm{Spitzer}=9.50$, which is marginally consistent with previous estimates \citep{finkelstein22}.
However, the posterior distribution also exhibits a secondary peak at $z\sim2$.
The best-fit redshift derived from the \textit{NIRCam+HST} photometry is $z_\mathrm{NIRCam}=2.17^{+0.17}_{-0.15}$, consistent with the secondary peak in the \textit{Spitzer+HST} fit.
These results suggest that this galaxy is likely to be a star-forming galaxy with a relatively strong ($\sim1\,\mathrm{mag}$) Balmer break.
These results are in good agreement with the fact that UDS\_7815 was excluded from the final $z\sim10$ sample in \citet{finkelstein22} (Section~\ref{sec:samples}).
\par

Among the three galaxies, none of their best-fit SEDs show any high-$z$ solutions, although they were previously selected as $z\sim10$ candidates.
Given that their estimated redshifts are $z\sim2\text{--}2.5$, we regard these galaxies as low-$z$ interlopers that were misclassified as $z\sim10$ sources. 
Note that the majority of high-$z$ galaxy candidates selected in \citet{Finkelstein22b} are confirmed or suggested to be at high-$z$ by \textit{JWST} observations (R. Larson et al. in prep.).
In the next section, we discuss the cause of these misclassifications.
\par

\begin{table}
    \caption{Results of the SED fitting}
    \label{tab:sed_fitting}
    \begin{flushleft}
    \renewcommand{\arraystretch}{1.2}
    \begin{tabular}{lccc}
    \hline
    Parameter & UDS\_18697 & COSMOS\_20646 & UDS\_7815 \\ \hline
    $z$ & $2.56^{+0.19}_{-0.30}$ & $2.17^{+0.34}_{-0.22}$ & $2.17^{+0.17}_{-0.15}$ \\ 
    $t_{\mathrm{SF}}\,[\mathrm{Gyr}]$ & $0.77^{+0.66}_{-0.36}$ & $2.19^{+0.48}_{-0.58}$ & $0.60^{+0.67}_{-0.39}$ \\ 
    $\tau\,[\mathrm{Gyr}]$ & $0.38^{+0.35}_{-0.31}$ & $0.01^{+0.07}_{-0.01}$ & $0.39^{+0.37}_{-0.37}$ \\
    $A_V\,\mathrm{[mag]}$ & $2.81^{+0.76}_{-0.45}$ & $1.55^{+0.58}_{-0.56}$ & $0.46^{+0.40}_{-0.32}$ \\ 
    $\mathrm{log}\,$\textit{U} & $-2.03^{+0.73}_{-0.96}$ & $-2.61^{+1.08}_{-0.92}$ & $-1.93^{+0.59}_{-0.75}$ \\
    $Z$ [$Z_{\odot}$] & $>0.003^\dagger$ & $0.18^{+0.37}_{-0.15}$ & $0.24^{+0.28}_{-0.12}$ \\ 
    $\mathrm{log}\,M_*\,[M_{\odot}]$ & $10.51^{+0.11}_{-0.20}$ & $10.50^{+0.08}_{-0.08}$ & $9.06^{+0.14}_{-0.17}$ \\ \hline
    \end{tabular}
    \par\raggedright
    $\dagger$: Since metallicity of UDS\_18697 is highly uncertain, $3\sigma$ value calculated as a single-sided $0.135$ percentile is adopted as a lower limit.
\end{flushleft}
\end{table}

\section{Discussion}
\label{sec:discussion}
\subsection{Effect of Photometry Measurement on Redshift Estimation}
\label{sec:sed-based discussion}
The SED fitting results and follow-up \textit{JWST}/NIRSpec observations strongly suggest that five out of the six in our sample are low-$z$ interlopers.
Motivated by the results, we discuss reasons why these probable low-$z$ interlopers were previously selected as $z\sim10$ candidates.
\par

The main reason for the misclassification of UDS\_18697 is likely an underestimation of the \textit{Spitzer}/IRAC [3.6] flux density.
This object has a bright nearby galaxy located about $3\,\mathrm{arcsec}$ away.
Because of the modest angular resolution of \textit{Spitzer}/IRAC \citep[$\sim1.5$ arcsec;][]{Fazio04}, light from nearby sources can be blended with that of targets.
To mitigate this effect, \citet{finkelstein22} separated the flux of each target (\textit{deblending}) to accurately measure the flux of galaxies with nearby objects.
Even after careful deblending on images, the [3.6] flux density was one-fifth of the NIRCam F356W flux density.
In addition, the previous F160W flux density was larger than that in the ASTRODEEP-JWST catalogue by a factor of 1.5.
These two factors made the apparent flat UV to optical continuum.
Furthermore, the continuum at $\lambda \lesssim 1.3\,\mu$m is heavily attenuated by dust, making this object undetected in F115W band.
\par

The change in the photo-$z$ estimates of COSMOS\_20646 can be explained by a combination of multiple reasons. 
The new NIRCam photometry continuously traces a red stellar continuum from $\lambda=1.1$ $\mu$m to 4.4 $\mu$m, ruling out the possibility of being an LBG.
Another possible factor changing the best-fit SED is difficulty of deblending.
Since COSMOS\_20646 also has a nearby galaxy about $2\,\mathrm{arcsec}$ apart, there could be a possible oversubtraction.
\par

The misclassification of UDS\_7815 as a $z\sim10$ galaxy was likely caused by its faintness and the Balmer break. 
In the \textit{Spitzer+HST} photometry, all bands blueward of the Balmer break are fainter than those in the ASTRODEEP-JWST catalogue by a factor of $2\text{--}3$.
This suppressed flux at shorter wavelength caused the Balmer break to resemble a Lyman break, resulting in the $z\sim10$ solution.
The Lyman break method requires stringent upper limits at wavelengths shorter than the Lyman break to select LBGs securely, but UDS\_7815, which is $0.5\,\mathrm{mag}$ fainter than UDS\_18697 and COSMOS\_20646 in \textit{Spitzer+HST} photometry, can be easily affected by noise fluctuation in photometric measurements \citep{merlin24}.
Indeed, UDS\_7815 was detected in \textit{HST}/WFC3 F814W with the significance level of $3.4\sigma$ in the ASTRODEEP-JWST catalogue, while it was reported to be below $2\sigma$ previously.
Such a fluctuation directly leads to the misclassification of the photometric redshifts, especially in faint objects like UDS\_7815.
\par

In summary, the difficulty of photometry deblending in low-resolution images, the faintness of targets, and the sparse photometry are the main causes of the inaccurate photo-$z$ estimation for the three galaxies.
For UDS\_18697, for example, the [3.6] flux density is significantly offset from the best-fit SED estimated from the \textit{NIRCam+HST} photometry.
If there were sufficiently deep observations at $2\text{--}3\,\mu$m, the object would likely not have been selected as a $z\sim10$ candidate.
Also, it is worth noting that \citet{Boueens19} already excluded UDS\_18697 from $z>9$ candidates based on ground-based $J$2-band data from ZFOURGE, showing the importance of deep ground-based data for rejecting lower-$z$ galaxies.
Taken together, these results suggest that robust high-$z$ galaxy selection requires not only careful deblending and deep observations at shorter wavelengths, but also broad wavelength coverage.
\par

\subsection{Revisiting Colour Selection Criteria}
\subsubsection{Comparison with JWST Observed Galaxies}
Follow-up spectroscopic observations and SED fitting suggested that most of our targets are likely to be low-$z$ interlopers rather than $z\sim10$ galaxies.
Given this situation, we revisit colour--colour selection criteria and discuss an appropriate criterion using NIRCam photometry of an archival spectroscopic high-$z$ sample.
\par

We compiled spectroscopically confirmed galaxies from the JWST Advanced Deep Extragalactic Survey \citep[JADES;][]{eisenstein23} and the Canadian NIRISS Unbiased Cluster Survey \citep[CANUCS;][]{Willott22} catalogues.
The $5\sigma$ depths in $0\farcs3$ diameter aperture are about $29.5\,\mathrm{mag}$ and $28.9\,\mathrm{mag}$ in F090W for JADES \citep{rieke23} and CANUCS \citep{sarrouh25}, respectively.
For JADES, we used NIRCam GOODS-N and GOODS-S v1.1 catalogue \citep{rieke23} and NIRSpec PRISM v1.0 (GOODS-N) and v2.0 (GOODS-S-Deep) catalogue \citep{bunker24,deugenio25}\footnote{\url{https://archive.stsci.edu/hlsp/jades}}.
The NIRCam and NIRSpec catalogues were matched based on \texttt{NIRCam\_ID} in the NIRSpec catalogues.
For CANUCS, we used catalogues of five cluster fields (not flanking fields) from CANUCS DR1 \citep{sarrouh25}\footnote{\url{https://niriss.github.io/data_release1.html}}.
To assess colour selection criteria using the spectroscopic redshift, we employed only galaxies with secure spectroscopic redshifts selected based on flags in catalogues, i.e., \texttt{z\_Spec\_flag} $\in$ [A, B, C] (\texttt{Z\_Q} $\in$ [1, 2]) for JADES (CANUCS) sample.
Colours were calculated using photometry measured in Kron radii after homogenising the point-spread functions (PSFs) to that of the F444W.
\par

On the basis of the correct redshifts of the JADES and CANUCS spectroscopic sample, we examined locations of our ALMA sample observed with NIRCam (UDS\_18697, COSMOS\_20646, and UDS\_7815) and the spectroscopic sample on the colour--colour diagram for the $z\sim10$ galaxy selection.
All the three galaxies were selected from \citet{finkelstein22}, who constructed a high-$z$ sample based on photo-$z$ probability distribution from the SED fitting with \textsc{eazy}.
Since \citet{finkelstein22} showed that all their $z>8.5$ candidates met or were within $1\sigma$ of the selection criteria employed in \citet{bouwens16b}, we use colour criteria similar to \citet{bouwens16b} ones and compare how colours change with deep, high-angular resolution NIRCam photometry.
In \citet{bouwens16b}, galaxies were required to show (1) $J_{125}-H_{160}>0.5$ and (2) $H_{160}-[3.6]<1.4$, where $J_{125}$ is the \textit{HST}/WFC3 F125W magnitude.
Instead of F125W, F160W, and $[3.6]$ filters, we calculated the colour with NIRCam F115W, F150W, and F356W filters, respectively;  
(1) $\mathrm{F115W}-\mathrm{F150W}>0.5$ and (2) $\mathrm{F150W}-\mathrm{F356W}<1.4$.
Since the Lyman break is traced by F115W instead of F125W with the criteria, we limit our analysis to $z>8.5$.
We also set a detection threshold of $S/N_\mathrm{F150W}>5$ to ensure that galaxies are robustly detected at longer wavelengths.
For a dropout criterion, then, we set $S/N_\mathrm{F090W}<2$.
\par

\begin{figure*}
\begin{minipage}[b]{0.49\linewidth}
    \centering
    \includegraphics[width=0.95\columnwidth]{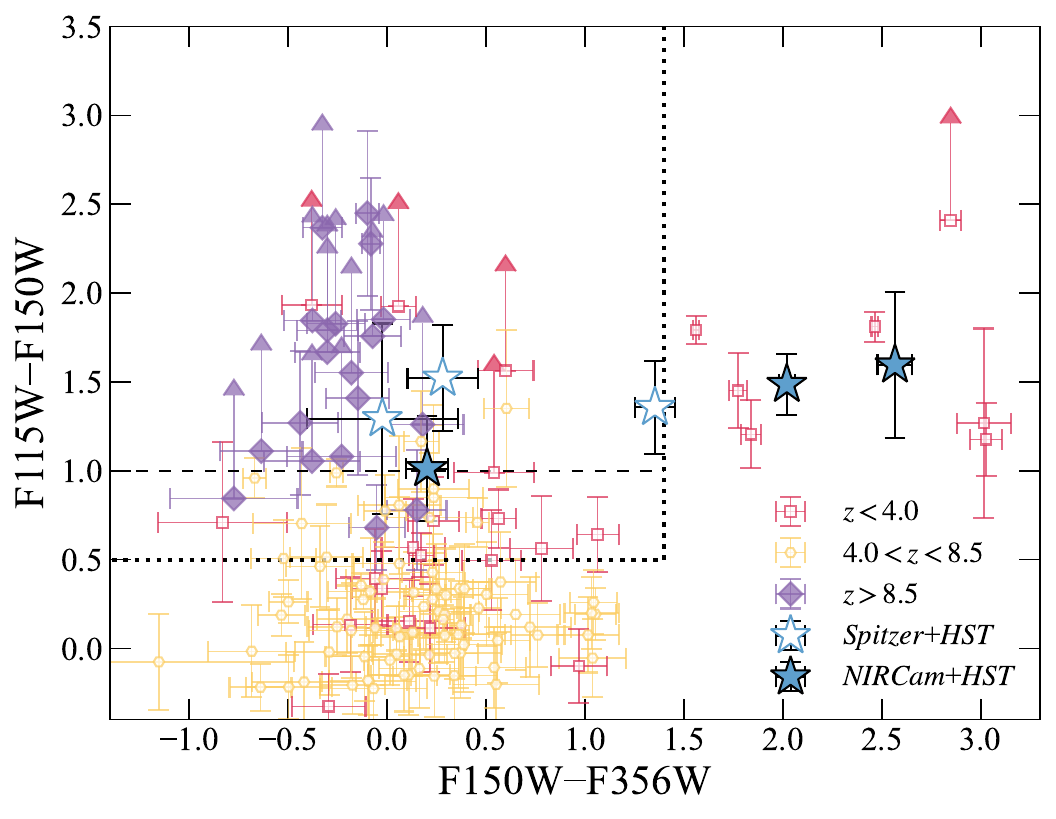}
\end{minipage}
\begin{minipage}[b]{0.49\linewidth}
    \centering
    \includegraphics[width=0.95\columnwidth]{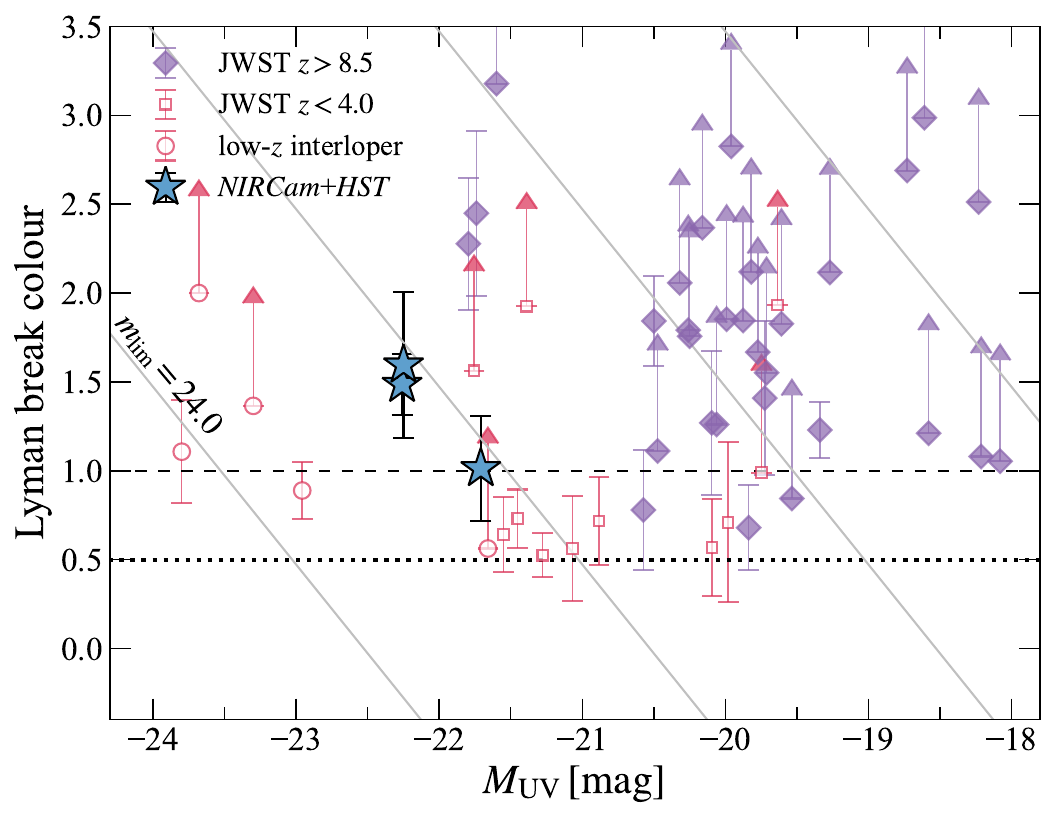}
\end{minipage}
\caption{\textit{Left:}
A colour--colour diagram showing the distribution of galaxies from the JADES and CANUCS spectroscopic catalogues, and our sample galaxies with NIRCam photometry.
The spectroscopically confirmed galaxies in different redshift ranges are shown in different colours and symbols.
Only galaxies that satisfy $S/N_{\mathrm{F090W}}<2$ and $S/N_{\mathrm{F150W}}>5$ are plotted.
Filled blue stars indicate our targets with colours measured based on NIRCam photometry, whereas open blue stars are the targets with colours based on \textit{HST} and \textit{Spitzer}.
Within our targets, UDS\_7815 does not satisfy the dropout criterion in F090W, despite satisfying the colour criteria with NIRCam photometry.
The black dotted lines show the colour selection criteria for $z\sim10$ galaxies equivalent to those employed by \citet{bouwens16b} who used \textit{HST} and \textit{Spitzer}.
The black dashed line is a more strict threshold value ($\mathrm{F115W}-\mathrm{F150W}=1.0$) for reference. 
\textit{Right:}
A relation between Lyman break colour, UV absolute magnitude $M_\mathrm{UV}$, and survey depth $m_\mathrm{lim}$.
In addition to galaxies from the JADES and CANUCS catalogues and our targets, other individual galaxies compiled from the literature are plotted.
All galaxies except for our targets are spectroscopically confirmed.
Symbols are the same as in the left panel, except that purple diamonds also include objects from the literature, and red open circles indicate low-$z$ interlopers from the literature.
The $M_{\mathrm{UV}}$ and the Lyman break colours of low-redshift contaminants are presented in the literature or calculated by fixing redshift to $z=10.0$ (see text for details).
The diagonal lines indicate the constraining power of the Lyman break colours at $z=10$ as functions of $5\sigma$ limiting magnitudes of $m_\mathrm{lim}=24.0,26.0, 28.0, 30.0$ from left to right.
The constraining power is calculated using flux densities corresponding to $2\sigma$ of each survey depth.
Lower limits of the data points are given as $2\sigma$.
The literature from which we compiled samples is follows; \citet{Bowler20, harikane22b,naidu22,arrabalharo23,castellano23,curtis-lake23, donnan23,finkelstein23, robertson23,tacchella23,carniani24,castellano24,sato24,helton25, napolitano25}.
}
\label{fig:color diagram}
\end{figure*}

The left panel of Figure~\ref{fig:color diagram} is a colour--colour diagram of the spectroscopic sample, together with our targets observed with NIRCam.
For our targets, colours measured with \textit{HST} and \textit{Spitzer}/IRAC ($\mathrm{F125W}-\mathrm{F160W}$ and $\mathrm{F150W}-[3.6]$; \textit{Spitzer}-based colours) are also shown by open blue stars for comparison.
The selection colour criteria for $z\sim9$ galaxies are indicated by the dotted lines.
With \textit{Spitzer}-based colours, all three galaxies indeed satisfy the criteria.
However, with colours based on NIRCam photometry (NIRCam-based colours), UDS\_18697 and COSMOS\_20646 do not satisfy the $\mathrm{F150W}-\mathrm{F356W}$ threshold, supporting the scenario that these galaxies are not at $z\sim10$.
These changes in colours were caused by the difficulty of deblending of nearby bright sources in \textit{Spitzer}/IRAC [3.6] (Section~\ref{sec:sed-based discussion}).
UDS\_7815 satisfies the colour criteria with NIRCam-based colours in the left panel of Figure~\ref{fig:color diagram}, but does not satisfy the dropout threshold ($S/N_\mathrm{F090W} = 3.9$). 
Thus, our three target galaxies are no longer selected as $z\sim10$ candidates not only from the SED fitting but also from the colour selection.
\par

The colour criteria have successfully selected all $z>8.5$ galaxies with $S/N_\mathrm{F090W}<2$ and $S/N_\mathrm{F150W}>5$, while rejecting most galaxies at $4<z_\mathrm{spec}<8.5$\footnote{Although some $4<z<8.5$ galaxies satisfy $\mathrm{F115W}-\mathrm{F150W}>0.5$, more than half of them are at $z=8\text{--}8.5$, where Lyman break is redshifted to shorter half of the F115W band.}$^,$\footnote{\textit{JWST} spectroscopic observations samples are not complete and may be biased to galaxies with blue spectra or strong breaks. Addressing this selection bias is beyond the scope of this paper.}.
However, there are two main populations of low-$z$ interlopers, represented by our target galaxies.
\par

The first population is dusty galaxies at $z\sim2\text{--}3$ exhibiting red $\mathrm{F115W}-\mathrm{F150W}$ and $\mathrm{F150W}-\mathrm{F356W}$ colours, like UDS\_18697 and COSMOS\_20646.
While the dusty galaxies are expected to be located on the right side of the selection area (the left panel of Figure~\ref{fig:color diagram}), they may nevertheless be misclassified by the following two reasons.
The first reason is strong emission lines that boost flux density in F150W, resulting in creating a flat continuum and a false break \citep{arrabalharo23,Zavala23,Adams25,Asada25}.
Including medium bands in the photometry is an efficient solution to reject dusty contaminants, as medium bands can trace emission lines \citep{Harikane24,Adams25,Asada25}.
The second reason is photometric discrepancies, which accidentally happen or arise from deblending issues, that make an LBG-like spectrum \citep[e.g., XMM3-3085;][]{Bowler20,Harikane25}.
Using as many bands as possible can be a solution, as they continuously trace a continuum and enable detecting a possible discrepancy in photometry (Section~\ref{sec:sed-based discussion}).
\par

The other population is $z\sim2\text{--}2.5$ Balmer break galaxies (BBGs) exhibiting red $\mathrm{F115W}-\mathrm{F150W}$ colours, like UDS\_7815.
In the spectroscopic sample from the catalogues, about $55\%$ of the $z\sim9$ candidates selected with the criterion of $\mathrm{F115W}-\mathrm{F150W}>0.5$ are actually interlopers at $z_\mathrm{spec}<8.5$ (i.e., the purity is $\simeq45\%$).
If the non-detection bands like F090W are deep enough, we may reject BBGs even with the modest Lyman break colour criterion because BBGs would be detected in the bands. 
Using a more stringent Lyman break colour criterion is also a solution to reject BBGs.
Elevating the criterion to $\mathrm{F115W}-\mathrm{F150W}>1.0$, for instance, can increase the purity of the $z>8.5$ sample to $\simeq75\%$.
Thus, adopting a stricter criterion for the Lyman break colour is an effective approach to select robust high-$z$ candidates.
The effectiveness of a stringent threshold has been shown with \textit{JWST} imaging survey.
For example, \citet{Atek23-uncover} employed a colour criterion of $\mathrm{F115W}-\mathrm{F150W}>1.5$ ($\mathrm{F150W}-\mathrm{F200W}>1.5$) to select $z\sim9\text{--}11$ ($z\sim12\text{--}15$) candidates, and follow-up spectroscopic observation for four of the candidates successfully confirmed that all of them were indeed at $z>9$ \citep{Fujimoto24}.
We note, however, that a stricter Lyman break colour threshold reduces completeness by missing true high-$z$ galaxies.
This is particularly severe for intrinsically faint galaxies because they cannot produce a large colour difference between dropout and detection bands.

\par

\subsubsection{Colour Threshold in the Context of Observation Depth}
\label{sec:Break strength and observation depth}
The Lyman break colour as a function of the UV magnitude enables exploration of another aspect of the selection colour criteria.
Although increasing the criterion for the Lyman break colour can improve the purity of high-$z$ galaxy samples, the maximum Lyman break colour that can be measured (hereafter, referred to as the \textit{constraining power}) is determined by the survey depth.
\par

The right panel of Figure~\ref{fig:color diagram} illustrates the relation between the Lyman break colour, absolute UV magnitude $M_{\mathrm{UV}}$, and observation depth $m_{\mathrm{lim}}$.
In addition to $z_\mathrm{spec}>8.5$ galaxies from JADES and CANUCS, several spectroscopically confirmed $z_\mathrm{spec}>8.5$ galaxies presented in the literature and $z_{\rm spec}<4.0$ galaxies from the catalogues are also shown in the figure.
The Lyman break colour here refers to F115W$-$F150W for $z\sim8.5\text{--}11$ galaxies and F150W$-$F200W for $z\sim11\text{--}16$ galaxies.\footnote{Since NIRCam data are unavailable for some low-$z$ interlopers, we used photometry measured by ground-based telescopes. For HD1 and HD2 (XMM3-3085), which were initially found as $H$-dropouts ($J$-dropout) by \citet{harikane22b} \citep{Bowler20} and confirmed as $z\sim3\text{--}4$ ($z\sim2.5$) by \citet{sato24} \citep{Harikane25}, we alternatively used VISTA or UKIRT $H-K$ ($J-H$) as their Lyman break colours.}
For galaxies from the catalogues, we calculated $M_{\mathrm{UV}}$ using photometry in
F150W (F200W) band for galaxies at $z\sim8.5\text{--}10.5$ ($\sim10.5\text{--}13$), corresponding to $\lambda_\mathrm{rest}\sim130\text{--}167\,\mathrm{nm}$ ($143\text{--}174\,\mathrm{nm}$).
$M_\mathrm{UV}$ of $z_\mathrm{spec}<4$ galaxies (red squares) was calculated in the F150W with a fixed redshift of $z=10$ to illustrate the brightness of low-$z$ interlopers if they were at $z=10$.
For galaxies from the literature, we adopted the published $M_\mathrm{UV}$ values except for HD3 \citep{harikane22b,sato24}.
For HD3, we calculated the $M_\mathrm{UV}$ value using the photometric redshift presented in \citet{harikane22b}.
\par

The right panel of Figure~\ref{fig:color diagram} indicates a difference between true high-$z$ galaxies and low-$z$ interlopers.
The diagonal solid lines in the figure indicate the constraining power of the Lyman break colour for each limiting magnitude $m_\text{lim}$.
The constraining power of the Lyman break colour was calculated using the flux density that corresponds to $2\sigma$ of each limiting magnitude, as we set the dropout threshold of $S/N_\mathrm{F090W}<2$ and consequently give $2\sigma$ lower limits on the Lyman break colour.
This means that data points, including measurements and lower limits, must be located below a diagonal line corresponding to each observation depth.
In other words, the faintest high-$z$ candidates that can be selected with a given limiting magnitude are located close to an intersection of a diagonal line (survey depth) and a horizontal line (colour criterion).
Therefore, the right panel of Figure~\ref{fig:color diagram} illustrates that many of the low-$z$ interlopers are barely selected with both the detection and selection criteria. 
Such data points are easily affected by noise fluctuations, and thus, it is not surprising that they are low-$z$ interlopers.
\par

These contaminations are more critical in shallower surveys.
For example, surveys with $m_\mathrm{lim}=26.0\,\mathrm{mag}$ will search for galaxies with $M_\mathrm{UV}\lesssim-21.5\,\mathrm{mag}$ with the Lyman break colour of $\gtrsim1.0$.
Such galaxies are brighter than the characteristic magnitude $M^*$ of the luminosity function at $z\sim8\text{--}10$, $M^*=-21.0\,~\mathrm{mag}$ \citep{PerezGonzalez23}, where the luminosity function follows an exponential or a steeper power-law function.
This rapid decrease in the number density further reduces the probability that the selected candidates are certainly at $z>8.5$.
Indeed, \citet{Harikane25} showed that the number density of bright ($M_\mathrm{UV}\lesssim-21.5\,\mathrm{mag}$) $z\sim10\text{--}12$ galaxies will be smaller than that of interlopers if only $0.2\%$ of passive galaxies at $z\sim2.5\text{--}4$ are misclassified as high-$z$ galaxies, indicating that the purity in apparently brighter galaxies can be lower than in fainter galaxies.
\par

\subsubsection{Implications to future imaging surveys}
These discussions will give some insights into current and future IR imaging surveys.
First, contaminations discussed above would not be a serious problem for \textit{JWST} observations. 
As shown in the right panel of Figure~\ref{fig:color diagram}, faint galaxies are selected as $z>8.5$ and indeed spectroscopically confirmed as high redshift, even though they show relatively small ($\sim0.7\,\mathrm{mag}$) Lyman break colour.
These galaxies benefit from high-resolution, high-sensitivity, multi-band imaging with NIRCam.
Low-$z$ interlopers ($z_\mathrm{spec}<4$) observed with \textit{JWST} are not as bright as low-$z$ interlopers found by ground-based facilities, reflecting the deeper survey depths.
Since the \textit{JWST} low-$z$ interlopers are fainter than the characteristic magnitude of luminosity functions at $z\sim8\text{--}10$\footnote{Although galaxies with $M_\mathrm{UV}\lesssim-21.0\,\mathrm{mag}$ are expected to be detected in F090W with a limiting magnitude of the JADES and CANUCS, several $z_\mathrm{spec}<4$ galaxies with that brightness were selected by the colour criteria.
This is due to their larger photometry error in F090W by a factor of $\sim2$, which is equivalent to degrading the observation depth by $\sim0.9\,\mathrm{mag}$.
This reasonably explains why they were not detected in F090W.}, their influence on purity is smaller than that on bright LBGs.
\par

Upcoming wide-field surveys, on the other hand, can be affected by the contamination discussed above.
Large imaging surveys with \textit{Euclid}, \textit{Roman}, and \textit{GREX-PLUS} are expected to find a large number of high-$z$ galaxy candidates.
As part of the Euclid Deep Fields, \textit{Euclid} will observe over $\sim50\,\mathrm{deg^2}$ with $5\sigma$ depths of about $26.4\,\mathrm{mag}$ in $Y_\mathrm{E}, J_\mathrm{E}$, and $H_\mathrm{E}$-band \citep{Euclid,EuclidDeepField} with angular resolutions of $0.18\,$arcsec.
This survey depth enables detections of galaxies with $M_\mathrm{UV}\lesssim-21\,\mathrm{mag}$ with the Lyman break colour criterion of $\gtrsim1.0$ (the right panel of Figure~\ref{fig:color diagram}).
As discussed above, the $z\sim8\text{--}10$ candidates selected with this absolute UV magnitude will be as bright as or brighter than the characteristic magnitude $M^*$, which can be highly affected by low-$z$ interlopers (Section \ref{sec:Break strength and observation depth}).
In addition, as the \textit{Euclid} reddest filter is the $H$-band, the Lyman break method has to rely on \textit{Spitzer}/IRAC photometry \citep{Euclid-IRAC}, which would suffer similar limitations to our target selection in terms of the number of filters and the difficulty of deblending.
Therefore, despite the tradeoff with completeness, increasing a purity with a stringent colour criterion and deep observation blueward of the Lyman break will be crucial to avoid heavy contamination in high-$z$ candidate samples.
\par

The Roman High-Latitude Wide-Area Survey (HLWAS, Deep Tier) will survey over $\sim19\,\mathrm{deg^2}$ sky with $5\sigma$ depths of $27.7\,\mathrm{mag}$ in $Y$, $J$, $H$-band and $25.9\,\mathrm{mag}$ in $K$-band \citep{roman}.
The depths in $J$, $H$-bands enable a search for galaxies $>1$ mag fainter than the characteristic UV magnitude at $z\sim8$--$10$ (the right panel of Figure~\ref{fig:color diagram}).
The high angular resolution ($\sim0\farcs1$) and redder $K$-band observations of the Roman HLWAS can reduce contaminations in the high-$z$ candidates.
\par

Finally, \textit{GREX-PLUS}, a mission concept of a space telescope, plans to offer DEEP imaging surveys at $2.0\text{--}4.5\,\mu$m ($26.5\,\mathrm{mag}$) and $4.5\text{--}7.7\,\mu$m ($\geq24.5\,\mathrm{mag}$) over $10\,\mathrm{deg^2}$ sky with angular resolution higher than \textit{Spitzer} by a factor of $1.5$ \citep{grex}.
The combinations of these IR wide-field surveys will continuously trace rest-frame UV to optical spectra of $z\sim10$ galaxies and relax the difficulty of deblending.
These deep, multi-band, high-resolution surveys, with the aid of careful target selections, will help to construct robust high-$z$ candidate samples at $z\sim10$ and beyond, which are essential for efficient follow-up spectroscopy.
In addition to \textit{JWST}, $30\text{--}40\,\textrm{m}$ optical-to-IR ground-based telescopes and ALMA with Wideband Sensitivity Upgrade \citep[WSU;][]{wsu} will play significant roles for confirming galaxies in the first $500\,\mathrm{Myr}$ of the Universe.
\par

\section{Conclusion}
\label{sec:conclusion}
In this paper, we have reported ALMA [\ion{O}{iii}]\,$88\,\mu$m observation for six $z\sim10$ candidates selected by \textit{\textit{HST}}, \textit{\textit{Spitzer}}, and ground-based facilities.
Our main findings are:
\begin{enumerate}
    \item We have detected possible line features in UDS\_18697 ($S/N=4.5$).
    However, \textit{JWST}/NIRSpec follow-up observations confirmed the redshifts as $z=2.54$ (R. Larson et al. in prep.), confirming the detected line as CO($J=9\text{--}8$) at $z=2.536$.
    The remaining five galaxies show no line emission.
    
    \item UDS\_18697 also shows a strong dust continuum emission with the significance level of $29.9\sigma$. The derived FIR luminosity is $L_\mathrm{FIR}=1.1^{+1.0}_{-0.3}\times10^{12}\,L_{\odot}$, classifying the galaxy as a ULIRG. 
    The FIR luminosity of UDS\_18697 is one order of magnitude fainter than previous CO($J=9\text{--}8$) detected $z>2$ DSFGs, making UDS\_18697 one of the faintest examples at $z>2$. 
    Together with the CO line luminosity, we have found that UDS\_18697 lies slightly below the local $L_\mathrm{FIR}-L'_\mathrm{CO}$ relation, though it is still consistent with the local relation within the uncertainties. 
    
    \item Although our targets were suggested to be at $z\sim10$ in previous photometry based on \textit{HST} and \textit{Spitzer}, SED fitting with newly available \textit{JWST}/NIRCam photometry suggests that all three galaxies observed with \textit{JWST}/NIRCam among the six target galaxies are at $z\sim2\text{--}2.5$. Based on SED fitting, they are likely to be dusty star-forming or Balmer break galaxies at low redshift. One of the three galaxies is UDS\_18697, whose best-fit redshift from the SED fitting is fully consistent with that confirmed with \textit{JWST}/NIRSpec ($z=2.54$).
    COSMOS-z10-1 and COSMOS-z10-2 are also suggested to be at $z\sim2.8$ and $z\sim1.7$, respectively, based on follow-up \textit{JWST}/NIRSpec observations (J. Weaver et al. in prep.; N. Nezhad et al. in prep.).
    J2140+0241 is still not conclusively identified due to a lack of new data.

    \item The reasons that caused the previous misclassification are summarised into the following three points: difficulty of deblending due to coarse spatial resolution of \textit{Spitzer}/IRAC, lack of multiple filters that continuously trace continuum emission, and insufficient deep observation at blueward of the Lyman break.
    
    \item Based on JADES and CANUCS data, we have found that increasing the $\mathrm{F115W}-\mathrm{F150W}$ threshold value, which traces the Lyman break colour, can reduce low-$z$ interlopers while keeping many of the true $z>8.5$ galaxies. 
    We have also confirmed that low-$z$ interlopers reported so far are brighter than the characteristic magnitude $M_*$ of UV luminosity functions, where the number density rapidly decreases. 
    Thus, low-$z$ interlopers can be much more abundant than the true high-$z$ galaxies. 
    Since upcoming wide-field surveys will search for bright Lyman break galaxies, constructing reliable high-$z$ candidate samples for efficient spectroscopic follow-up observations requires stringent Lyman break criteria, high-resolution imaging, wide wavelength coverage, and sufficiently deep observations at short wavelengths.
\end{enumerate}

\section*{Acknowledgements}
We appreciate the anonymous referee’s constructive feedback, which significantly improved the manuscript.
We thank Yoshihisa Asada and Georgios Magdis for insightful discussions.
We also acknowledge ALMA Helpdesk staff members for calibrating the ALMA datasets.
This work was supported by JSPS KAKENHI Grant Numbers JP22H04939, JP23H00131, JP24H00002, JP26H02069, and JP26K17200.
T.H. was supported by Leading Initiative for Excellent Young Researchers, MEXT, Japan (HJH02007) and by JSPS KAKENHI grant Nos. JP23K22529 and JP25K00020.
Y.S. and K.M. acknowledge the Waseda University Grant for Special Research Projects (Project number: 2025E-019 and 2025C-484).
J.R.W. acknowledges that support for this work was provided by The Brinson Foundation through a Brinson Prize Fellowship grant.
This paper makes use of the following ALMA data: ADS/JAO.ALMA\#2019.1.00397.S and ADS/JAO.ALMA\#2021.1.01396.S.
ALMA is a partnership of ESO (representing its member states), NSF (USA) and NINS (Japan), together with NRC (Canada), MOST and ASIAA (Taiwan), and KASI (Republic of Korea), in cooperation with the Republic of Chile.
The Joint ALMA Observatory is operated by ESO, AUI/NRAO and NAOJ. 

\section*{Data Availability}
All data utilised in this work is available upon reasonable request to the corresponding author.



\bibliographystyle{mnras}
\bibliography{references} 

@ARTICLE{schouws25,
       author = {{Schouws}, Sander and {Bouwens}, Rychard J. and {Ormerod}, Katherine and {Smit}, Renske and {Algera}, Hiddo and {Sommovigo}, Laura and {Hodge}, Jacqueline and {Ferrara}, Andrea and {Oesch}, Pascal A. and {Rowland}, Lucie E. and {van Leeuwen}, Ivana and {Stefanon}, Mauro and {Herard-Demanche}, Thomas and {Fudamoto}, Yoshinobu and {R{\"o}ttgering}, Huub and {van der Werf}, Paul},
        title = "{Detection of [O III]$_{88 {\ensuremath{\mu}}m}$ in JADES-GS-z14-0 at z = 14.1793}",
      journal = {\apj},
         year = 2025,
        month = jul,
       volume = {988},
       number = {1},
          eid = {19},
        pages = {19},
          doi = {10.3847/1538-4357/adbf1b},
archivePrefix = {arXiv},
       eprint = {2409.20549},
 primaryClass = {astro-ph.GA},
       adsurl = {https://ui.adsabs.harvard.edu/abs/2025ApJ...988...19S}
}

@ARTICLE{Harikane25,
       author = {{Harikane}, Yuichi and {Inoue}, Akio K. and {Ellis}, Richard S. and {Ouchi}, Masami and {Nakazato}, Yurina and {Yoshida}, Naoki and {Ono}, Yoshiaki and {Sun}, Fengwu and {Sato}, Riku A. and {Ferrami}, Giovanni and {Fujimoto}, Seiji and {Kashikawa}, Nobunari and {McLeod}, Derek J. and {P{\'e}rez-Gonz{\'a}lez}, Pablo G. and {Sawicki}, Marcin and {Sugahara}, Yuma and {Xu}, Yi and {Yamanaka}, Satoshi and {Carnall}, Adam C. and {Cullen}, Fergus and {Dunlop}, James S. and {Egami}, Eiichi and {Grogin}, Norman and {Isobe}, Yuki and {Koekemoer}, Anton M. and {Laporte}, Nicolas and {Lee}, Chien-Hsiu and {Magee}, Dan and {Matsuo}, Hiroshi and {Matsuoka}, Yoshiki and {Mawatari}, Ken and {Nakajima}, Kimihiko and {Nakane}, Minami and {Tamura}, Yoichi and {Umeda}, Hiroya and {Yanagisawa}, Hiroto},
        title = "{JWST, ALMA, and Keck Spectroscopic Constraints on the UV Luminosity Functions at z {\ensuremath{\sim}} 7{\textendash}14: Clumpiness and Compactness of the Brightest Galaxies in the Early Universe}",
      journal = {\apj},
         year = 2025,
        month = feb,
       volume = {980},
       number = {1},
          eid = {138},
        pages = {138},
          doi = {10.3847/1538-4357/ad9b2c},
archivePrefix = {arXiv},
       eprint = {2406.18352},
 primaryClass = {astro-ph.GA},
       adsurl = {https://ui.adsabs.harvard.edu/abs/2025ApJ...980..138H}
}

@ARTICLE{merlin24,
       author = {{Merlin}, E. and {Santini}, P. and {Paris}, D. and {Castellano}, M. and {Fontana}, A. and {Treu}, T. and {Finkelstein}, S.~L. and {Dunlop}, J.~S. and {Arrabal Haro}, P. and {Bagley}, M. and {Boyett}, K. and {Calabr{\`o}}, A. and {Correnti}, M. and {Davis}, K. and {Dickinson}, M. and {Donnan}, C.~T. and {Ferguson}, H.~C. and {Fortuni}, F. and {Giavalisco}, M. and {Glazebrook}, K. and {Grazian}, A. and {Grogin}, N.~A. and {Hathi}, N. and {Hirschmann}, M. and {Kartaltepe}, J.~S. and {Kewley}, L.~J. and {Kirkpatrick}, A. and {Kocevski}, D.~D. and {Koekemoer}, A.~M. and {Leung}, G. and {Lotz}, J.~M. and {Lucas}, R.~A. and {Magee}, D.~K. and {Marchesini}, D. and {Mascia}, S. and {McLeod}, D.~J. and {McLure}, R.~J. and {Nanayakkara}, T. and {Napolitano}, L. and {Nonino}, M. and {Papovich}, C. and {Pentericci}, L. and {P{\'e}rez-Gonz{\'a}lez}, P.~G. and {Pirzkal}, N. and {Ravindranath}, S. and {Roberts-Borsani}, G. and {Somerville}, R.~S. and {Trenti}, M. and {Trump}, J.~R. and {Vulcani}, B. and {Wang}, X. and {Watson}, P.~J. and {Wilkins}, S.~M. and {Yang}, G. and {Yung}, L.~Y.~A.},
        title = "{ASTRODEEP-JWST: NIRCam-HST multi-band photometry and redshifts for half a million sources in six extragalactic deep fields}",
      journal = {\aap},
         year = 2024,
        month = nov,
       volume = {691},
          eid = {A240},
        pages = {A240},
          doi = {10.1051/0004-6361/202451409},
archivePrefix = {arXiv},
       eprint = {2409.00169},
 primaryClass = {astro-ph.GA},
       adsurl = {https://ui.adsabs.harvard.edu/abs/2024A&A...691A.240M}
}

@ARTICLE{sato24,
       author = {{Sato}, Riku A. and {Inoue}, Akio K. and {Harikane}, Yuichi and {Shimakawa}, Rhythm and {Sugahara}, Yuma and {Tamura}, Yoichi and {Hashimoto}, Takuya and {Ito}, Kei and {Yamanaka}, Satoshi and {Mawatari}, Ken and {Fudamoto}, Yoshinobu and {Ren}, Yi W.},
        title = "{JWST/NIRSpec spectroscopy of intermediate-mass quiescent galaxies at z   3-4}",
      journal = {\mnras},
         year = 2024,
        month = nov,
       volume = {534},
       number = {4},
        pages = {3552-3564},
          doi = {10.1093/mnras/stae2300},
archivePrefix = {arXiv},
       eprint = {2410.08745},
 primaryClass = {astro-ph.GA},
       adsurl = {https://ui.adsabs.harvard.edu/abs/2024MNRAS.534.3552S}
}

@ARTICLE{harikane22b,
       author = {{Harikane}, Yuichi and {Inoue}, Akio K. and {Mawatari}, Ken and {Hashimoto}, Takuya and {Yamanaka}, Satoshi and {Fudamoto}, Yoshinobu and {Matsuo}, Hiroshi and {Tamura}, Yoichi and {Dayal}, Pratika and {Yung}, L.~Y. Aaron and {Hutter}, Anne and {Pacucci}, Fabio and {Sugahara}, Yuma and {Koekemoer}, Anton M.},
        title = "{A Search for H-Dropout Lyman Break Galaxies at z 12-16}",
      journal = {\apj},
         year = 2022,
        month = apr,
       volume = {929},
       number = {1},
          eid = {1},
        pages = {1},
          doi = {10.3847/1538-4357/ac53a9},
archivePrefix = {arXiv},
       eprint = {2112.09141},
 primaryClass = {astro-ph.GA},
       adsurl = {https://ui.adsabs.harvard.edu/abs/2022ApJ...929....1H}
}

@ARTICLE{weaver22,
       author = {{Weaver}, J.~R. and {Kauffmann}, O.~B. and {Ilbert}, O. and {McCracken}, H.~J. and {Moneti}, A. and {Toft}, S. and {Brammer}, G. and {Shuntov}, M. and {Davidzon}, I. and {Hsieh}, B.~C. and {Laigle}, C. and {Anastasiou}, A. and {Jespersen}, C.~K. and {Vinther}, J. and {Capak}, P. and {Casey}, C.~M. and {McPartland}, C.~J.~R. and {Milvang-Jensen}, B. and {Mobasher}, B. and {Sanders}, D.~B. and {Zalesky}, L. and {Arnouts}, S. and {Aussel}, H. and {Dunlop}, J.~S. and {Faisst}, A. and {Franx}, M. and {Furtak}, L.~J. and {Fynbo}, J.~P.~U. and {Gould}, K.~M.~L. and {Greve}, T.~R. and {Gwyn}, S. and {Kartaltepe}, J.~S. and {Kashino}, D. and {Koekemoer}, A.~M. and {Kokorev}, V. and {Le F{\`e}vre}, O. and {Lilly}, S. and {Masters}, D. and {Magdis}, G. and {Mehta}, V. and {Peng}, Y. and {Riechers}, D.~A. and {Salvato}, M. and {Sawicki}, M. and {Scarlata}, C. and {Scoville}, N. and {Shirley}, R. and {Silverman}, J.~D. and {Sneppen}, A. and {Smolc̆i{\'c}}, V. and {Steinhardt}, C. and {Stern}, D. and {Tanaka}, M. and {Taniguchi}, Y. and {Teplitz}, H.~I. and {Vaccari}, M. and {Wang}, W. -H. and {Zamorani}, G.},
        title = "{COSMOS2020: A Panchromatic View of the Universe to z{\ensuremath{\sim}}10 from Two Complementary Catalogs}",
      journal = {\apjs},
         year = 2022,
        month = jan,
       volume = {258},
       number = {1},
          eid = {11},
        pages = {11},
          doi = {10.3847/1538-4365/ac3078},
archivePrefix = {arXiv},
       eprint = {2110.13923},
 primaryClass = {astro-ph.GA},
       adsurl = {https://ui.adsabs.harvard.edu/abs/2022ApJS..258...11W}
}

@ARTICLE{planck20,
       author = {{Planck Collaboration} and {Aghanim}, N. and {Akrami}, Y. and {Ashdown}, M. and {Aumont}, J. and {Baccigalupi}, C. and {Ballardini}, M. and {Banday}, A.~J. and {Barreiro}, R.~B. and {Bartolo}, N. and {Basak}, S. and {Battye}, R. and {Benabed}, K. and {Bernard}, J. -P. and {Bersanelli}, M. and {Bielewicz}, P. and {Bock}, J.~J. and {Bond}, J.~R. and {Borrill}, J. and {Bouchet}, F.~R. and {Boulanger}, F. and {Bucher}, M. and {Burigana}, C. and {Butler}, R.~C. and {Calabrese}, E. and {Cardoso}, J. -F. and {Carron}, J. and {Challinor}, A. and {Chiang}, H.~C. and {Chluba}, J. and {Colombo}, L.~P.~L. and {Combet}, C. and {Contreras}, D. and {Crill}, B.~P. and {Cuttaia}, F. and {de Bernardis}, P. and {de Zotti}, G. and {Delabrouille}, J. and {Delouis}, J. -M. and {Di Valentino}, E. and {Diego}, J.~M. and {Dor{\'e}}, O. and {Douspis}, M. and {Ducout}, A. and {Dupac}, X. and {Dusini}, S. and {Efstathiou}, G. and {Elsner}, F. and {En{\ss}lin}, T.~A. and {Eriksen}, H.~K. and {Fantaye}, Y. and {Farhang}, M. and {Fergusson}, J. and {Fernandez-Cobos}, R. and {Finelli}, F. and {Forastieri}, F. and {Frailis}, M. and {Fraisse}, A.~A. and {Franceschi}, E. and {Frolov}, A. and {Galeotta}, S. and {Galli}, S. and {Ganga}, K. and {G{\'e}nova-Santos}, R.~T. and {Gerbino}, M. and {Ghosh}, T. and {Gonz{\'a}lez-Nuevo}, J. and {G{\'o}rski}, K.~M. and {Gratton}, S. and {Gruppuso}, A. and {Gudmundsson}, J.~E. and {Hamann}, J. and {Handley}, W. and {Hansen}, F.~K. and {Herranz}, D. and {Hildebrandt}, S.~R. and {Hivon}, E. and {Huang}, Z. and {Jaffe}, A.~H. and {Jones}, W.~C. and {Karakci}, A. and {Keih{\"a}nen}, E. and {Keskitalo}, R. and {Kiiveri}, K. and {Kim}, J. and {Kisner}, T.~S. and {Knox}, L. and {Krachmalnicoff}, N. and {Kunz}, M. and {Kurki-Suonio}, H. and {Lagache}, G. and {Lamarre}, J. -M. and {Lasenby}, A. and {Lattanzi}, M. and {Lawrence}, C.~R. and {Le Jeune}, M. and {Lemos}, P. and {Lesgourgues}, J. and {Levrier}, F. and {Lewis}, A. and {Liguori}, M. and {Lilje}, P.~B. and {Lilley}, M. and {Lindholm}, V. and {L{\'o}pez-Caniego}, M. and {Lubin}, P.~M. and {Ma}, Y. -Z. and {Mac{\'\i}as-P{\'e}rez}, J.~F. and {Maggio}, G. and {Maino}, D. and {Mandolesi}, N. and {Mangilli}, A. and {Marcos-Caballero}, A. and {Maris}, M. and {Martin}, P.~G. and {Martinelli}, M. and {Mart{\'\i}nez-Gonz{\'a}lez}, E. and {Matarrese}, S. and {Mauri}, N. and {McEwen}, J.~D. and {Meinhold}, P.~R. and {Melchiorri}, A. and {Mennella}, A. and {Migliaccio}, M. and {Millea}, M. and {Mitra}, S. and {Miville-Desch{\^e}nes}, M. -A. and {Molinari}, D. and {Montier}, L. and {Morgante}, G. and {Moss}, A. and {Natoli}, P. and {N{\o}rgaard-Nielsen}, H.~U. and {Pagano}, L. and {Paoletti}, D. and {Partridge}, B. and {Patanchon}, G. and {Peiris}, H.~V. and {Perrotta}, F. and {Pettorino}, V. and {Piacentini}, F. and {Polastri}, L. and {Polenta}, G. and {Puget}, J. -L. and {Rachen}, J.~P. and {Reinecke}, M. and {Remazeilles}, M. and {Renzi}, A. and {Rocha}, G. and {Rosset}, C. and {Roudier}, G. and {Rubi{\~n}o-Mart{\'\i}n}, J.~A. and {Ruiz-Granados}, B. and {Salvati}, L. and {Sandri}, M. and {Savelainen}, M. and {Scott}, D. and {Shellard}, E.~P.~S. and {Sirignano}, C. and {Sirri}, G. and {Spencer}, L.~D. and {Sunyaev}, R. and {Suur-Uski}, A. -S. and {Tauber}, J.~A. and {Tavagnacco}, D. and {Tenti}, M. and {Toffolatti}, L. and {Tomasi}, M. and {Trombetti}, T. and {Valenziano}, L. and {Valiviita}, J. and {Van Tent}, B. and {Vibert}, L. and {Vielva}, P. and {Villa}, F. and {Vittorio}, N. and {Wandelt}, B.~D. and {Wehus}, I.~K. and {White}, M. and {White}, S.~D.~M. and {Zacchei}, A. and {Zonca}, A.},
        title = "{Planck 2018 results. VI. Cosmological parameters}",
      journal = {\aap},
         year = 2020,
        month = sep,
       volume = {641},
          eid = {A6},
        pages = {A6},
          doi = {10.1051/0004-6361/201833910},
archivePrefix = {arXiv},
       eprint = {1807.06209},
 primaryClass = {astro-ph.CO},
       adsurl = {https://ui.adsabs.harvard.edu/abs/2020A&A...641A...6P}
}

@ARTICLE{hashimoto19,
       author = {{Hashimoto}, Takuya and {Inoue}, Akio K. and {Mawatari}, Ken and {Tamura}, Yoichi and {Matsuo}, Hiroshi and {Furusawa}, Hisanori and {Harikane}, Yuichi and {Shibuya}, Takatoshi and {Knudsen}, Kirsten K. and {Kohno}, Kotaro and {Ono}, Yoshiaki and {Zackrisson}, Erik and {Okamoto}, Takashi and {Kashikawa}, Nobunari and {Oesch}, Pascal A. and {Ouchi}, Masami and {Ota}, Kazuaki and {Shimizu}, Ikkoh and {Taniguchi}, Yoshiaki and {Umehata}, Hideki and {Watson}, Darach},
        title = "{Big Three Dragons: A z = 7.15 Lyman-break galaxy detected in [O III] 88 {\ensuremath{\mu}}m, [C II] 158 {\ensuremath{\mu}}m, and dust continuum with ALMA}",
      journal = {\pasj},
         year = 2019,
        month = aug,
       volume = {71},
       number = {4},
          eid = {71},
        pages = {71},
          doi = {10.1093/pasj/psz049},
archivePrefix = {arXiv},
       eprint = {1806.00486},
 primaryClass = {astro-ph.GA},
       adsurl = {https://ui.adsabs.harvard.edu/abs/2019PASJ...71...71H}
}

@ARTICLE{morishita18,
       author = {{Morishita}, T. and {Trenti}, M. and {Stiavelli}, M. and {Bradley}, L.~D. and {Coe}, D. and {Oesch}, P.~A. and {Mason}, C.~A. and {Bridge}, J.~S. and {Holwerda}, B.~W. and {Livermore}, R.~C. and {Salmon}, B. and {Schmidt}, K.~B. and {Shull}, J.~M. and {Treu}, T.},
        title = "{The Bright-end Galaxy Candidates at z {\ensuremath{\sim}} 9 from 79 Independent HST Fields}",
      journal = {\apj},
         year = 2018,
        month = nov,
       volume = {867},
       number = {2},
          eid = {150},
        pages = {150},
          doi = {10.3847/1538-4357/aae68c},
archivePrefix = {arXiv},
       eprint = {1809.07604},
 primaryClass = {astro-ph.GA},
       adsurl = {https://ui.adsabs.harvard.edu/abs/2018ApJ...867..150M}
}

@ARTICLE{carnall18,
       author = {{Carnall}, A.~C. and {McLure}, R.~J. and {Dunlop}, J.~S. and {Dav{\'e}}, R.},
        title = "{Inferring the star formation histories of massive quiescent galaxies with BAGPIPES: evidence for multiple quenching mechanisms}",
      journal = {\mnras},
         year = 2018,
        month = nov,
       volume = {480},
       number = {4},
        pages = {4379-4401},
          doi = {10.1093/mnras/sty2169},
archivePrefix = {arXiv},
       eprint = {1712.04452},
 primaryClass = {astro-ph.GA},
       adsurl = {https://ui.adsabs.harvard.edu/abs/2018MNRAS.480.4379C}
}

@ARTICLE{inoue16,
       author = {{Inoue}, Akio K. and {Tamura}, Yoichi and {Matsuo}, Hiroshi and {Mawatari}, Ken and {Shimizu}, Ikkoh and {Shibuya}, Takatoshi and {Ota}, Kazuaki and {Yoshida}, Naoki and {Zackrisson}, Erik and {Kashikawa}, Nobunari and {Kohno}, Kotaro and {Umehata}, Hideki and {Hatsukade}, Bunyo and {Iye}, Masanori and {Matsuda}, Yuichi and {Okamoto}, Takashi and {Yamaguchi}, Yuki},
        title = "{Detection of an oxygen emission line from a high-redshift galaxy in the reionization epoch}",
      journal = {Science},
         year = 2016,
        month = jun,
       volume = {352},
       number = {6293},
        pages = {1559-1562},
          doi = {10.1126/science.aaf0714},
archivePrefix = {arXiv},
       eprint = {1606.04989},
 primaryClass = {astro-ph.GA},
       adsurl = {https://ui.adsabs.harvard.edu/abs/2016Sci...352.1559I}
}

@ARTICLE{brammer08,
       author = {{Brammer}, Gabriel B. and {van Dokkum}, Pieter G. and {Coppi}, Paolo},
        title = "{EAZY: A Fast, Public Photometric Redshift Code}",
      journal = {\apj},
         year = 2008,
        month = oct,
       volume = {686},
       number = {2},
        pages = {1503-1513},
          doi = {10.1086/591786},
archivePrefix = {arXiv},
       eprint = {0807.1533},
 primaryClass = {astro-ph},
       adsurl = {https://ui.adsabs.harvard.edu/abs/2008ApJ...686.1503B}
}

@INPROCEEDINGS{mullin07,
       author = {{McMullin}, J.~P. and {Waters}, B. and {Schiebel}, D. and {Young}, W. and {Golap}, K.},
        title = "{CASA Architecture and Applications}",
    booktitle = {Astronomical Data Analysis Software and Systems XVI},
         year = 2007,
       editor = {{Shaw}, R.~A. and {Hill}, F. and {Bell}, D.~J.},
       series = {Astronomical Society of the Pacific Conference Series},
       volume = {376},
        month = oct,
        pages = {127},
       adsurl = {https://ui.adsabs.harvard.edu/abs/2007ASPC..376..127M}
}

@ARTICLE{calzetti00,
       author = {{Calzetti}, Daniela and {Armus}, Lee and {Bohlin}, Ralph C. and {Kinney}, Anne L. and {Koornneef}, Jan and {Storchi-Bergmann}, Thaisa},
        title = "{The Dust Content and Opacity of Actively Star-forming Galaxies}",
      journal = {\apj},
         year = 2000,
        month = apr,
       volume = {533},
       number = {2},
        pages = {682-695},
          doi = {10.1086/308692},
archivePrefix = {arXiv},
       eprint = {astro-ph/9911459},
 primaryClass = {astro-ph},
       adsurl = {https://ui.adsabs.harvard.edu/abs/2000ApJ...533..682C}
}

@ARTICLE{oke83,
       author = {{Oke}, J.~B. and {Gunn}, J.~E.},
        title = "{Secondary standard stars for absolute spectrophotometry.}",
      journal = {\apj},
         year = 1983,
        month = mar,
       volume = {266},
        pages = {713-717},
          doi = {10.1086/160817},
       adsurl = {https://ui.adsabs.harvard.edu/abs/1983ApJ...266..713O}
}

@ARTICLE{finkelstein22,
       author = {{Finkelstein}, Steven L. and {Bagley}, Micaela and {Song}, Mimi and {Larson}, Rebecca and {Papovich}, Casey and {Dickinson}, Mark and {Finkelstein}, Keely D. and {Koekemoer}, Anton M. and {Pirzkal}, Norbert and {Somerville}, Rachel S. and {Yung}, L.~Y. Aaron and {Behroozi}, Peter and {Ferguson}, Harry and {Giavalisco}, Mauro and {Grogin}, Norman and {Hathi}, Nimish and {Hutchison}, Taylor A. and {Jung}, Intae and {Kocevski}, Dale and {Kawinwanichakij}, Lalitwadee and {Rojas-Ruiz}, Sof{\'\i}a and {Ryan}, Russell and {Snyder}, Gregory F. and {Tacchella}, Sandro},
        title = "{A Census of the Bright z = 8.5-11 Universe with the Hubble and Spitzer Space Telescopes in the CANDELS Fields}",
      journal = {\apj},
         year = 2022,
        month = mar,
       volume = {928},
       number = {1},
          eid = {52},
        pages = {52},
          doi = {10.3847/1538-4357/ac3aed},
archivePrefix = {arXiv},
       eprint = {2106.13813},
 primaryClass = {astro-ph.GA},
       adsurl = {https://ui.adsabs.harvard.edu/abs/2022ApJ...928...52F}
}

@ARTICLE{carniani25,
       author = {{Carniani}, Stefano and {D'Eugenio}, Francesco and {Ji}, Xihan and {Parlanti}, Eleonora and {Scholtz}, Jan and {Sun}, Fengwu and {Venturi}, Giacomo and {Bakx}, Tom J.~L.~C. and {Curti}, Mirko and {Maiolino}, Roberto and {Tacchella}, Sandro and {Zavala}, Jorge A. and {Hainline}, Kevin and {Witstok}, Joris and {Johnson}, Benjamin D. and {Alberts}, Stacey and {Bunker}, Andrew J. and {Charlot}, St{\'e}phane and {Eisenstein}, Daniel J. and {Helton}, Jakob M. and {Jakobsen}, Peter and {Kumari}, Nimisha and {Robertson}, Brant and {Saxena}, Aayush and {{\"U}bler}, Hannah and {Williams}, Christina C. and {Willmer}, Christopher N.~A. and {Willott}, Chris},
        title = "{The eventful life of a luminous galaxy at z = 14: metal enrichment, feedback, and low gas fraction?}",
      journal = {\aap},
         year = 2025,
        month = apr,
       volume = {696},
          eid = {A87},
        pages = {A87},
          doi = {10.1051/0004-6361/202452451},
archivePrefix = {arXiv},
       eprint = {2409.20533},
 primaryClass = {astro-ph.GA},
       adsurl = {https://ui.adsabs.harvard.edu/abs/2025A&A...696A..87C}
}

@ARTICLE{bouwens16b,
       author = {{Bouwens}, R.~J. and {Oesch}, P.~A. and {Labb{\'e}}, I. and {Illingworth}, G.~D. and {Fazio}, G.~G. and {Coe}, D. and {Holwerda}, B. and {Smit}, R. and {Stefanon}, M. and {van Dokkum}, P.~G. and {Trenti}, M. and {Ashby}, M.~L.~N. and {Huang}, J. -S. and {Spitler}, L. and {Straatman}, C. and {Bradley}, L. and {Magee}, D.},
        title = "{The Bright End of the z {\ensuremath{\sim}} 9 and z {\ensuremath{\sim}} 10 UV Luminosity Functions Using All Five CANDELS Fields*}",
      journal = {\apj},
         year = 2016,
        month = oct,
       volume = {830},
       number = {2},
          eid = {67},
        pages = {67},
          doi = {10.3847/0004-637X/830/2/67},
archivePrefix = {arXiv},
       eprint = {1506.01035},
 primaryClass = {astro-ph.GA},
       adsurl = {https://ui.adsabs.harvard.edu/abs/2016ApJ...830...67B}
}

@ARTICLE{deugenio25,
       author = {{D'Eugenio}, Francesco and {Cameron}, Alex J. and {Scholtz}, Jan and {Carniani}, Stefano and {Willott}, Chris J. and {Curtis-Lake}, Emma and {Bunker}, Andrew J. and {Parlanti}, Eleonora and {Maiolino}, Roberto and {Willmer}, Christopher N.~A. and {Jakobsen}, Peter and {Robertson}, Brant E. and {Johnson}, Benjamin D. and {Tacchella}, Sandro and {Cargile}, Phillip A. and {Rawle}, Tim and {Arribas}, Santiago and {Chevallard}, Jacopo and {Curti}, Mirko and {Egami}, Eiichi and {Eisenstein}, Daniel J. and {Kumari}, Nimisha and {Looser}, Tobias J. and {Rieke}, Marcia J. and {Rodr{\'\i}guez Del Pino}, Bruno and {Saxena}, Aayush and {{\"U}bler}, Hannah and {Venturi}, Giacomo and {Witstok}, Joris and {Baker}, William M. and {Bhatawdekar}, Rachana and {Bonaventura}, Nina and {Boyett}, Kristan and {Charlot}, Stephane and {Danhaive}, A. Lola and {Hainline}, Kevin N. and {Hausen}, Ryan and {Helton}, Jakob M. and {Ji}, Xihan and {Ji}, Zhiyuan and {Jones}, Gareth C. and {Juod{\v{z}}balis}, Ignas and {Maseda}, Michael V. and {P{\'e}rez-Gonz{\'a}lez}, Pablo G. and {Perna}, Michele and {Pusk{\'a}s}, D{\'a}vid and {Shivaei}, Irene and {Silcock}, Maddie S. and {Simmonds}, Charlotte and {Smit}, Renske and {Sun}, Fengwu and {Villanueva}, Natalia C. and {Williams}, Christina C. and {Zhu}, Yongda},
        title = "{JADES Data Release 3: NIRSpec/Microshutter Assembly Spectroscopy for 4000 Galaxies in the GOODS Fields}",
      journal = {\apjs},
         year = 2025,
        month = mar,
       volume = {277},
       number = {1},
          eid = {4},
        pages = {4},
          doi = {10.3847/1538-4365/ada148},
archivePrefix = {arXiv},
       eprint = {2404.06531},
 primaryClass = {astro-ph.GA},
       adsurl = {https://ui.adsabs.harvard.edu/abs/2025ApJS..277....4D}
}

@ARTICLE{bunker24,
       author = {{Bunker}, Andrew J. and {Cameron}, Alex J. and {Curtis-Lake}, Emma and {Jakobsen}, Peter and {Carniani}, Stefano and {Curti}, Mirko and {Witstok}, Joris and {Maiolino}, Roberto and {D'Eugenio}, Francesco and {Looser}, Tobias J. and {Willott}, Chris and {Bonaventura}, Nina and {Hainline}, Kevin and {{\"U}bler}, Hannah and {Willmer}, Christopher N.~A. and {Saxena}, Aayush and {Smit}, Renske and {Alberts}, Stacey and {Arribas}, Santiago and {Baker}, William M. and {Baum}, Stefi and {Bhatawdekar}, Rachana and {Bowler}, Rebecca A.~A. and {Boyett}, Kristan and {Charlot}, Stephane and {Chen}, Zuyi and {Chevallard}, Jacopo and {Circosta}, Chiara and {DeCoursey}, Christa and {de Graaff}, Anna and {Egami}, Eiichi and {Eisenstein}, Daniel J. and {Endsley}, Ryan and {Ferruit}, Pierre and {Giardino}, Giovanna and {Hausen}, Ryan and {Helton}, Jakob M. and {Hviding}, Raphael E. and {Ji}, Zhiyuan and {Johnson}, Benjamin D. and {Jones}, Gareth C. and {Kumari}, Nimisha and {Laseter}, Isaac and {L{\"u}tzgendorf}, Nora and {Maseda}, Michael V. and {Nelson}, Erica and {Parlanti}, Eleonora and {Perna}, Michele and {Rauscher}, Bernard J. and {Rawle}, Tim and {Rix}, Hans-Walter and {Rieke}, Marcia and {Robertson}, Brant and {Rodr{\'\i}guez Del Pino}, Bruno and {Sandles}, Lester and {Scholtz}, Jan and {Sharpe}, Katherine and {Skarbinski}, Maya and {Stark}, Daniel P. and {Sun}, Fengwu and {Tacchella}, Sandro and {Topping}, Michael W. and {Villanueva}, Natalia C. and {Wallace}, Imaan E.~B. and {Williams}, Christina C. and {Woodrum}, Charity},
        title = "{JADES NIRSpec initial data release for the Hubble Ultra Deep Field: Redshifts and line fluxes of distant galaxies from the deepest JWST Cycle 1 NIRSpec multi-object spectroscopy}",
      journal = {\aap},
         year = 2024,
        month = oct,
       volume = {690},
          eid = {A288},
        pages = {A288},
          doi = {10.1051/0004-6361/202347094},
archivePrefix = {arXiv},
       eprint = {2306.02467},
 primaryClass = {astro-ph.GA},
       adsurl = {https://ui.adsabs.harvard.edu/abs/2024A&A...690A.288B}
}

@ARTICLE{eisenstein23,
       author = {{Eisenstein}, Daniel J. and {Willott}, Chris and {Alberts}, Stacey and {Arribas}, Santiago and {Bonaventura}, Nina and {Bunker}, Andrew J. and {Cameron}, Alex J. and {Carniani}, Stefano and {Charlot}, Stephane and {Curtis-Lake}, Emma and {D'Eugenio}, Francesco and {Endsley}, Ryan and {Ferruit}, Pierre and {Giardino}, Giovanna and {Hainline}, Kevin and {Hausen}, Ryan and {Jakobsen}, Peter and {Johnson}, Benjamin D. and {Maiolino}, Roberto and {Rieke}, Marcia and {Rieke}, George and {Rix}, Hans-Walter and {Robertson}, Brant and {Stark}, Daniel P. and {Tacchella}, Sandro and {Williams}, Christina C. and {Willmer}, Christopher N.~A. and {Baker}, William M. and {Baum}, Stefi and {Bhatawdekar}, Rachana and {Boyett}, Kristan and {Chen}, Zuyi and {Chevallard}, Jacopo and {Circosta}, Chiara and {Curti}, Mirko and {Danhaive}, A. Lola and {DeCoursey}, Christa and {de Graaff}, Anna and {Dressler}, Alan and {Egami}, Eiichi and {Helton}, Jakob M. and {Hviding}, Raphael E. and {Ji}, Zhiyuan and {Jones}, Gareth C. and {Kumari}, Nimisha and {L{\"u}tzgendorf}, Nora and {Laseter}, Isaac and {Looser}, Tobias J. and {Lyu}, Jianwei and {Maseda}, Michael V. and {Nelson}, Erica and {Parlanti}, Eleonora and {Perna}, Michele and {Pusk{\'a}s}, D{\'a}vid and {Rawle}, Tim and {Rodr{\'\i}guez Del Pino}, Bruno and {Sandles}, Lester and {Saxena}, Aayush and {Scholtz}, Jan and {Sharpe}, Katherine and {Shivaei}, Irene and {Silcock}, Maddie S. and {Simmonds}, Charlotte and {Skarbinski}, Maya and {Smit}, Renske and {Stone}, Meredith and {Suess}, Katherine A. and {Sun}, Fengwu and {Tang}, Mengtao and {Topping}, Michael W. and {{\"U}bler}, Hannah and {Villanueva}, Natalia C. and {Wallace}, Imaan E.~B. and {Whitler}, Lily and {Witstok}, Joris and {Woodrum}, Charity},
        title = "{Overview of the JWST Advanced Deep Extragalactic Survey (JADES)}",
      journal = {arXiv e-prints},
         year = 2023,
        month = jun,
          eid = {arXiv:2306.02465},
        pages = {arXiv:2306.02465},
          doi = {10.48550/arXiv.2306.02465},
archivePrefix = {arXiv},
       eprint = {2306.02465},
 primaryClass = {astro-ph.GA},
       adsurl = {https://ui.adsabs.harvard.edu/abs/2023arXiv230602465E}
}

@ARTICLE{sarrouh25,
       author = {{Sarrouh}, Ghassan T.~E. and {Asada}, Yoshihisa and {Martis}, Nicholas S. and {Willott}, Chris J. and {Iyer}, Kartheik G. and {Noirot}, Ga{\"e}l and {Muzzin}, Adam and {Sawicki}, Marcin and {Brammer}, Gabriel and {Desprez}, Guillaume and {Rihtar{\v{s}}i{\v{c}}}, Gregor and {Zabl}, Johannes and {Abraham}, Roberto and {Brada{\v{c}}}, Maru{\v{s}}a and {Doyon}, Ren{\'e} and {Antwi-Danso}, Jacqueline and {Berek}, Samantha and {Brown}, Westley and {Estrada-Carpenter}, Vince and {Favaro}, Jeremy and {Felicioni}, Giordano and {Forrest}, Ben and {Gaspar}, Gaia and {Gould}, Katriona M.~L. and {Gledhill}, Rachel and {Harshan}, Anishya and {Jahan}, Nusrath and {Jagga}, Naadiyah and {Jude{\v{z}}}, Jon and {Marchesini}, Danilo and {Markov}, Vladan and {Matharu}, Jasleen and {MacFarland}, Shannon and {Merchant}, Maya and {M{\'e}rida}, Rosa M. and {Mowla}, Lamiya and {Myers}, Katherine and {Omori}, Kiyoaki C. and {Pacifici}, Camilla and {Ravindranath}, Swara and {Robbins}, Luke and {Strait}, Victoria and {Sok}, Visal and {Tan}, Vivian Yun Yan and {Tripodi}, Roberta and {Wilson}, Gillian and {Withers}, Sunna},
        title = "{CANUCS/Technicolor Data Release 1: Imaging, Photometry, Slit Spectroscopy, and Stellar Population Parameters}",
      journal = {arXiv e-prints},
         year = 2025,
        month = jun,
          eid = {arXiv:2506.21685},
        pages = {arXiv:2506.21685},
          doi = {10.48550/arXiv.2506.21685},
archivePrefix = {arXiv},
       eprint = {2506.21685},
 primaryClass = {astro-ph.GA},
       adsurl = {https://ui.adsabs.harvard.edu/abs/2025arXiv250621685S}
}

@ARTICLE{arrabalharo23,
       author = {{Arrabal Haro}, Pablo and {Dickinson}, Mark and {Finkelstein}, Steven L. and {Kartaltepe}, Jeyhan S. and {Donnan}, Callum T. and {Burgarella}, Denis and {Carnall}, Adam C. and {Cullen}, Fergus and {Dunlop}, James S. and {Fern{\'a}ndez}, Vital and {Fujimoto}, Seiji and {Jung}, Intae and {Krips}, Melanie and {Larson}, Rebecca L. and {Papovich}, Casey and {P{\'e}rez-Gonz{\'a}lez}, Pablo G. and {Amor{\'\i}n}, Ricardo O. and {Bagley}, Micaela B. and {Buat}, V{\'e}ronique and {Casey}, Caitlin M. and {Chworowsky}, Katherine and {Cohen}, Seth H. and {Ferguson}, Henry C. and {Giavalisco}, Mauro and {Huertas-Company}, Marc and {Hutchison}, Taylor A. and {Kocevski}, Dale D. and {Koekemoer}, Anton M. and {Lucas}, Ray A. and {McLeod}, Derek J. and {McLure}, Ross J. and {Pirzkal}, Norbert and {Seill{\'e}}, Lise-Marie and {Trump}, Jonathan R. and {Weiner}, Benjamin J. and {Wilkins}, Stephen M. and {Zavala}, Jorge A.},
        title = "{Confirmation and refutation of very luminous galaxies in the early Universe}",
      journal = {\nat},
         year = 2023,
        month = oct,
       volume = {622},
       number = {7984},
        pages = {707-711},
          doi = {10.1038/s41586-023-06521-7},
archivePrefix = {arXiv},
       eprint = {2303.15431},
 primaryClass = {astro-ph.GA},
       adsurl = {https://ui.adsabs.harvard.edu/abs/2023Natur.622..707A}
}

@ARTICLE{finkelstein23,
       author = {{Finkelstein}, Steven L. and {Bagley}, Micaela B. and {Ferguson}, Henry C. and {Wilkins}, Stephen M. and {Kartaltepe}, Jeyhan S. and {Papovich}, Casey and {Yung}, L.~Y. Aaron and {Arrabal Haro}, Pablo and {Behroozi}, Peter and {Dickinson}, Mark and {Kocevski}, Dale D. and {Koekemoer}, Anton M. and {Larson}, Rebecca L. and {Le Bail}, Aur{\'e}lien and {Morales}, Alexa M. and {P{\'e}rez-Gonz{\'a}lez}, Pablo G. and {Burgarella}, Denis and {Dav{\'e}}, Romeel and {Hirschmann}, Michaela and {Somerville}, Rachel S. and {Wuyts}, Stijn and {Bromm}, Volker and {Casey}, Caitlin M. and {Fontana}, Adriano and {Fujimoto}, Seiji and {Gardner}, Jonathan P. and {Giavalisco}, Mauro and {Grazian}, Andrea and {Grogin}, Norman A. and {Hathi}, Nimish P. and {Hutchison}, Taylor A. and {Jha}, Saurabh W. and {Jogee}, Shardha and {Kewley}, Lisa J. and {Kirkpatrick}, Allison and {Long}, Arianna S. and {Lotz}, Jennifer M. and {Pentericci}, Laura and {Pierel}, Justin D.~R. and {Pirzkal}, Nor and {Ravindranath}, Swara and {Ryan}, Russell E. and {Trump}, Jonathan R. and {Yang}, Guang and {Bhatawdekar}, Rachana and {Bisigello}, Laura and {Buat}, V{\'e}ronique and {Calabr{\`o}}, Antonello and {Castellano}, Marco and {Cleri}, Nikko J. and {Cooper}, M.~C. and {Croton}, Darren and {Daddi}, Emanuele and {Dekel}, Avishai and {Elbaz}, David and {Franco}, Maximilien and {Gawiser}, Eric and {Holwerda}, Benne W. and {Huertas-Company}, Marc and {Jaskot}, Anne E. and {Leung}, Gene C.~K. and {Lucas}, Ray A. and {Mobasher}, Bahram and {Pandya}, Viraj and {Tacchella}, Sandro and {Weiner}, Benjamin J. and {Zavala}, Jorge A.},
        title = "{CEERS Key Paper. I. An Early Look into the First 500 Myr of Galaxy Formation with JWST}",
      journal = {\apjl},
         year = 2023,
        month = mar,
       volume = {946},
       number = {1},
          eid = {L13},
        pages = {L13},
          doi = {10.3847/2041-8213/acade4},
archivePrefix = {arXiv},
       eprint = {2211.05792},
 primaryClass = {astro-ph.GA},
       adsurl = {https://ui.adsabs.harvard.edu/abs/2023ApJ...946L..13F}
}

@ARTICLE{donnan23,
       author = {{Donnan}, C.~T. and {McLeod}, D.~J. and {Dunlop}, J.~S. and {McLure}, R.~J. and {Carnall}, A.~C. and {Begley}, R. and {Cullen}, F. and {Hamadouche}, M.~L. and {Bowler}, R.~A.~A. and {Magee}, D. and {McCracken}, H.~J. and {Milvang-Jensen}, B. and {Moneti}, A. and {Targett}, T.},
        title = "{The evolution of the galaxy UV luminosity function at redshifts z ≃ 8 - 15 from deep JWST and ground-based near-infrared imaging}",
      journal = {\mnras},
         year = 2023,
        month = feb,
       volume = {518},
       number = {4},
        pages = {6011-6040},
          doi = {10.1093/mnras/stac3472},
archivePrefix = {arXiv},
       eprint = {2207.12356},
 primaryClass = {astro-ph.GA},
       adsurl = {https://ui.adsabs.harvard.edu/abs/2023MNRAS.518.6011D}
}

@ARTICLE{helton25,
       author = {{Helton}, Jakob M. and {Rieke}, George H. and {Alberts}, Stacey and {Wu}, Zihao and {Eisenstein}, Daniel J. and {Hainline}, Kevin N. and {Carniani}, Stefano and {Ji}, Zhiyuan and {Baker}, William M. and {Bhatawdekar}, Rachana and {Bunker}, Andrew J. and {Cargile}, Phillip A. and {Charlot}, St{\'e}phane and {Chevallard}, Jacopo and {D'Eugenio}, Francesco and {Egami}, Eiichi and {Johnson}, Benjamin D. and {Jones}, Gareth C. and {Lyu}, Jianwei and {Maiolino}, Roberto and {P{\'e}rez-Gonz{\'a}lez}, Pablo G. and {Rieke}, Marcia J. and {Robertson}, Brant and {Saxena}, Aayush and {Scholtz}, Jan and {Shivaei}, Irene and {Sun}, Fengwu and {Tacchella}, Sandro and {Whitler}, Lily and {Williams}, Christina C. and {Willmer}, Christopher N.~A. and {Willott}, Chris and {Witstok}, Joris and {Zhu}, Yongda},
        title = "{Photometric detection at 7.7 {\ensuremath{\mu}}m of a galaxy beyond redshift 14 with JWST/MIRI}",
      journal = {Nature Astronomy},
         year = 2025,
        month = may,
       volume = {9},
        pages = {729-740},
          doi = {10.1038/s41550-025-02503-z},
archivePrefix = {arXiv},
       eprint = {2405.18462},
 primaryClass = {astro-ph.GA},
       adsurl = {https://ui.adsabs.harvard.edu/abs/2025NatAs...9..729H}
}

@ARTICLE{carniani24,
       author = {{Carniani}, Stefano and {Hainline}, Kevin and {D'Eugenio}, Francesco and {Eisenstein}, Daniel J. and {Jakobsen}, Peter and {Witstok}, Joris and {Johnson}, Benjamin D. and {Chevallard}, Jacopo and {Maiolino}, Roberto and {Helton}, Jakob M. and {Willott}, Chris and {Robertson}, Brant and {Alberts}, Stacey and {Arribas}, Santiago and {Baker}, William M. and {Bhatawdekar}, Rachana and {Boyett}, Kristan and {Bunker}, Andrew J. and {Cameron}, Alex J. and {Cargile}, Phillip A. and {Charlot}, St{\'e}phane and {Curti}, Mirko and {Curtis-Lake}, Emma and {Egami}, Eiichi and {Giardino}, Giovanna and {Isaak}, Kate and {Ji}, Zhiyuan and {Jones}, Gareth C. and {Kumari}, Nimisha and {Maseda}, Michael V. and {Parlanti}, Eleonora and {P{\'e}rez-Gonz{\'a}lez}, Pablo G. and {Rawle}, Tim and {Rieke}, George and {Rieke}, Marcia and {Del Pino}, Bruno Rodr{\'\i}guez and {Saxena}, Aayush and {Scholtz}, Jan and {Smit}, Renske and {Sun}, Fengwu and {Tacchella}, Sandro and {{\"U}bler}, Hannah and {Venturi}, Giacomo and {Williams}, Christina C. and {Willmer}, Christopher N.~A.},
        title = "{Spectroscopic confirmation of two luminous galaxies at a redshift of 14}",
      journal = {\nat},
         year = 2024,
        month = sep,
       volume = {633},
       number = {8029},
        pages = {318-322},
          doi = {10.1038/s41586-024-07860-9},
archivePrefix = {arXiv},
       eprint = {2405.18485},
 primaryClass = {astro-ph.GA},
       adsurl = {https://ui.adsabs.harvard.edu/abs/2024Natur.633..318C}
}

@ARTICLE{robertson23,
       author = {{Robertson}, B.~E. and {Tacchella}, S. and {Johnson}, B.~D. and {Hainline}, K. and {Whitler}, L. and {Eisenstein}, D.~J. and {Endsley}, R. and {Rieke}, M. and {Stark}, D.~P. and {Alberts}, S. and {Dressler}, A. and {Egami}, E. and {Hausen}, R. and {Rieke}, G. and {Shivaei}, I. and {Williams}, C.~C. and {Willmer}, C.~N.~A. and {Arribas}, S. and {Bonaventura}, N. and {Bunker}, A. and {Cameron}, A.~J. and {Carniani}, S. and {Charlot}, S. and {Chevallard}, J. and {Curti}, M. and {Curtis-Lake}, E. and {D'Eugenio}, F. and {Jakobsen}, P. and {Looser}, T.~J. and {L{\"u}tzgendorf}, N. and {Maiolino}, R. and {Maseda}, M.~V. and {Rawle}, T. and {Rix}, H. -W. and {Smit}, R. and {{\"U}bler}, H. and {Willott}, C. and {Witstok}, J. and {Baum}, S. and {Bhatawdekar}, R. and {Boyett}, K. and {Chen}, Z. and {de Graaff}, A. and {Florian}, M. and {Helton}, J.~M. and {Hviding}, R.~E. and {Ji}, Z. and {Kumari}, N. and {Lyu}, J. and {Nelson}, E. and {Sandles}, L. and {Saxena}, A. and {Suess}, K.~A. and {Sun}, F. and {Topping}, M. and {Wallace}, I.~E.~B.},
        title = "{Identification and properties of intense star-forming galaxies at redshifts z > 10}",
      journal = {Nature Astronomy},
         year = 2023,
        month = may,
       volume = {7},
        pages = {611-621},
          doi = {10.1038/s41550-023-01921-1},
archivePrefix = {arXiv},
       eprint = {2212.04480},
 primaryClass = {astro-ph.GA},
       adsurl = {https://ui.adsabs.harvard.edu/abs/2023NatAs...7..611R}
}

@ARTICLE{curtis-lake23,
       author = {{Curtis-Lake}, Emma and {Carniani}, Stefano and {Cameron}, Alex and {Charlot}, Stephane and {Jakobsen}, Peter and {Maiolino}, Roberto and {Bunker}, Andrew and {Witstok}, Joris and {Smit}, Renske and {Chevallard}, Jacopo and {Willott}, Chris and {Ferruit}, Pierre and {Arribas}, Santiago and {Bonaventura}, Nina and {Curti}, Mirko and {D'Eugenio}, Francesco and {Franx}, Marijn and {Giardino}, Giovanna and {Looser}, Tobias J. and {L{\"u}tzgendorf}, Nora and {Maseda}, Michael V. and {Rawle}, Tim and {Rix}, Hans-Walter and {Rodr{\'\i}guez del Pino}, Bruno and {{\"U}bler}, Hannah and {Sirianni}, Marco and {Dressler}, Alan and {Egami}, Eiichi and {Eisenstein}, Daniel J. and {Endsley}, Ryan and {Hainline}, Kevin and {Hausen}, Ryan and {Johnson}, Benjamin D. and {Rieke}, Marcia and {Robertson}, Brant and {Shivaei}, Irene and {Stark}, Daniel P. and {Tacchella}, Sandro and {Williams}, Christina C. and {Willmer}, Christopher N.~A. and {Bhatawdekar}, Rachana and {Bowler}, Rebecca and {Boyett}, Kristan and {Chen}, Zuyi and {de Graaff}, Anna and {Helton}, Jakob M. and {Hviding}, Raphael E. and {Jones}, Gareth C. and {Kumari}, Nimisha and {Lyu}, Jianwei and {Nelson}, Erica and {Perna}, Michele and {Sandles}, Lester and {Saxena}, Aayush and {Suess}, Katherine A. and {Sun}, Fengwu and {Topping}, Michael W. and {Wallace}, Imaan E.~B. and {Whitler}, Lily},
        title = "{Spectroscopic confirmation of four metal-poor galaxies at z = 10.3-13.2}",
      journal = {Nature Astronomy},
         year = 2023,
        month = may,
       volume = {7},
        pages = {622-632},
          doi = {10.1038/s41550-023-01918-w},
archivePrefix = {arXiv},
       eprint = {2212.04568},
 primaryClass = {astro-ph.GA},
       adsurl = {https://ui.adsabs.harvard.edu/abs/2023NatAs...7..622C}
}

@ARTICLE{napolitano25,
       author = {{Napolitano}, L. and {Castellano}, M. and {Pentericci}, L. and {Arrabal Haro}, P. and {Fontana}, A. and {Treu}, T. and {Bergamini}, P. and {Calabr{\`o}}, A. and {Mascia}, S. and {Morishita}, T. and {Roberts-Borsani}, G. and {Santini}, P. and {Vanzella}, E. and {Vulcani}, B. and {Zakharova}, D. and {Bakx}, T. and {Dickinson}, M. and {Grillo}, C. and {Leethochawalit}, N. and {Llerena}, M. and {Merlin}, E. and {Paris}, D. and {Rojas-Ruiz}, S. and {Rosati}, P. and {Wang}, X. and {Yoon}, I. and {Zavala}, J.},
        title = "{Seven wonders of Cosmic Dawn: JWST confirms a high abundance of galaxies and AGN at z ≃ 9{\textendash}11 in the GLASS field}",
      journal = {\aap},
         year = 2025,
        month = jan,
       volume = {693},
          eid = {A50},
        pages = {A50},
          doi = {10.1051/0004-6361/202452090},
archivePrefix = {arXiv},
       eprint = {2410.10967},
 primaryClass = {astro-ph.GA},
       adsurl = {https://ui.adsabs.harvard.edu/abs/2025A&A...693A..50N}
}

@ARTICLE{naidu22,
       author = {{Naidu}, Rohan P. and {Oesch}, Pascal A. and {van Dokkum}, Pieter and {Nelson}, Erica J. and {Suess}, Katherine A. and {Brammer}, Gabriel and {Whitaker}, Katherine E. and {Illingworth}, Garth and {Bouwens}, Rychard and {Tacchella}, Sandro and {Matthee}, Jorryt and {Allen}, Natalie and {Bezanson}, Rachel and {Conroy}, Charlie and {Labbe}, Ivo and {Leja}, Joel and {Leonova}, Ecaterina and {Magee}, Dan and {Price}, Sedona H. and {Setton}, David J. and {Strait}, Victoria and {Stefanon}, Mauro and {Toft}, Sune and {Weaver}, John R. and {Weibel}, Andrea},
        title = "{Two Remarkably Luminous Galaxy Candidates at z {\ensuremath{\approx}} 10-12 Revealed by JWST}",
      journal = {\apjl},
         year = 2022,
        month = nov,
       volume = {940},
       number = {1},
          eid = {L14},
        pages = {L14},
          doi = {10.3847/2041-8213/ac9b22},
archivePrefix = {arXiv},
       eprint = {2207.09434},
 primaryClass = {astro-ph.GA},
       adsurl = {https://ui.adsabs.harvard.edu/abs/2022ApJ...940L..14N}
}

@ARTICLE{castellano23,
       author = {{Castellano}, Marco and {Fontana}, Adriano and {Treu}, Tommaso and {Merlin}, Emiliano and {Santini}, Paola and {Bergamini}, Pietro and {Grillo}, Claudio and {Rosati}, Piero and {Acebron}, Ana and {Leethochawalit}, Nicha and {Paris}, Diego and {Bonchi}, Andrea and {Belfiori}, Davide and {Calabr{\`o}}, Antonello and {Correnti}, Matteo and {Nonino}, Mario and {Polenta}, Gianluca and {Trenti}, Michele and {Boyett}, Kristan and {Brammer}, G. and {Broadhurst}, Tom and {Caminha}, Gabriel B. and {Chen}, Wenlei and {Filippenko}, Alexei V. and {Fortuni}, Flaminia and {Glazebrook}, Karl and {Mascia}, Sara and {Mason}, Charlotte A. and {Menci}, Nicola and {Meneghetti}, Massimo and {Mercurio}, Amata and {Metha}, Benjamin and {Morishita}, Takahiro and {Nanayakkara}, Themiya and {Pentericci}, Laura and {Roberts-Borsani}, Guido and {Roy}, Namrata and {Vanzella}, Eros and {Vulcani}, Benedetta and {Yang}, Lilan and {Wang}, Xin},
        title = "{Early Results from GLASS-JWST. XIX. A High Density of Bright Galaxies at z {\ensuremath{\approx}} 10 in the A2744 Region}",
      journal = {\apjl},
         year = 2023,
        month = may,
       volume = {948},
       number = {2},
          eid = {L14},
        pages = {L14},
          doi = {10.3847/2041-8213/accea5},
archivePrefix = {arXiv},
       eprint = {2212.06666},
 primaryClass = {astro-ph.GA},
       adsurl = {https://ui.adsabs.harvard.edu/abs/2023ApJ...948L..14C}
}

@ARTICLE{tacchella23,
       author = {{Tacchella}, Sandro and {Eisenstein}, Daniel J. and {Hainline}, Kevin and {Johnson}, Benjamin D. and {Baker}, William M. and {Helton}, Jakob M. and {Robertson}, Brant and {Suess}, Katherine A. and {Chen}, Zuyi and {Nelson}, Erica and {Pusk{\'a}s}, D{\'a}vid and {Sun}, Fengwu and {Alberts}, Stacey and {Egami}, Eiichi and {Hausen}, Ryan and {Rieke}, George and {Rieke}, Marcia and {Shivaei}, Irene and {Williams}, Christina C. and {Willmer}, Christopher N.~A. and {Bunker}, Andrew and {Cameron}, Alex J. and {Carniani}, Stefano and {Charlot}, Stephane and {Curti}, Mirko and {Curtis-Lake}, Emma and {Looser}, Tobias J. and {Maiolino}, Roberto and {Maseda}, Michael V. and {Rawle}, Tim and {Rix}, Hans-Walter and {Smit}, Renske and {{\"U}bler}, Hannah and {Willott}, Chris and {Witstok}, Joris and {Baum}, Stefi and {Bhatawdekar}, Rachana and {Boyett}, Kristan and {Danhaive}, A. Lola and {de Graaff}, Anna and {Endsley}, Ryan and {Ji}, Zhiyuan and {Lyu}, Jianwei and {Sandles}, Lester and {Saxena}, Aayush and {Scholtz}, Jan and {Topping}, Michael W. and {Whitler}, Lily},
        title = "{JADES Imaging of GN-z11: Revealing the Morphology and Environment of a Luminous Galaxy 430 Myr after the Big Bang}",
      journal = {\apj},
         year = 2023,
        month = jul,
       volume = {952},
       number = {1},
          eid = {74},
        pages = {74},
          doi = {10.3847/1538-4357/acdbc6},
archivePrefix = {arXiv},
       eprint = {2302.07234},
 primaryClass = {astro-ph.GA},
       adsurl = {https://ui.adsabs.harvard.edu/abs/2023ApJ...952...74T}
}

@ARTICLE{castellano24,
       author = {{Castellano}, Marco and {Napolitano}, Lorenzo and {Fontana}, Adriano and {Roberts-Borsani}, Guido and {Treu}, Tommaso and {Vanzella}, Eros and {Zavala}, Jorge A. and {Arrabal Haro}, Pablo and {Calabr{\`o}}, Antonello and {Llerena}, Mario and {Mascia}, Sara and {Merlin}, Emiliano and {Paris}, Diego and {Pentericci}, Laura and {Santini}, Paola and {Bakx}, Tom J.~L.~C. and {Bergamini}, Pietro and {Cupani}, Guido and {Dickinson}, Mark and {Filippenko}, Alexei V. and {Glazebrook}, Karl and {Grillo}, Claudio and {Kelly}, Patrick L. and {Malkan}, Matthew A. and {Mason}, Charlotte A. and {Morishita}, Takahiro and {Nanayakkara}, Themiya and {Rosati}, Piero and {Sani}, Eleonora and {Wang}, Xin and {Yoon}, Ilsang},
        title = "{JWST NIRSpec Spectroscopy of the Remarkable Bright Galaxy GHZ2/GLASS-z12 at Redshift 12.34}",
      journal = {\apj},
         year = 2024,
        month = sep,
       volume = {972},
       number = {2},
          eid = {143},
        pages = {143},
          doi = {10.3847/1538-4357/ad5f88},
archivePrefix = {arXiv},
       eprint = {2403.10238},
 primaryClass = {astro-ph.GA},
       adsurl = {https://ui.adsabs.harvard.edu/abs/2024ApJ...972..143C}
}

@ARTICLE{witstok25,
       author = {{Witstok}, Joris and {Smit}, Renske and {Baker}, William M. and {Rinaldi}, Pierluigi and {Hainline}, Kevin N. and {Algera}, Hiddo S.~B. and {Arribas}, Santiago and {Bakx}, Tom J.~L.~C. and {Bunker}, Andrew J. and {Carniani}, Stefano and {Charlot}, St{\'e}phane and {Chevallard}, Jacopo and {Curti}, Mirko and {Curtis-Lake}, Emma and {Eisenstein}, Daniel J. and {Heintz}, Kasper E. and {Helton}, Jakob M. and {Jones}, Gareth C. and {Maiolino}, Roberto and {Maseda}, Michael V. and {P{\'e}rez-Gonz{\'a}lez}, Pablo G. and {Pollock}, Clara L. and {Robertson}, Brant E. and {Saxena}, Aayush and {Scholtz}, Jan and {Shivaei}, Irene and {Sun}, Fengwu and {Tacchella}, Sandro and {{\"U}bler}, Hannah and {Watson}, Darach and {Willott}, Chris J. and {Wu}, Zihao},
        title = "{On the origins of oxygen: ALMA and JWST characterise the multi-phase, metal-enriched, star-bursting medium within a 'normal' $z > 11$ galaxy}",
      journal = {arXiv e-prints},
         year = 2025,
        month = jul,
          eid = {arXiv:2507.22888},
        pages = {arXiv:2507.22888},
          doi = {10.48550/arXiv.2507.22888},
archivePrefix = {arXiv},
       eprint = {2507.22888},
 primaryClass = {astro-ph.GA},
       adsurl = {https://ui.adsabs.harvard.edu/abs/2025arXiv250722888W}
}

@ARTICLE{zavala24,
       author = {{Zavala}, Jorge A. and {Bakx}, Tom and {Mitsuhashi}, Ikki and {Castellano}, Marco and {Calabro}, Antonello and {Akins}, Hollis and {Buat}, Veronique and {Casey}, Caitlin M. and {Fernandez-Arenas}, David and {Franco}, Maximilien and {Fontana}, Adriano and {Hatsukade}, Bunyo and {Ho}, Luis C. and {Ikeda}, Ryota and {Kartaltepe}, Jeyhan and {Koekemoer}, Anton M. and {McKinney}, Jed and {Napolitano}, Lorenzo and {P{\'e}rez-Gonz{\'a}lez}, Pablo G. and {Santini}, Paola and {Serjeant}, Stephen and {Terlevich}, Elena and {Terlevich}, Roberto and {Yung}, L.~Y. Aaron},
        title = "{ALMA Detection of [O III] 88 {\ensuremath{\mu}}m at z = 12.33: Exploring the Nature and Evolution of GHZ2 as a Massive Compact Stellar System}",
      journal = {\apjl},
         year = 2024,
        month = dec,
       volume = {977},
       number = {1},
          eid = {L9},
        pages = {L9},
          doi = {10.3847/2041-8213/ad8f38},
archivePrefix = {arXiv},
       eprint = {2411.03593},
 primaryClass = {astro-ph.GA},
       adsurl = {https://ui.adsabs.harvard.edu/abs/2024ApJ...977L...9Z}
}

@ARTICLE{rieke23,
       author = {{Rieke}, Marcia J. and {Robertson}, Brant and {Tacchella}, Sandro and {Hainline}, Kevin and {Johnson}, Benjamin D. and {Hausen}, Ryan and {Ji}, Zhiyuan and {Willmer}, Christopher N.~A. and {Eisenstein}, Daniel J. and {Pusk{\'a}s}, D{\'a}vid and {Alberts}, Stacey and {Arribas}, Santiago and {Baker}, William M. and {Baum}, Stefi and {Bhatawdekar}, Rachana and {Bonaventura}, Nina and {Boyett}, Kristan and {Bunker}, Andrew J. and {Cameron}, Alex J. and {Carniani}, Stefano and {Charlot}, Stephane and {Chevallard}, Jacopo and {Chen}, Zuyi and {Curti}, Mirko and {Curtis-Lake}, Emma and {Danhaive}, A. Lola and {DeCoursey}, Christa and {Dressler}, Alan and {Egami}, Eiichi and {Endsley}, Ryan and {Helton}, Jakob M. and {Hviding}, Raphael E. and {Kumari}, Nimisha and {Looser}, Tobias J. and {Lyu}, Jianwei and {Maiolino}, Roberto and {Maseda}, Michael V. and {Nelson}, Erica J. and {Rieke}, George and {Rix}, Hans-Walter and {Sandles}, Lester and {Saxena}, Aayush and {Sharpe}, Katherine and {Shivaei}, Irene and {Skarbinski}, Maya and {Smit}, Renske and {Stark}, Daniel P. and {Stone}, Meredith and {Suess}, Katherine A. and {Sun}, Fengwu and {Topping}, Michael and {{\"U}bler}, Hannah and {Villanueva}, Natalia C. and {Wallace}, Imaan E.~B. and {Williams}, Christina C. and {Willott}, Chris and {Whitler}, Lily and {Witstok}, Joris and {Woodrum}, Charity},
        title = "{JADES Initial Data Release for the Hubble Ultra Deep Field: Revealing the Faint Infrared Sky with Deep JWST NIRCam Imaging}",
      journal = {\apjs},
         year = 2023,
        month = nov,
       volume = {269},
       number = {1},
          eid = {16},
        pages = {16},
          doi = {10.3847/1538-4365/acf44d},
archivePrefix = {arXiv},
       eprint = {2306.02466},
 primaryClass = {astro-ph.GA},
       adsurl = {https://ui.adsabs.harvard.edu/abs/2023ApJS..269...16R}
}

@ARTICLE{castellano25,
       author = {{Castellano}, M. and {Fontana}, A. and {Merlin}, E. and {Santini}, P. and {Napolitano}, L. and {Menci}, N. and {Calabr{\`o}}, A. and {Paris}, D. and {Pentericci}, L. and {Zavala}, J. and {Dickinson}, M. and {Finkelstein}, S.~L. and {Treu}, T. and {Amorin}, R.~O. and {Arrabal Haro}, P. and {Bergamini}, P. and {Bisigello}, L. and {Daddi}, E. and {Dayal}, P. and {Dekel}, A. and {Ferrara}, A. and {Fortuni}, F. and {Gandolfi}, G. and {Giavalisco}, M. and {Grillo}, C. and {Guida}, S.~T. and {Hathi}, N.~P. and {Holwerda}, B.~W. and {Koekemoer}, A.~M. and {Kokorev}, V. and {Li}, Z. and {Llerena}, M. and {Lucas}, R.~A. and {Mascia}, S. and {Metha}, B. and {Morishita}, T. and {Nanayakkara}, T. and {Pacucci}, F. and {P{\'e}rez-Gonz{\'a}lez}, P.~G. and {Roberts-Borsani}, G. and {Rodighiero}, G. and {Rosati}, P. and {Salazar}, V. and {Schneider}, R. and {Somerville}, R.~S. and {Taylor}, A. and {Trenti}, M. and {Trinca}, A. and {Wang}, X. and {Watson}, P.~J. and {Yang}, L. and {Yung}, L.~Y.~A.},
        title = "{Pushing JWST to the extremes: search and scrutiny of bright galaxy candidates at z$\simeq$15-30}",
      journal = {arXiv e-prints},
         year = 2025,
        month = apr,
          eid = {arXiv:2504.05893},
        pages = {arXiv:2504.05893},
          doi = {10.48550/arXiv.2504.05893},
archivePrefix = {arXiv},
       eprint = {2504.05893},
 primaryClass = {astro-ph.GA},
       adsurl = {https://ui.adsabs.harvard.edu/abs/2025arXiv250405893C}
}

@ARTICLE{Liu15,
       author = {{Liu}, Daizhong and {Gao}, Yu and {Isaak}, Kate and {Daddi}, Emanuele and {Yang}, Chentao and {Lu}, Nanyao and {van der Werf}, Paul},
        title = "{High-J CO versus Far-infrared Relations in Normal and Starburst Galaxies}",
      journal = {\apjl},
         year = 2015,
        month = sep,
       volume = {810},
       number = {2},
          eid = {L14},
        pages = {L14},
          doi = {10.1088/2041-8205/810/2/L14},
archivePrefix = {arXiv},
       eprint = {1504.05897},
 primaryClass = {astro-ph.GA},
       adsurl = {https://ui.adsabs.harvard.edu/abs/2015ApJ...810L..14L}
}

@ARTICLE{Butler23,
       author = {{Butler}, Kirsty M. and {van der Werf}, Paul P. and {Omont}, Alain and {Cox}, Pierre},
        title = "{Neutral outflows in high-z QSOs}",
      journal = {\aap},
         year = 2023,
        month = jun,
       volume = {674},
          eid = {L5},
        pages = {L5},
          doi = {10.1051/0004-6361/202346271},
archivePrefix = {arXiv},
       eprint = {2305.04098},
 primaryClass = {astro-ph.GA},
       adsurl = {https://ui.adsabs.harvard.edu/abs/2023A&A...674L...5B}
}

@ARTICLE{Riechers25,
       author = {{Riechers}, Dominik A.},
        title = "{Do Unusually Cold Starburst Galaxies Exist? A Case Study}",
      journal = {\apj},
         year = 2025,
        month = feb,
       volume = {980},
       number = {1},
          eid = {59},
        pages = {59},
          doi = {10.3847/1538-4357/ada4ac},
archivePrefix = {arXiv},
       eprint = {2401.05487},
 primaryClass = {astro-ph.GA},
       adsurl = {https://ui.adsabs.harvard.edu/abs/2025ApJ...980...59R}
}

@ARTICLE{Solomon92,
       author = {{Solomon}, P.~M. and {Downes}, D. and {Radford}, S.~J.~E.},
        title = "{Warm Molecular Gas in the Primeval Galaxy IRAS 10214+4724}",
      journal = {\apjl},
         year = 1992,
        month = oct,
       volume = {398},
        pages = {L29},
          doi = {10.1086/186569},
       adsurl = {https://ui.adsabs.harvard.edu/abs/1992ApJ...398L..29S}
}

@ARTICLE{Riechers21a,
       author = {{Riechers}, Dominik A. and {Cooray}, Asantha and {P{\'e}rez-Fournon}, Ismael and {Neri}, Roberto},
        title = "{The GADOT Galaxy Survey: Dense Gas and Feedback in Herschel-selected Starburst Galaxies at Redshifts 2 to 6}",
      journal = {\apj},
         year = 2021,
        month = jun,
       volume = {913},
       number = {2},
          eid = {141},
        pages = {141},
          doi = {10.3847/1538-4357/abf6d7},
archivePrefix = {arXiv},
       eprint = {2101.11006},
 primaryClass = {astro-ph.GA},
       adsurl = {https://ui.adsabs.harvard.edu/abs/2021ApJ...913..141R}
}

@ARTICLE{Rosenberg15,
       author = {{Rosenberg}, M.~J.~F. and {van der Werf}, P.~P. and {Aalto}, S. and {Armus}, L. and {Charmandaris}, V. and {D{\'\i}az-Santos}, T. and {Evans}, A.~S. and {Fischer}, J. and {Gao}, Y. and {Gonz{\'a}lez-Alfonso}, E. and {Greve}, T.~R. and {Harris}, A.~I. and {Henkel}, C. and {Israel}, F.~P. and {Isaak}, K.~G. and {Kramer}, C. and {Meijerink}, R. and {Naylor}, D.~A. and {Sanders}, D.~B. and {Smith}, H.~A. and {Spaans}, M. and {Spinoglio}, L. and {Stacey}, G.~J. and {Veenendaal}, I. and {Veilleux}, S. and {Walter}, F. and {Wei{\ss}}, A. and {Wiedner}, M.~C. and {van der Wiel}, M.~H.~D. and {Xilouris}, E.~M.},
        title = "{The Herschel Comprehensive (U)LIRG Emission Survey (HERCULES): CO Ladders, Fine Structure Lines, and Neutral Gas Cooling}",
      journal = {\apj},
         year = 2015,
        month = mar,
       volume = {801},
       number = {2},
          eid = {72},
        pages = {72},
          doi = {10.1088/0004-637X/801/2/72},
archivePrefix = {arXiv},
       eprint = {1501.02985},
 primaryClass = {astro-ph.GA},
       adsurl = {https://ui.adsabs.harvard.edu/abs/2015ApJ...801...72R}
}

@ARTICLE{Canameras15,
       author = {{Ca{\~n}ameras}, R. and {Nesvadba}, N.~P.~H. and {Guery}, D. and {McKenzie}, T. and {K{\"o}nig}, S. and {Petitpas}, G. and {Dole}, H. and {Frye}, B. and {Flores-Cacho}, I. and {Montier}, L. and {Negrello}, M. and {Beelen}, A. and {Boone}, F. and {Dicken}, D. and {Lagache}, G. and {Le Floc'h}, E. and {Altieri}, B. and {B{\'e}thermin}, M. and {Chary}, R. and {de Zotti}, G. and {Giard}, M. and {Kneissl}, R. and {Krips}, M. and {Malhotra}, S. and {Martinache}, C. and {Omont}, A. and {Pointecouteau}, E. and {Puget}, J. -L. and {Scott}, D. and {Soucail}, G. and {Valtchanov}, I. and {Welikala}, N. and {Yan}, L.},
        title = "{Planck's dusty GEMS: The brightest gravitationally lensed galaxies discovered with the Planck all-sky survey}",
      journal = {\aap},
         year = 2015,
        month = sep,
       volume = {581},
          eid = {A105},
        pages = {A105},
          doi = {10.1051/0004-6361/201425128},
archivePrefix = {arXiv},
       eprint = {1506.01962},
 primaryClass = {astro-ph.GA},
       adsurl = {https://ui.adsabs.harvard.edu/abs/2015A&A...581A.105C}
}

@ARTICLE{Canameras18,
       author = {{Ca{\~n}ameras}, R. and {Yang}, C. and {Nesvadba}, N.~P.~H. and {Beelen}, A. and {Kneissl}, R. and {Koenig}, S. and {Le Floc'h}, E. and {Limousin}, M. and {Malhotra}, S. and {Omont}, A. and {Scott}, D.},
        title = "{Planck's dusty GEMS. VI. Multi-J CO excitation and interstellar medium conditions in dusty starburst galaxies at z = 2-4}",
      journal = {\aap},
         year = 2018,
        month = nov,
       volume = {620},
          eid = {A61},
        pages = {A61},
          doi = {10.1051/0004-6361/201833625},
archivePrefix = {arXiv},
       eprint = {1811.11215},
 primaryClass = {astro-ph.GA},
       adsurl = {https://ui.adsabs.harvard.edu/abs/2018A&A...620A..61C}
}

@ARTICLE{Sommovigo22,
       author = {{Sommovigo}, L. and {Ferrara}, A. and {Pallottini}, A. and {Dayal}, P. and {Bouwens}, R.~J. and {Smit}, R. and {da Cunha}, E. and {De Looze}, I. and {Bowler}, R.~A.~A. and {Hodge}, J. and {Inami}, H. and {Oesch}, P. and {Endsley}, R. and {Gonzalez}, V. and {Schouws}, S. and {Stark}, D. and {Stefanon}, M. and {Aravena}, M. and {Graziani}, L. and {Riechers}, D. and {Schneider}, R. and {van der Werf}, P. and {Algera}, H. and {Barrufet}, L. and {Fudamoto}, Y. and {Hygate}, A.~P.~S. and {Labb{\'e}}, I. and {Li}, Y. and {Nanayakkara}, T. and {Topping}, M.},
        title = "{The ALMA REBELS Survey: cosmic dust temperature evolution out to z   7}",
      journal = {\mnras},
         year = 2022,
        month = jul,
       volume = {513},
       number = {3},
        pages = {3122-3135},
          doi = {10.1093/mnras/stac302},
archivePrefix = {arXiv},
       eprint = {2202.01227},
 primaryClass = {astro-ph.GA},
       adsurl = {https://ui.adsabs.harvard.edu/abs/2022MNRAS.513.3122S}
}

@ARTICLE{Zavala23,
       author = {{Zavala}, Jorge A. and {Buat}, V{\'e}ronique and {Casey}, Caitlin M. and {Finkelstein}, Steven L. and {Burgarella}, Denis and {Bagley}, Micaela B. and {Ciesla}, Laure and {Daddi}, Emanuele and {Dickinson}, Mark and {Ferguson}, Henry C. and {Franco}, Maximilien and {Jim{\'e}nez-Andrade}, E.~F. and {Kartaltepe}, Jeyhan S. and {Koekemoer}, Anton M. and {Le Bail}, Aur{\'e}lien and {Murphy}, E.~J. and {Papovich}, Casey and {Tacchella}, Sandro and {Wilkins}, Stephen M. and {Aretxaga}, Itziar and {Behroozi}, Peter and {Champagne}, Jaclyn B. and {Fontana}, Adriano and {Giavalisco}, Mauro and {Grazian}, Andrea and {Grogin}, Norman A. and {Kewley}, Lisa J. and {Kocevski}, Dale D. and {Kirkpatrick}, Allison and {Lotz}, Jennifer M. and {Pentericci}, Laura and {P{\'e}rez-Gonz{\'a}lez}, Pablo G. and {Pirzkal}, Nor and {Ravindranath}, Swara and {Somerville}, Rachel S. and {Trump}, Jonathan R. and {Yang}, Guang and {Yung}, L.~Y. Aaron and {Almaini}, Omar and {Amor{\'\i}n}, Ricardo O. and {Annunziatella}, Marianna and {Arrabal Haro}, Pablo and {Backhaus}, Bren E. and {Barro}, Guillermo and {Bell}, Eric F. and {Bhatawdekar}, Rachana and {Bisigello}, Laura and {Buitrago}, Fernando and {Calabr{\`o}}, Antonello and {Castellano}, Marco and {Ch{\'a}vez Ortiz}, {\'O}scar A. and {Chworowsky}, Katherine and {Cleri}, Nikko J. and {Cohen}, Seth H. and {Cole}, Justin W. and {Cooke}, Kevin C. and {Cooper}, M.~C. and {Cooray}, Asantha R. and {Costantin}, Luca and {Cox}, Isabella G. and {Croton}, Darren and {Dav{\'e}}, Romeel and {de La Vega}, Alexander and {Dekel}, Avishai and {Elbaz}, David and {Estrada-Carpenter}, Vicente and {Fern{\'a}ndez}, Vital and {Finkelstein}, Keely D. and {Freundlich}, Jonathan and {Fujimoto}, Seiji and {Garc{\'\i}a-Argum{\'a}nez}, {\'A}ngela and {Gardner}, Jonathan P. and {Gawiser}, Eric and {G{\'o}mez-Guijarro}, Carlos and {Guo}, Yuchen and {Hamilton}, Timothy S. and {Hathi}, Nimish P. and {Holwerda}, Benne W. and {Hirschmann}, Michaela and {Huertas-Company}, Marc and {Hutchison}, Taylor A. and {Iyer}, Kartheik G. and {Jaskot}, Anne E. and {Jha}, Saurabh W. and {Jogee}, Shardha and {Juneau}, St{\'e}phanie and {Jung}, Intae and {Kassin}, Susan A. and {Kurczynski}, Peter and {Larson}, Rebecca L. and {Leung}, Gene C.~K. and {Long}, Arianna S. and {Lucas}, Ray A. and {Magnelli}, Benjamin and {Mantha}, Kameswara Bharadwaj and {Matharu}, Jasleen and {McGrath}, Elizabeth J. and {McIntosh}, Daniel H. and {Medrano}, Aubrey and {Merlin}, Emiliano and {Mobasher}, Bahram and {Morales}, Alexa M. and {Newman}, Jeffrey A. and {Nicholls}, David C. and {Pandya}, Viraj and {Rafelski}, Marc and {Ronayne}, Kaila and {Rose}, Caitlin and {Ryan}, Russell E. and {Santini}, Paola and {Seill{\'e}}, Lise-Marie and {Shah}, Ekta A. and {Shen}, Lu and {Simons}, Raymond C. and {Snyder}, Gregory F. and {Stanway}, Elizabeth R. and {Straughn}, Amber N. and {Teplitz}, Harry I. and {Vanderhoof}, Brittany N. and {Vega-Ferrero}, Jes{\'u}s and {Wang}, Weichen and {Weiner}, Benjamin J. and {Willmer}, Christopher N.~A. and {Wuyts}, Stijn and {Ceers Team}},
        title = "{Dusty Starbursts Masquerading as Ultra-high Redshift Galaxies in JWST CEERS Observations}",
      journal = {\apjl},
         year = 2023,
        month = feb,
       volume = {943},
       number = {2},
          eid = {L9},
        pages = {L9},
          doi = {10.3847/2041-8213/acacfe},
archivePrefix = {arXiv},
       eprint = {2208.01816},
 primaryClass = {astro-ph.GA},
       adsurl = {https://ui.adsabs.harvard.edu/abs/2023ApJ...943L...9Z}
}

@ARTICLE{Asada25,
       author = {{Asada}, Yoshihisa and {Willott}, Chris J. and {Muzzin}, Adam and {Brada{\v{c}}}, Maru{\v{s}}a and {Brammer}, Gabriel and {Desprez}, Guillaume and {Iyer}, Kartheik G. and {Marchesini}, Danilo and {Martis}, Nicholas S. and {Noirot}, Ga{\"e}l and {Sarrouh}, Ghassan T.~E. and {Sawicki}, Marcin and {Withers}, Sunna and {Fujimoto}, Seiji and {Felicioni}, Giordano and {Goovaerts}, Ilias and {Jude{\v{z}}}, Jon and {Jagga}, Naadiyah and {Merchant}, Maya and {M{\'e}rida}, Rosa M. and {Robbins}, Luke},
        title = "{Earliest Galaxy Evolution in the CANUCS+Technicolor Fields: Galaxy Properties at z {\ensuremath{\sim}} 10─16 Seen with the Full NIRCam Medium- and Broadband Filters}",
      journal = {\apj},
         year = 2026,
        month = jan,
       volume = {996},
       number = {2},
          eid = {115},
        pages = {115},
          doi = {10.3847/1538-4357/ae1f8d},
archivePrefix = {arXiv},
       eprint = {2507.03124},
 primaryClass = {astro-ph.GA},
       adsurl = {https://ui.adsabs.harvard.edu/abs/2026ApJ...996..115A}
}

@ARTICLE{Stark16,
       author = {{Stark}, Daniel P.},
        title = "{Galaxies in the First Billion Years After the Big Bang}",
      journal = {\araa},
         year = 2016,
        month = sep,
       volume = {54},
        pages = {761-803},
          doi = {10.1146/annurev-astro-081915-023417},
       adsurl = {https://ui.adsabs.harvard.edu/abs/2016ARA&A..54..761S}
}

@ARTICLE{Algera24,
       author = {{Algera}, Hiddo S.~B. and {Inami}, Hanae and {Sommovigo}, Laura and {Fudamoto}, Yoshinobu and {Schneider}, Raffaella and {Graziani}, Luca and {Dayal}, Pratika and {Bouwens}, Rychard and {Aravena}, Manuel and {da Cunha}, Elisabete and {Ferrara}, Andrea and {Hygate}, Alexander P.~S. and {van Leeuwen}, Ivana and {De Looze}, Ilse and {Palla}, Marco and {Pallottini}, Andrea and {Smit}, Renske and {Stefanon}, Mauro and {Topping}, Michael and {van der Werf}, Paul P.},
        title = "{Cold dust and low [O III]/[C II] ratios: an evolved star-forming population at redshift 7}",
      journal = {\mnras},
         year = 2024,
        month = jan,
       volume = {527},
       number = {3},
        pages = {6867-6887},
          doi = {10.1093/mnras/stad3111},
archivePrefix = {arXiv},
       eprint = {2301.09659},
 primaryClass = {astro-ph.GA},
       adsurl = {https://ui.adsabs.harvard.edu/abs/2024MNRAS.527.6867A}
}

@ARTICLE{Algera25b,
       author = {{Algera}, Hiddo and {Rowland}, Lucie and {Smit}, Renske and {Fisher}, Rebecca and {Ramambason}, Lise and {Kumari}, Nimisha and {Vallini}, Livia and {Inami}, Hanae and {Nanayakkara}, Themiya and {Stefanon}, Mauro and {Aravena}, Manuel and {Bakx}, Tom and {Bouwens}, Rychard and {Bowler}, Rebecca and {Cescon}, Karin and {Chen}, Chian-Chou and {Dayal}, Pratika and {De Looze}, Ilse and {Ferrara}, Andrea and {Fudamoto}, Yoshinobu and {Komarova}, Lena and {van Leeuwen}, Ivana and {Ormerod}, Katherine and {Schouws}, Sander and {Sommovigo}, Laura and {Vijayan}, Aswin and {Wang}, Wei-Hao and {van der Werf}, Paul and {Witstok}, Joris},
        title = "{REBELS-IFU: on the origin of the elevated [OIII]/[CII] ratios in the early Universe}",
      journal = {arXiv e-prints},
         year = 2025,
        month = sep,
          eid = {arXiv:2509.16071},
        pages = {arXiv:2509.16071},
          doi = {10.48550/arXiv.2509.16071},
archivePrefix = {arXiv},
       eprint = {2509.16071},
 primaryClass = {astro-ph.GA},
       adsurl = {https://ui.adsabs.harvard.edu/abs/2025arXiv250916071A}
}

@ARTICLE{Inoue14,
       author = {{Inoue}, A.~K. and {Shimizu}, I. and {Tamura}, Y. and {Matsuo}, H. and {Okamoto}, T. and {Yoshida}, N.},
        title = "{ALMA Will Determine the Spectroscopic Redshift z > 8 with FIR [O III] Emission Lines}",
      journal = {\apjl},
         year = 2014,
        month = jan,
       volume = {780},
       number = {2},
          eid = {L18},
        pages = {L18},
          doi = {10.1088/2041-8205/780/2/L18},
archivePrefix = {arXiv},
       eprint = {1312.0684},
 primaryClass = {astro-ph.GA},
       adsurl = {https://ui.adsabs.harvard.edu/abs/2014ApJ...780L..18I}
}

@ARTICLE{Tamura19,
       author = {{Tamura}, Yoichi and {Mawatari}, Ken and {Hashimoto}, Takuya and {Inoue}, Akio K. and {Zackrisson}, Erik and {Christensen}, Lise and {Binggeli}, Christian and {Matsuda}, Yuichi and {Matsuo}, Hiroshi and {Takeuchi}, Tsutomu T. and {Asano}, Ryosuke S. and {Sunaga}, Kaho and {Shimizu}, Ikkoh and {Okamoto}, Takashi and {Yoshida}, Naoki and {Lee}, Minju M. and {Shibuya}, Takatoshi and {Taniguchi}, Yoshiaki and {Umehata}, Hideki and {Hatsukade}, Bunyo and {Kohno}, Kotaro and {Ota}, Kazuaki},
        title = "{Detection of the Far-infrared [O III] and Dust Emission in a Galaxy at Redshift 8.312: Early Metal Enrichment in the Heart of the Reionization Era}",
      journal = {\apj},
         year = 2019,
        month = mar,
       volume = {874},
       number = {1},
          eid = {27},
        pages = {27},
          doi = {10.3847/1538-4357/ab0374},
archivePrefix = {arXiv},
       eprint = {1806.04132},
 primaryClass = {astro-ph.GA},
       adsurl = {https://ui.adsabs.harvard.edu/abs/2019ApJ...874...27T}
}

@ARTICLE{Hashimoto18,
       author = {{Hashimoto}, Takuya and {Laporte}, Nicolas and {Mawatari}, Ken and {Ellis}, Richard S. and {Inoue}, Akio K. and {Zackrisson}, Erik and {Roberts-Borsani}, Guido and {Zheng}, Wei and {Tamura}, Yoichi and {Bauer}, Franz E. and {Fletcher}, Thomas and {Harikane}, Yuichi and {Hatsukade}, Bunyo and {Hayatsu}, Natsuki H. and {Matsuda}, Yuichi and {Matsuo}, Hiroshi and {Okamoto}, Takashi and {Ouchi}, Masami and {Pell{\'o}}, Roser and {Rydberg}, Claes-Erik and {Shimizu}, Ikkoh and {Taniguchi}, Yoshiaki and {Umehata}, Hideki and {Yoshida}, Naoki},
        title = "{The onset of star formation 250 million years after the Big Bang}",
      journal = {\nat},
         year = 2018,
        month = may,
       volume = {557},
       number = {7705},
        pages = {392-395},
          doi = {10.1038/s41586-018-0117-z},
archivePrefix = {arXiv},
       eprint = {1805.05966},
 primaryClass = {astro-ph.GA},
       adsurl = {https://ui.adsabs.harvard.edu/abs/2018Natur.557..392H}
}

@ARTICLE{Steidel92,
       author = {{Steidel}, Charles C. and {Hamilton}, Donald},
        title = "{Deep Imaging of redshift QSO Fields Below the Lyman Limit. I. The Field of Q0000-263 and galaxies at Z= 3.4}",
      journal = {\aj},
         year = 1992,
        month = sep,
       volume = {104},
        pages = {941},
          doi = {10.1086/116287},
       adsurl = {https://ui.adsabs.harvard.edu/abs/1992AJ....104..941S}
}

@ARTICLE{Inoue20,
       author = {{Inoue}, Akio K. and {Hashimoto}, Takuya and {Chihara}, Hiroki and {Koike}, Chiyoe},
        title = "{Radiative equilibrium estimates of dust temperature and mass in high-redshift galaxies}",
      journal = {\mnras},
         year = 2020,
        month = jun,
       volume = {495},
       number = {2},
        pages = {1577-1592},
          doi = {10.1093/mnras/staa1203},
archivePrefix = {arXiv},
       eprint = {2004.12612},
 primaryClass = {astro-ph.GA},
       adsurl = {https://ui.adsabs.harvard.edu/abs/2020MNRAS.495.1577I}
}

@ARTICLE{Castellano22,
       author = {{Castellano}, Marco and {Fontana}, Adriano and {Treu}, Tommaso and {Santini}, Paola and {Merlin}, Emiliano and {Leethochawalit}, Nicha and {Trenti}, Michele and {Vanzella}, Eros and {Mestric}, Uros and {Bonchi}, Andrea and {Belfiori}, Davide and {Nonino}, Mario and {Paris}, Diego and {Polenta}, Gianluca and {Roberts-Borsani}, Guido and {Boyett}, Kristan and {Brada{\v{c}}}, Maru{\v{s}}a and {Calabr{\`o}}, Antonello and {Glazebrook}, Karl and {Grillo}, Claudio and {Mascia}, Sara and {Mason}, Charlotte and {Mercurio}, Amata and {Morishita}, Takahiro and {Nanayakkara}, Themiya and {Pentericci}, Laura and {Rosati}, Piero and {Vulcani}, Benedetta and {Wang}, Xin and {Yang}, Lilan},
        title = "{Early Results from GLASS-JWST. III. Galaxy Candidates at z  9-15}",
      journal = {\apjl},
         year = 2022,
        month = oct,
       volume = {938},
       number = {2},
          eid = {L15},
        pages = {L15},
          doi = {10.3847/2041-8213/ac94d0},
archivePrefix = {arXiv},
       eprint = {2207.09436},
 primaryClass = {astro-ph.GA},
       adsurl = {https://ui.adsabs.harvard.edu/abs/2022ApJ...938L..15C}
}

@ARTICLE{Naidu22-LBG,
       author = {{Naidu}, Rohan P. and {Oesch}, Pascal A. and {van Dokkum}, Pieter and {Nelson}, Erica J. and {Suess}, Katherine A. and {Brammer}, Gabriel and {Whitaker}, Katherine E. and {Illingworth}, Garth and {Bouwens}, Rychard and {Tacchella}, Sandro and {Matthee}, Jorryt and {Allen}, Natalie and {Bezanson}, Rachel and {Conroy}, Charlie and {Labbe}, Ivo and {Leja}, Joel and {Leonova}, Ecaterina and {Magee}, Dan and {Price}, Sedona H. and {Setton}, David J. and {Strait}, Victoria and {Stefanon}, Mauro and {Toft}, Sune and {Weaver}, John R. and {Weibel}, Andrea},
        title = "{Two Remarkably Luminous Galaxy Candidates at z {\ensuremath{\approx}} 10-12 Revealed by JWST}",
      journal = {\apjl},
         year = 2022,
        month = nov,
       volume = {940},
       number = {1},
          eid = {L14},
        pages = {L14},
          doi = {10.3847/2041-8213/ac9b22},
archivePrefix = {arXiv},
       eprint = {2207.09434},
 primaryClass = {astro-ph.GA},
       adsurl = {https://ui.adsabs.harvard.edu/abs/2022ApJ...940L..14N}
}

@ARTICLE{Adams23,
       author = {{Adams}, N.~J. and {Conselice}, C.~J. and {Ferreira}, L. and {Austin}, D. and {Trussler}, J.~A.~A. and {Juod{\v{z}}balis}, I. and {Wilkins}, S.~M. and {Caruana}, J. and {Dayal}, P. and {Verma}, A. and {Vijayan}, A.~P.},
        title = "{Discovery and properties of ultra-high redshift galaxies (9 < z < 12) in the JWST ERO SMACS 0723 Field}",
      journal = {\mnras},
         year = 2023,
        month = jan,
       volume = {518},
       number = {3},
        pages = {4755-4766},
          doi = {10.1093/mnras/stac3347},
archivePrefix = {arXiv},
       eprint = {2207.11217},
 primaryClass = {astro-ph.GA},
       adsurl = {https://ui.adsabs.harvard.edu/abs/2023MNRAS.518.4755A}
}

@ARTICLE{Atek23,
       author = {{Atek}, Hakim and {Shuntov}, Marko and {Furtak}, Lukas J. and {Richard}, Johan and {Kneib}, Jean-Paul and {Mahler}, Guillaume and {Zitrin}, Adi and {McCracken}, H.~J. and {Charlot}, St{\'e}phane and {Chevallard}, Jacopo and {Chemerynska}, Iryna},
        title = "{Revealing galaxy candidates out to z   16 with JWST observations of the lensing cluster SMACS0723}",
      journal = {\mnras},
         year = 2023,
        month = feb,
       volume = {519},
       number = {1},
        pages = {1201-1220},
          doi = {10.1093/mnras/stac3144},
archivePrefix = {arXiv},
       eprint = {2207.12338},
 primaryClass = {astro-ph.GA},
       adsurl = {https://ui.adsabs.harvard.edu/abs/2023MNRAS.519.1201A}
}

@ARTICLE{Finkelstein22b,
       author = {{Finkelstein}, Steven L. and {Bagley}, Micaela B. and {Arrabal Haro}, Pablo and {Dickinson}, Mark and {Ferguson}, Henry C. and {Kartaltepe}, Jeyhan S. and {Papovich}, Casey and {Burgarella}, Denis and {Kocevski}, Dale D. and {Huertas-Company}, Marc and {Iyer}, Kartheik G. and {Koekemoer}, Anton M. and {Larson}, Rebecca L. and {P{\'e}rez-Gonz{\'a}lez}, Pablo G. and {Rose}, Caitlin and {Tacchella}, Sandro and {Wilkins}, Stephen M. and {Chworowsky}, Katherine and {Medrano}, Aubrey and {Morales}, Alexa M. and {Somerville}, Rachel S. and {Yung}, L.~Y. Aaron and {Fontana}, Adriano and {Giavalisco}, Mauro and {Grazian}, Andrea and {Grogin}, Norman A. and {Kewley}, Lisa J. and {Kirkpatrick}, Allison and {Kurczynski}, Peter and {Lotz}, Jennifer M. and {Pentericci}, Laura and {Pirzkal}, Nor and {Ravindranath}, Swara and {Ryan}, Russell E. and {Trump}, Jonathan R. and {Yang}, Guang and {Almaini}, Omar and {Amor{\'\i}n}, Ricardo O. and {Annunziatella}, Marianna and {Backhaus}, Bren E. and {Barro}, Guillermo and {Behroozi}, Peter and {Bell}, Eric F. and {Bhatawdekar}, Rachana and {Bisigello}, Laura and {Bromm}, Volker and {Buat}, V{\'e}ronique and {Buitrago}, Fernando and {Calabr{\`o}}, Antonello and {Casey}, Caitlin M. and {Castellano}, Marco and {Ch{\'a}vez Ortiz}, {\'O}scar A. and {Ciesla}, Laure and {Cleri}, Nikko J. and {Cohen}, Seth H. and {Cole}, Justin W. and {Cooke}, Kevin C. and {Cooper}, M.~C. and {Cooray}, Asantha R. and {Costantin}, Luca and {Cox}, Isabella G. and {Croton}, Darren and {Daddi}, Emanuele and {Dav{\'e}}, Romeel and {de La Vega}, Alexander and {Dekel}, Avishai and {Elbaz}, David and {Estrada-Carpenter}, Vicente and {Faber}, Sandra M. and {Fern{\'a}ndez}, Vital and {Finkelstein}, Keely D. and {Freundlich}, Jonathan and {Fujimoto}, Seiji and {Garc{\'\i}a-Argum{\'a}nez}, {\'A}ngela and {Gardner}, Jonathan P. and {Gawiser}, Eric and {G{\'o}mez-Guijarro}, Carlos and {Guo}, Yuchen and {Hamblin}, Kurt and {Hamilton}, Timothy S. and {Hathi}, Nimish P. and {Holwerda}, Benne W. and {Hirschmann}, Michaela and {Hutchison}, Taylor A. and {Jaskot}, Anne E. and {Jha}, Saurabh W. and {Jogee}, Shardha and {Juneau}, St{\'e}phanie and {Jung}, Intae and {Kassin}, Susan A. and {Le Bail}, Aur{\'e}lien and {Leung}, Gene C.~K. and {Lucas}, Ray A. and {Magnelli}, Benjamin and {Mantha}, Kameswara Bharadwaj and {Matharu}, Jasleen and {McGrath}, Elizabeth J. and {McIntosh}, Daniel H. and {Merlin}, Emiliano and {Mobasher}, Bahram and {Newman}, Jeffrey A. and {Nicholls}, David C. and {Pandya}, Viraj and {Rafelski}, Marc and {Ronayne}, Kaila and {Santini}, Paola and {Seill{\'e}}, Lise-Marie and {Shah}, Ekta A. and {Shen}, Lu and {Simons}, Raymond C. and {Snyder}, Gregory F. and {Stanway}, Elizabeth R. and {Straughn}, Amber N. and {Teplitz}, Harry I. and {Vanderhoof}, Brittany N. and {Vega-Ferrero}, Jes{\'u}s and {Wang}, Weichen and {Weiner}, Benjamin J. and {Willmer}, Christopher N.~A. and {Wuyts}, Stijn and {Zavala}, Jorge A. and {Ceers Team}},
        title = "{A Long Time Ago in a Galaxy Far, Far Away: A Candidate z {\ensuremath{\sim}} 12 Galaxy in Early JWST CEERS Imaging}",
      journal = {\apjl},
         year = 2022,
        month = dec,
       volume = {940},
       number = {2},
          eid = {L55},
        pages = {L55},
          doi = {10.3847/2041-8213/ac966e},
archivePrefix = {arXiv},
       eprint = {2207.12474},
 primaryClass = {astro-ph.GA},
       adsurl = {https://ui.adsabs.harvard.edu/abs/2022ApJ...940L..55F}
}

@ARTICLE{Gandolfi25,
       author = {{Gandolfi}, G. and {Rodighiero}, G. and {Bisigello}, L. and {Grazian}, A. and {Finkelstein}, S.~L. and {Dickinson}, M. and {Castellano}, M. and {Merlin}, E. and {Calabr{\`o}}, A. and {Papovich}, C. and {Bianchetti}, A. and {Ba{\~n}ados}, E. and {Benotto}, P. and {Buitrago}, F. and {Daddi}, E. and {Girardi}, G. and {Giulietti}, M. and {Hirschmann}, M. and {Holwerda}, B.~W. and {Arrabal Haro}, P. and {Lapi}, A. and {Lucas}, R.~A. and {Lyu}, Y. and {Massardi}, M. and {Pacucci}, F. and {P{\'e}rez-Gonz{\'a}lez}, P.~G. and {Ronconi}, T. and {Tarrasse}, M. and {Wilkins}, S. and {Vulcani}, B. and {Yung}, L.~Y.~A. and {Zavala}, J.~A. and {Backhaus}, B. and {Bagley}, M. and {Buat}, V. and {Burgarella}, D. and {Kartaltepe}, J. and {Khusanova}, Y. and {Kirkpatrick}, A. and {Kocevski}, D. and {Koekemoer}, A.~M. and {Lambrides}, E. and {Pirzkal}, N. and {Yang}, G.},
        title = "{Ultra High-Redshift or Closer-by, Dust-Obscured Galaxies? Deciphering the Nature of Faint, Previously Missed F200W-Dropouts in CEERS}",
      journal = {arXiv e-prints},
         year = 2025,
        month = feb,
          eid = {arXiv:2502.02637},
        pages = {arXiv:2502.02637},
          doi = {10.48550/arXiv.2502.02637},
archivePrefix = {arXiv},
       eprint = {2502.02637},
 primaryClass = {astro-ph.GA},
       adsurl = {https://ui.adsabs.harvard.edu/abs/2025arXiv250202637G}
}

@ARTICLE{Rodighiero23,
       author = {{Rodighiero}, Giulia and {Bisigello}, Laura and {Iani}, Edoardo and {Marasco}, Antonino and {Grazian}, Andrea and {Sinigaglia}, Francesco and {Cassata}, Paolo and {Gruppioni}, Carlotta},
        title = "{JWST unveils heavily obscured (active and passive) sources up to z   13}",
      journal = {\mnras},
         year = 2023,
        month = jan,
       volume = {518},
       number = {1},
        pages = {L19-L24},
          doi = {10.1093/mnrasl/slac115},
archivePrefix = {arXiv},
       eprint = {2208.02825},
 primaryClass = {astro-ph.GA},
       adsurl = {https://ui.adsabs.harvard.edu/abs/2023MNRAS.518L..19R}
}

@ARTICLE{Ishigaki15,
       author = {{Ishigaki}, Masafumi and {Kawamata}, Ryota and {Ouchi}, Masami and {Oguri}, Masamune and {Shimasaku}, Kazuhiro and {Ono}, Yoshiaki},
        title = "{Hubble Frontier Fields First Complete Cluster Data: Faint Galaxies at z \raisebox{-0.5ex}\textasciitilde 5-10 for UV Luminosity Functions and Cosmic Reionization}",
      journal = {\apj},
         year = 2015,
        month = jan,
       volume = {799},
       number = {1},
          eid = {12},
        pages = {12},
          doi = {10.1088/0004-637X/799/1/12},
archivePrefix = {arXiv},
       eprint = {1408.6903},
 primaryClass = {astro-ph.GA},
       adsurl = {https://ui.adsabs.harvard.edu/abs/2015ApJ...799...12I}
}

@ARTICLE{Kawamata16,
       author = {{Kawamata}, Ryota and {Oguri}, Masamune and {Ishigaki}, Masafumi and {Shimasaku}, Kazuhiro and {Ouchi}, Masami},
        title = "{Precise Strong Lensing Mass Modeling of Four Hubble Frontier Field Clusters and a Sample of Magnified High-redshift Galaxies}",
      journal = {\apj},
         year = 2016,
        month = mar,
       volume = {819},
       number = {2},
          eid = {114},
        pages = {114},
          doi = {10.3847/0004-637X/819/2/114},
archivePrefix = {arXiv},
       eprint = {1510.06400},
 primaryClass = {astro-ph.GA},
       adsurl = {https://ui.adsabs.harvard.edu/abs/2016ApJ...819..114K}
}

@ARTICLE{Zheng12,
       author = {{Zheng}, Wei and {Postman}, Marc and {Zitrin}, Adi and {Moustakas}, John and {Shu}, Xinwen and {Jouvel}, Stephanie and {H{\o}st}, Ole and {Molino}, Alberto and {Bradley}, Larry and {Coe}, Dan and {Moustakas}, Leonidas A. and {Carrasco}, Mauricio and {Ford}, Holland and {Ben{\'\i}tez}, Narciso and {Lauer}, Tod R. and {Seitz}, Stella and {Bouwens}, Rychard and {Koekemoer}, Anton and {Medezinski}, Elinor and {Bartelmann}, Matthias and {Broadhurst}, Tom and {Donahue}, Megan and {Grillo}, Claudio and {Infante}, Leopoldo and {Jha}, Saurabh W. and {Kelson}, Daniel D. and {Lahav}, Ofer and {Lemze}, Doron and {Melchior}, Peter and {Meneghetti}, Massimo and {Merten}, Julian and {Nonino}, Mario and {Ogaz}, Sara and {Rosati}, Piero and {Umetsu}, Keiichi and {van der Wel}, Arjen},
        title = "{A magnified young galaxy from about 500 million years after the Big Bang}",
      journal = {\nat},
         year = 2012,
        month = sep,
       volume = {489},
       number = {7416},
        pages = {406-408},
          doi = {10.1038/nature11446},
archivePrefix = {arXiv},
       eprint = {1204.2305},
 primaryClass = {astro-ph.CO},
       adsurl = {https://ui.adsabs.harvard.edu/abs/2012Natur.489..406Z}
}

@ARTICLE{Perez-Gonzalez25,
       author = {{P{\'e}rez-Gonz{\'a}lez}, Pablo G. and {{\"O}stlin}, G{\"o}ran and {Costantin}, Luca and {Melinder}, Jens and {Finkelstein}, Steven L. and {Somerville}, Rachel S. and {Annunziatella}, Marianna and {{\'A}lvarez-M{\'a}rquez}, Javier and {Colina}, Luis and {Dekel}, Avishai and {Ferguson}, Henry C. and {Li}, Zhaozhou and {Yung}, L.~Y. Aaron and {Bagley}, Micaela B. and {Boogaard}, Leindert A. and {Burgarella}, Denis and {Calabr{\`o}}, Antonello and {Caputi}, Karina I. and {Cheng}, Yingjie and {Dickinson}, Mark and {Eckart}, Andreas and {Giavalisco}, Mauro and {Gillman}, Steven and {Greve}, Thomas R. and {Hamed}, Mahmoud and {Hathi}, Nimish P. and {Hjorth}, Jens and {Huertas-Company}, Marc and {Kartaltepe}, Jeyhan S. and {Koekemoer}, Anton M. and {Kokorev}, Vasily and {Labiano}, {\'A}lvaro and {Langeroodi}, Danial and {Leung}, Gene C.~K. and {Natarajan}, Priyamvada and {Papovich}, Casey and {Peissker}, Florian and {Pentericci}, Laura and {Pirzkal}, Nor and {Rinaldi}, Pierluigi and {van der Werf}, Paul and {Walter}, Fabian},
        title = "{The Rise of the Galactic Empire: Ultraviolet Luminosity Functions at z {\ensuremath{\sim}} 17 and z {\ensuremath{\sim}} 25 Estimated with the MIDIS+NGDEEP Ultra-deep JWST/NIRCam Data Set}",
      journal = {\apj},
         year = 2025,
        month = oct,
       volume = {991},
       number = {2},
          eid = {179},
        pages = {179},
          doi = {10.3847/1538-4357/adf8c9},
archivePrefix = {arXiv},
       eprint = {2503.15594},
 primaryClass = {astro-ph.GA},
       adsurl = {https://ui.adsabs.harvard.edu/abs/2025ApJ...991..179P}
}

@ARTICLE{Willott24,
       author = {{Willott}, Chris J. and {Desprez}, Guillaume and {Asada}, Yoshihisa and {Sarrouh}, Ghassan T.~E. and {Abraham}, Roberto and {Brada{\v{c}}}, Maru{\v{s}}a and {Brammer}, Gabe and {Estrada-Carpenter}, Vince and {Iyer}, Kartheik G. and {Martis}, Nicholas S. and {Matharu}, Jasleen and {Mowla}, Lamiya and {Muzzin}, Adam and {Noirot}, Ga{\"e}l and {Sawicki}, Marcin and {Strait}, Victoria and {Rihtar{\v{s}}i{\v{c}}}, Gregor and {Withers}, Sunna},
        title = "{A Steep Decline in the Galaxy Space Density beyond Redshift 9 in the CANUCS UV Luminosity Function}",
      journal = {\apj},
         year = 2024,
        month = may,
       volume = {966},
       number = {1},
          eid = {74},
        pages = {74},
          doi = {10.3847/1538-4357/ad35bc},
archivePrefix = {arXiv},
       eprint = {2311.12234},
 primaryClass = {astro-ph.GA},
       adsurl = {https://ui.adsabs.harvard.edu/abs/2024ApJ...966...74W}
}

@ARTICLE{Oesch14,
       author = {{Oesch}, P.~A. and {Bouwens}, R.~J. and {Illingworth}, G.~D. and {Labb{\'e}}, I. and {Smit}, R. and {Franx}, M. and {van Dokkum}, P.~G. and {Momcheva}, I. and {Ashby}, M.~L.~N. and {Fazio}, G.~G. and {Huang}, J.-S. and {Willner}, S.~P. and {Gonzalez}, V. and {Magee}, D. and {Trenti}, M. and {Brammer}, G.~B. and {Skelton}, R.~E. and {Spitler}, L.~R.},
        title = "{The Most Luminous z \raisebox{-0.5ex}\textasciitilde 9-10 Galaxy Candidates Yet Found: The Luminosity Function, Cosmic Star-formation Rate, and the First Mass Density Estimate at 500 Myr}",
      journal = {\apj},
         year = 2014,
        month = may,
       volume = {786},
       number = {2},
          eid = {108},
        pages = {108},
          doi = {10.1088/0004-637X/786/2/108},
archivePrefix = {arXiv},
       eprint = {1309.2280},
 primaryClass = {astro-ph.CO},
       adsurl = {https://ui.adsabs.harvard.edu/abs/2014ApJ...786..108O}
}

@ARTICLE{Vanzella11,
       author = {{Vanzella}, E. and {Pentericci}, L. and {Fontana}, A. and {Grazian}, A. and {Castellano}, M. and {Boutsia}, K. and {Cristiani}, S. and {Dickinson}, M. and {Gallozzi}, S. and {Giallongo}, E. and {Giavalisco}, M. and {Maiolino}, R. and {Moorwood}, A. and {Paris}, D. and {Santini}, P.},
        title = "{Spectroscopic Confirmation of Two Lyman Break Galaxies at Redshift Beyond 7}",
      journal = {\apjl},
         year = 2011,
        month = apr,
       volume = {730},
       number = {2},
          eid = {L35},
        pages = {L35},
          doi = {10.1088/2041-8205/730/2/L35},
archivePrefix = {arXiv},
       eprint = {1011.5500},
 primaryClass = {astro-ph.CO},
       adsurl = {https://ui.adsabs.harvard.edu/abs/2011ApJ...730L..35V}
}

@ARTICLE{Zitrin15,
       author = {{Zitrin}, Adi and {Labb{\'e}}, Ivo and {Belli}, Sirio and {Bouwens}, Rychard and {Ellis}, Richard S. and {Roberts-Borsani}, Guido and {Stark}, Daniel P. and {Oesch}, Pascal A. and {Smit}, Renske},
        title = "{Lyman{\ensuremath{\alpha}} Emission from a Luminous z = 8.68 Galaxy: Implications for Galaxies as Tracers of Cosmic Reionization}",
      journal = {\apjl},
         year = 2015,
        month = sep,
       volume = {810},
       number = {1},
          eid = {L12},
        pages = {L12},
          doi = {10.1088/2041-8205/810/1/L12},
archivePrefix = {arXiv},
       eprint = {1507.02679},
 primaryClass = {astro-ph.GA},
       adsurl = {https://ui.adsabs.harvard.edu/abs/2015ApJ...810L..12Z}
}

@ARTICLE{Bouwens15,
       author = {{Bouwens}, R.~J. and {Illingworth}, G.~D. and {Oesch}, P.~A. and {Trenti}, M. and {Labb{\'e}}, I. and {Bradley}, L. and {Carollo}, M. and {van Dokkum}, P.~G. and {Gonzalez}, V. and {Holwerda}, B. and {Franx}, M. and {Spitler}, L. and {Smit}, R. and {Magee}, D.},
        title = "{UV Luminosity Functions at Redshifts z {\ensuremath{\sim}} 4 to z {\ensuremath{\sim}} 10: 10,000 Galaxies from HST Legacy Fields}",
      journal = {\apj},
         year = 2015,
        month = apr,
       volume = {803},
       number = {1},
          eid = {34},
        pages = {34},
          doi = {10.1088/0004-637X/803/1/34},
archivePrefix = {arXiv},
       eprint = {1403.4295},
 primaryClass = {astro-ph.CO},
       adsurl = {https://ui.adsabs.harvard.edu/abs/2015ApJ...803...34B}
}



\appendix
\section{The full spectra and continuum image of all six galaxies}
\label{appendix:full data}
\begin{figure*}
    \includegraphics[width=\linewidth]{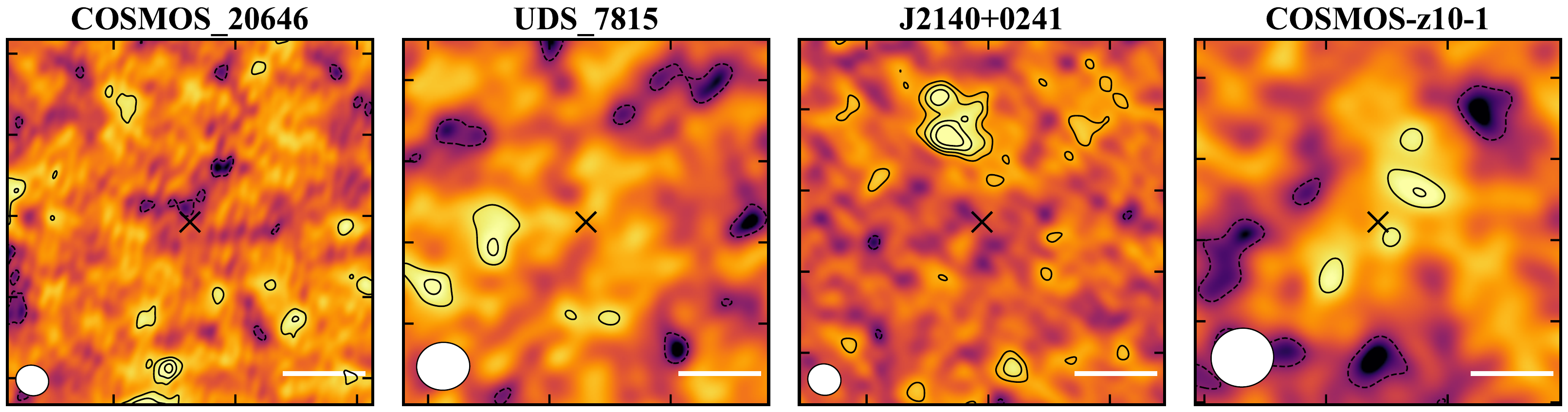}
    \caption{Continuum images of COSMOS\_20646, UDS\_7815, J2140+0241, and COSMOS-z10-1, along with the beam size (white ellipse) and the object position (black cross).
        Solid line contours show significance levels of $(2,3,4,5)\times\sigma$, while the $-2, -3\sigma$ contour is shown with the dashed lines.
    The scale bars in the lower-right corner indicate $1.0\,\mathrm{arcsec}$.
    }
    \label{fig:continuum_all}
\end{figure*}

\begin{figure*}
    \includegraphics[width=\linewidth]{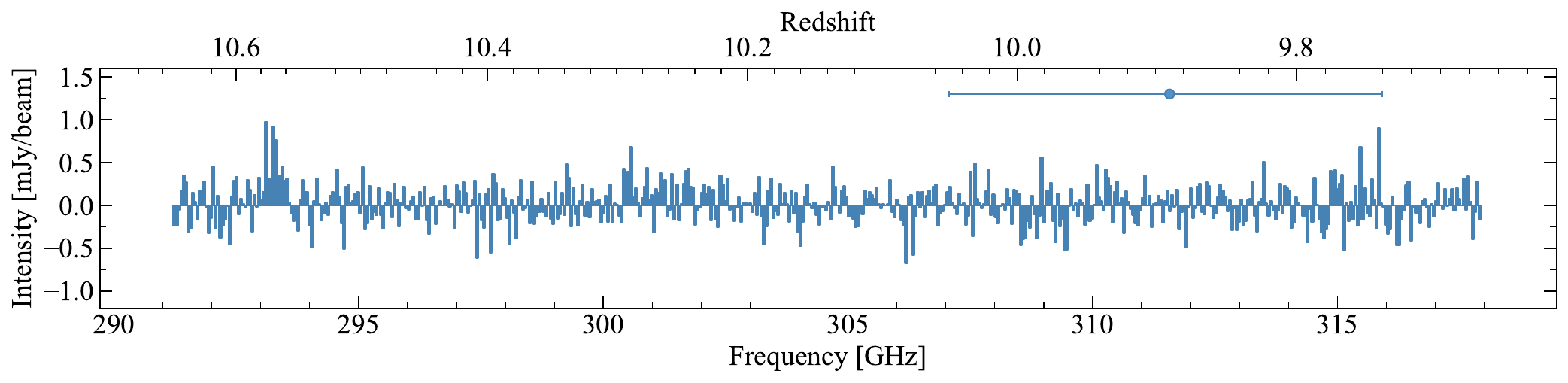}
    \caption{Continuum-subtracted full spectrum for UDS\_18697 together with photo-$z$ probability distribution from \citet{finkelstein22}.
    }
    \label{fig:full_spectrum_UDS18697}
\end{figure*}

\begin{figure*}
    \includegraphics[width=\linewidth]{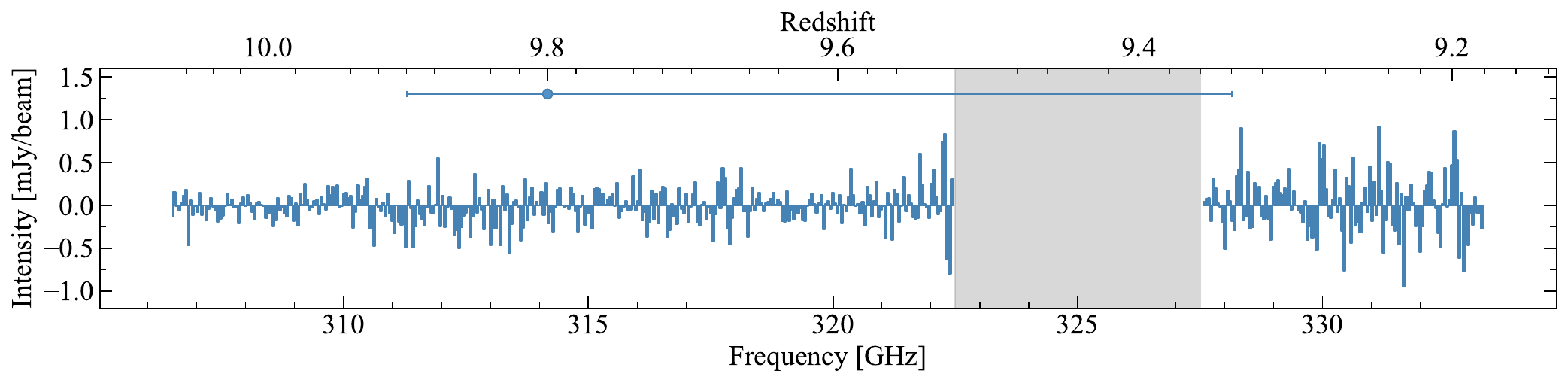}
   \caption{Same as the Figure~\ref{fig:full_spectrum_UDS18697} but for COSMOS\_20646.
   The spectrum between 322.5 and 327.5\,GHz is excluded because of strong atmospheric absorption, and this range is indicated by a grey shading.}
   \label{fig:full_spectrum_cosmos20646}
\end{figure*}

\begin{figure*}
    \includegraphics[width=\linewidth]{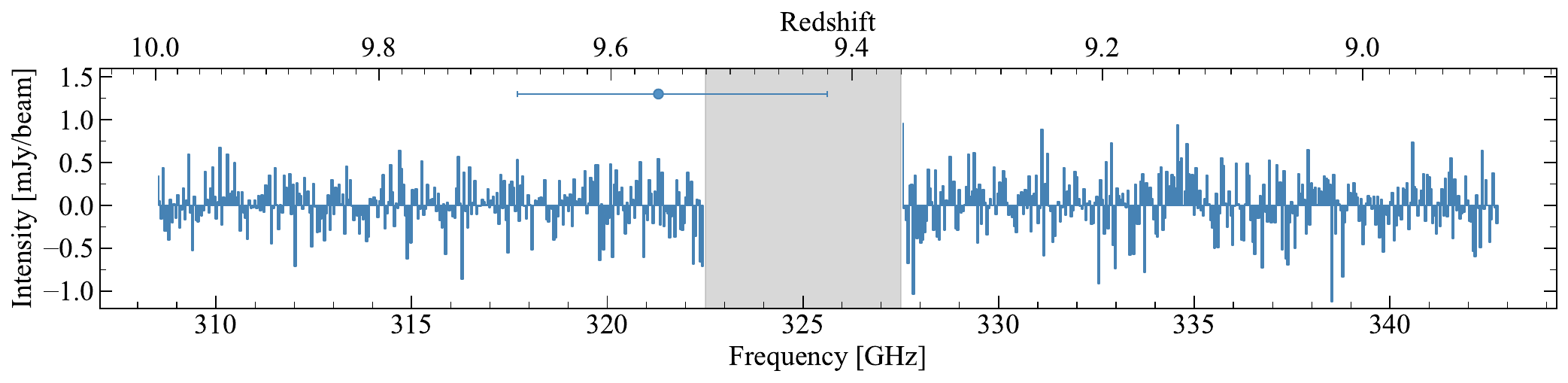}
    \caption{Same as the Figure~\ref{fig:full_spectrum_cosmos20646} but for COSMOS-z10-2.
    The photo-$z$ probability distribution is adopted from the COSMOS2020 catalogue.}
    \label{fig:full_spectrum_COSMOS-z10-2}
\end{figure*}

This appendix presents the continuum maps for COSMOS\_20646, UDS\_7815, J2140+0241, and COSMOS-z10-1 (Figure~\ref{fig:continuum_all}), as well as full-band spectra for all six targets (Figures~\ref{fig:full_spectrum_UDS18697}--\ref{fig:full_spectrum_COSMOS-z10-2}).
No continuum or line emission is detected in any source other than UDS\_18697 and COSMOS-z10-2.
Although a $\sim3\sigma$ signal is found near the position of COSMOS-z10-1, we do not regard it as a continuum emission from the source, because it is spatially offset from the reported object position and other signals with comparable significance level are also found within the same field of view.

\begin{figure*}
      \includegraphics[width=\linewidth]{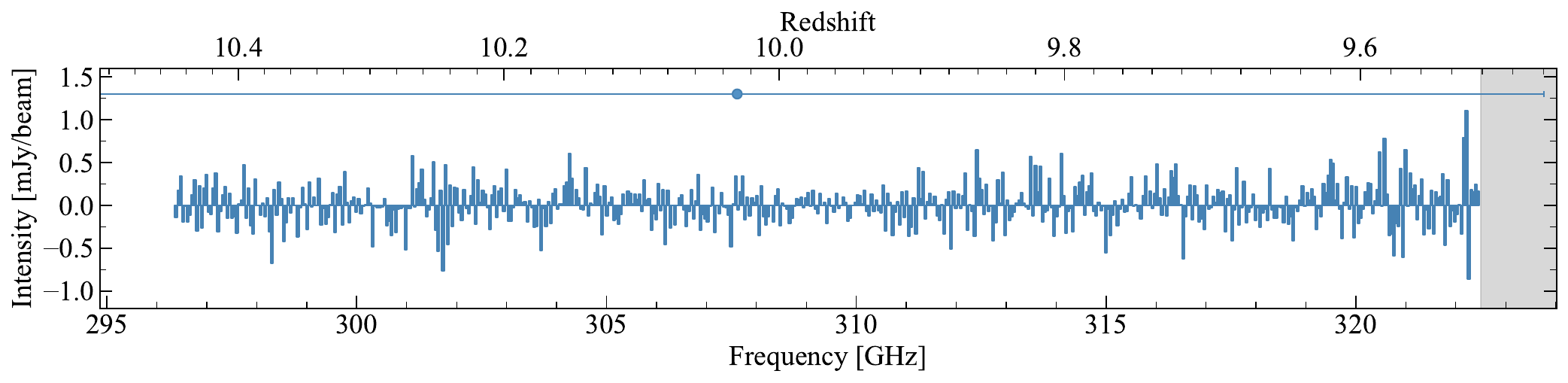}
   \caption{Same as the Figure~\ref{fig:full_spectrum_cosmos20646} but for UDS\_7815.}  
\end{figure*}

\begin{figure*}
    \includegraphics[width=\linewidth]{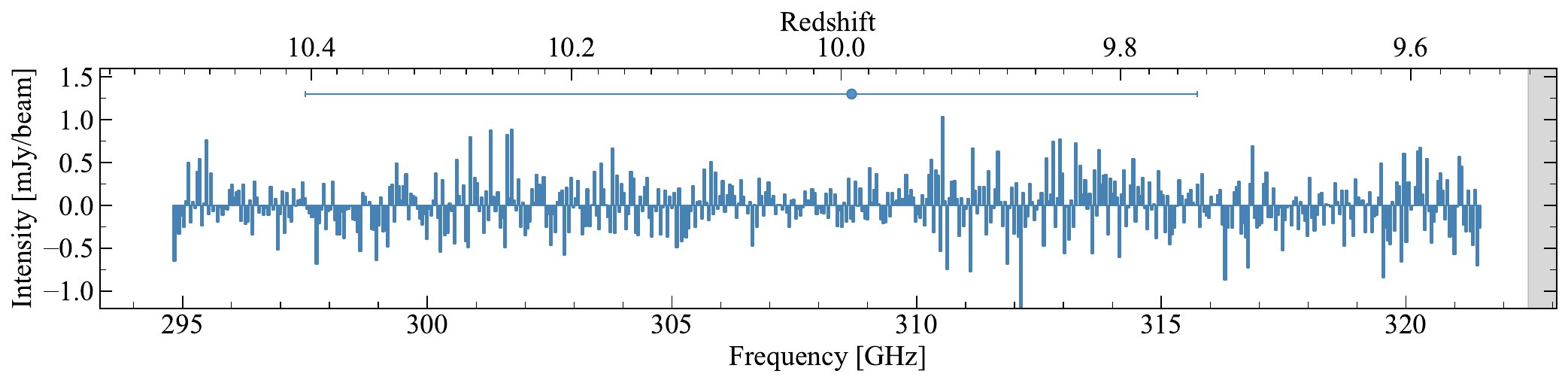}
    \caption{Same as the Figure~\ref{fig:full_spectrum_cosmos20646} but for J2140+0241.
    The photo-$z$ probability distribution is adopted from \citet{morishita18}.}
\end{figure*}

\begin{figure*}
    \includegraphics[width=\linewidth]{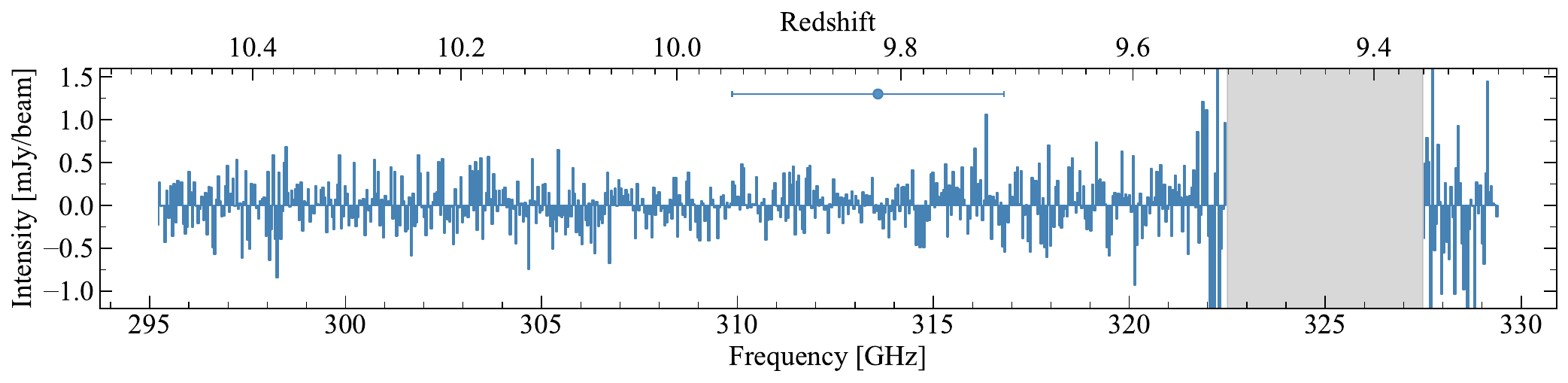}
    \caption{Same as the Figure~\ref{fig:full_spectrum_cosmos20646} but for COSMOS-z10-1.
    The photo-$z$ probability distribution is adopted from the COSMOS2020 catalogue.}
\end{figure*}

\section{Reliability of the line in UDS\_18697}
\label{appendix:line search}
To evaluate the reliability of the line features in UDS\_18697 and COSMOS-z10-2, we searched for clumps with the same or higher $S/N$ as the line features, as follows.
First, we created moment-0 maps over the entire continuum-subtracted cube using a velocity width corresponding to approximately twice the FWHM of the lines.
We then performed a line search over the full field of view (FoV) of each moment-0 map using the \texttt{find\_peaks} in photutils \citep{photutils}, and recorded the number of positive clumps $N_\mathrm{pos}$.
We also searched for peaks on the inverted moment-0 maps to count the number of spurious sources.
Since we expect no real sources with negative signals, the number of clumps in inverted maps ($N_\mathrm{neg}$) provides an estimate of the number of false positives.
The expected number of clumps within a circle from the target was then derived by scaling the number of clumps by the ratio of the area of the circle to the FoV area.
Radii of the circles were set to 1~arcsec for UDS\_18697 and 1.5~arcsec for COSMOS-z10-2, respectively, taking the larger beam size in COSMOS-z10-2 data into account.
\par
We obtained $N_\mathrm{pos}(r<1\,\mathrm{arcsec})=0.018$ and $N_\mathrm{neg}(r<1\,\mathrm{arcsec})=0.046$ for UDS\_18697, and $N_\mathrm{pos}(r<1.5\,\mathrm{arcsec})=0.10$ and $N_\mathrm{neg}(r<1.5\,\mathrm{arcsec})=0.05$ for COSMOS-z10-2.
Although the origin of the factor of $2\text{--}3$ difference between $N_\mathrm{neg}$ and $N_\mathrm{pos}$ remains unclear, it may be due to statistical fluctuations.
To evaluate the probability of obtaining spurious detection, we assume that the number of such signals follows a Poisson distribution.
This gives probabilities of $4.5\%$ for UDS\_18697 and $10\%$ for COSMOS-z10-2.
For UDS\_18697, given that the redshift inferred from the line is consistent with the NIRSpec redshift, we identify the line as the CO($J=9\text{--}8$) at $z=2.54$.
For COSMOS-z10-2, on the other hand, we do not claim the detection because the probability of a spurious signal is non-negligible, as well as the signal is spatially offset from the target position and the redshift remains uncertain.

\bsp	
\label{lastpage}
\end{document}